\documentclass[11pt]{article}
\usepackage{jheppub}
\usepackage[toc,page]{appendix}
\usepackage{graphicx}
\usepackage{subcaption}
\usepackage{amssymb, amsmath, amssymb}
\usepackage{slashed}
\usepackage{hyperref}
\hypersetup{colorlinks=true, allcolors=blue}
\usepackage{caption}
\usepackage{xcolor}
\usepackage{dsfont}
\usepackage{verbatim}
\usepackage{float}
\usepackage{physics}
\usepackage{tikz}
\usepackage{tikz-feynman}
\usepackage{multirow}
\usepackage{mathtools}
\usepackage{afterpage}

\title{Out-of-time-order correlators and Lyapunov exponents in sparse SYK}

\preprint{UTWI-21-2023, LLNL-JRNL-849296}

\author[1]{Elena C\'{a}ceres,}
\author[1,2]{Tyler Guglielmo,}
\author[1]{Brian Kent,}
\author[1]{Anderson Misobuchi}

\affiliation[1]{University of Texas, Austin, Physics Department, Austin TX 78712, USA}
\affiliation[2]{Lawrence Livermore National Laboratory, Livermore CA 94550, USA}

\emailAdd{elenac@utexas.edu, guglielmo2@llnl.gov, anderson.misobuchi@utexas.edu, brian\_kent@utexas.edu}

\abstract{We use a combination of analytical and numerical methods to study out-of-time order correlators (OTOCs) in the sparse Sachdev-Ye-Kitaev (SYK) model.  We find that at a given order of $N$,  the standard result for the $q$-local, all-to-all SYK,  obtained through the sum over ladder diagrams, is corrected by a series in the sparsity parameter, $k$. We present an algorithm to sum the diagrams at any given order of $1/(kq)^n$. We also study  OTOCs numerically as a function of the sparsity parameter and determine the Lyapunov exponent.  We find that numerical stability when extracting the Lyapunov exponent requires averaging over a massive number of realizations. This trade-off between the efficiency of the sparse model  and  consistent behavior at finite $N$ becomes more significant for larger values of $N$.}

\begin{document}
\maketitle
\section{Introduction}

Understanding the underlying mechanisms of the spread of information in strongly interacting quantum systems is a challenging problem that connects quantum information, condensed matter theory and quantum gravity. Out-of-time order correlators (OTOCs) provide a useful measure of how fast a local perturbation spreads across a quantum many-body system under time evolution \cite{Shenker_2014, Shenker:2014cwa, Maldacena:2015waa, Hosur:2015ylk,  Stanford:2015owe, Jensen:2016pah, Maldacena:2016hyu, Roberts:2016hpo, Swingle:2018ekw, Garcia-Garcia:2017bkg, Gu_2019, Fischler:2021rxy, Xu:2022vko, Garcia-Mata:2022voo}. Originally introduced as a semiclassical approach to superconductivity \cite{Larkin1969}, OTOCs experienced a renaissance with the discovery of a deep connection between quantum chaos and black holes, where black holes have been shown to be fast scramblers \cite{Shenker_2014, Sekino:2008he} and a bound on the rate of growth has been conjectured \cite{Maldacena:2015waa}. 

The Sachdev-Ye-Kitaev (SYK) model \cite{Kitaev:2015, Maldacena:2016hyu, Almheiri:2014cka}, a system of $N$  Majorana fermions with $q$-body interactions, {\it i.e.} $q$-local, has been extensively explored as a solvable model of quantum chaos and holography. At low temperatures, the system displays an approximate conformal symmetry and, as in generic black holes, it scrambles information at the fastest possible rate. Among its many variants, the sparse-SYK model \cite{Xu:2020shn, Garcia-Garcia:2020cdo} stands out as a simplified version with fewer interactions which allows one to probe finite $N$ corrections via numerical simulations. Remarkably, the sparse-SYK model with sparseness parameter $k$ and  only $k\,N$ interaction terms  in the Hamiltonian, is still a fast scrambler \cite{Xu:2020shn, Garcia-Garcia:2020cdo} and it can be shown to approximately reproduce properties of the original SYK model while allowing to uncover other features as in the spectral form factor studied in \cite{Caceres:2022kyr}. Additionally, the sparse-SYK model can serve as a replacement for the original SYK model when modeling  traversable wormholes \cite{Caceres:2021nsa, Jafferis:2022crx} including, remarkably, its simulation on a quantum processor \cite{Jafferis:2022crx}.

In this work, we  study out-of-time order correlators in the sparse-SYK model, both analytically and numerically. Even though  the sparse Hamiltonian contains fewer terms than the all-to-all SYK, there are  more diagrams contributing to the OTOCs at  order $\mathcal{O}(\frac{1}{N})$. We investigate ladder diagrams contributing to the OTOCs and  present an algorithm to sum the diagrams at any given order of $1/(kq)^n.$ However, an analytic expression for the sum of the diagrams  to all orders of $1/(kq)^n$ is out of reach because the methods  used in the all-to-all SYK are not applicable here.  Another of our main results is the numerical study of the OTOCs. We investigate them as a function of the sparsity parameter $k,$ with the goal of determining the Lyapunov exponent. We also explore to what extent the symmetry describing the large $N$ behavior of the OTOCs in SYK is present in the sparse model. 

\section{Preliminaries}
	
\subsection{Chaos and OTOCs}
\label{subsec:chaos}
 
Scrambling in quantum systems can be quantified via out-of-time order correlators (OTOCs).  In a chaotic system, OTOCs display an exponential growth at intermediate times characterized by a Lyapunov exponent $\lambda_L$, and the fast scrambling property translates into a Lyapunov exponent that saturates the chaos bound $\lambda_L\leq 2\pi/\beta$ \cite{Maldacena:2015waa}. The definition of OTOCs can be motivated as the quantum analog of the classical characterization of chaos using Poisson brackets, where the brackets are promoted to commutators (or anticommutators for fermionic operators) 
\begin{equation}
    C(t)=-\langle[W(t),V(0)]^2\rangle_\beta,
\end{equation}
where $V$ and $W$ are generic local Hermitian operators. The operators evolve according to the Hamiltonian $H$ of the system via $W(t)=e^{iHt}W(0)e^{-iHt}$. The brackets denote the thermal expectation value $\expval{\cdot}_\beta = Z^{-1}\text{tr}[.\,e^{-\beta H}]$ at inverse temperature $\beta=1/T$ and $Z=\text{tr}[e^{-\beta H}]$ is the partition function. Intuitively, this correlator measures how much an early perturbation $V$ affects the later measurement of $W$. After the expansion of the commutator squared, we see the appearance of a four-point correlator that is out-of-time ordered
\begin{equation}
    F^{(u)}(t) = \langle W(t)V(0)W(t)V(0)\rho\rangle_\beta,
\end{equation}
where $\rho=e^{-\beta H}$. In the SYK model, we typically consider $W$ and $V$ as two distinct Majorana operators. The above expression is the standard OTOC that is also referred to in the literature as `unregularized' OTOC. In the study of quantum chaos at finite $N$, a slightly modification of this correlator have been considered, named `regularized' OTOC
\begin{equation}
    F^{(r)}(t) = \langle W(t)\rho^{1/4}V(0)\rho^{1/4}W(t)\rho^{1/4}V(0)\rho^{1/4}\rangle_\beta,
\end{equation}
in which we split $\rho$ evenly between the operators. One of the reasons to work with the regularized version is because this version is less sensitive to finite-size effects, and it can be shown that the extracted Lyapunov exponent is the same as in the unregularized version \cite{Kobrin:2020xms}.

From the OTOC, one could derive the Lyapunov exponent by fitting an exponential function $a+b\,e^{\lambda_\text{fit}t}$ to the region of exponential growth. However, this method is known to lead to incorrect results in the SYK model for several reasons. The main subtlety is that, due to strong finite-size effects, the OTOC does not display a well-defined exponential growth window at intermediate times and finite size $N$. An alternative method was proposed \cite{Kobrin:2020xms}, where it is assumed that the OTOC admits the following large $N$ expansion based upon analytical arguments \cite{Stanford:2015owe, Gu_2019}
\begin{equation} \label{eq:otoc_expansion}
    F(t) = C_0 + C_1\left(\frac{e^{\lambda_L t}}{N}\right) + C_2\left(\frac{e^{\lambda_L t}}{N}\right)^2 +\ldots
\end{equation}
for $t\lesssim \frac{1}{\lambda_L}\log N$, which means that $F(t)$ obeys the rescaling symmetry
\begin{equation}
    N\to c\,N,\qquad t\to t+\frac{1}{\lambda_L}\log c.
\end{equation}
This symmetry is expected to hold for any many-body chaotic model governed by ladder diagrams close to the semiclassical limit.
    
\subsection{Sparse SYK}

The Sachdev-Ye-Kitaev (SYK) model \cite{Kitaev:2015, Maldacena:2016hyu, Almheiri:2014cka} is a toy model of lower dimensional quantum black holes consisting of a system of Majorana fermions with all-to-all interactions. The sparse-SYK model  \cite{Xu:2020shn, Garcia-Garcia:2020cdo} is a variant of the original model, which we will refer to as all-to-all SYK, with the advantage of allowing for more efficient computer simulation while preserving important black hole physics behavior. The Hamiltonian of the sparse-SYK model with $N$ Majorana fermions $\chi_j$, $j=1,\ldots, N$, and $q$-body interactions is defined as
\begin{align} \label{eq:H_single}
	H & = i^{q/2}\sum_{j_1<\ldots <j_q}J_{j_1\ldots j_q}x_{j_1\ldots j_q}\chi_{j_1}\ldots\chi_{j_q}.
\end{align}
The parameters $x_{j_1...j_q}$ are either 0 or 1 and they can be defined in different ways leading to different sparse models. The couplings $J_{j_1...j_q}$ are drawn from a Gaussian distribution with zero mean and variance given by\footnote{Sparse models with non-Gaussian couplings have also been considered \cite{Tezuka:2022mrr}.}
\begin{equation} 
	\langle \left(J_{j_1\ldots j_q}\right)^2\rangle = \frac{(q-1)!J^2}{pN^{q-1}},
\end{equation}
where $p$ is the fraction of terms in the Hamiltonian, i.e., the number of the terms such that $x_{j_1...j_q}=1$ divided by $\binom{N}{q}$, the number of terms in the all-to-all version. The parameter $J$ has dimensions of energy and sets the energy scale of the theory. It will be convenient to define the parameter
\begin{equation} \label{eq:k}
    k \equiv \frac{p}{N}\binom{N}{q},
\end{equation}
which is such that the Hamiltonian is a sum of $kN$ independent terms. Due to the random nature of the couplings, physical observables are obtained after performing an average over different disorder realizations of the system. If we choose all $x_{j_1\ldots j_q}$ to be 1, we recover the original all-to-all SYK Hamiltonian
\begin{equation} \label{eq:hamiltonian_all}
    H_\text{all-to-all} = i^{q/2}\sum_{1\leq j_1<\ldots<j_q\leq N}J_{j_1\ldots j_q}\chi_{j_1}\ldots\chi_{j_q}.
\end{equation}
In this case we have $p=1$ and we also recover the variance of the all-to-all SYK model
\begin{equation}
	\langle \left(J_{j_1\ldots j_q}\right)^2\rangle_\text{all-to-all} = \frac{(q-1)!J^2}{N^{q-1}}.
\end{equation}
One possible implementation of sparseness in the SYK model consists of taking $x_{ijkl}=1$ with probability $p$ and $x_{ijkl}=0$ with probability $1-p$. The system becomes more sparse as $p\to 0$ and we recover the all-to-all SYK when $p=1$. This procedure, known as \emph{random pruning}, is perhaps the simplest to implement. However, it makes necessary to take an additional average since $x_{ijkl}$ are treated as random variables and can potentially lead to disconnected clusters of Majorana fermions that do not interact with the rest of the system \cite{Garcia-Garcia:2020cdo}. 

\begin{figure}
    \centering
    \includegraphics[width=0.4\linewidth]{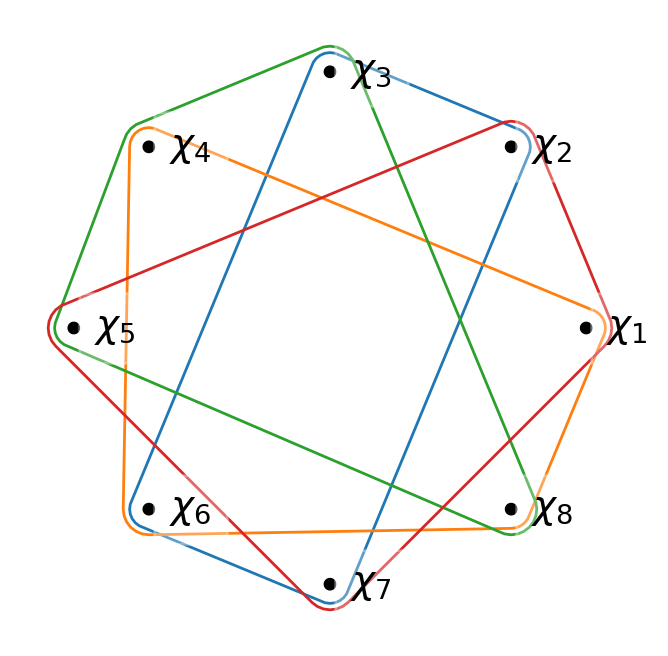}
    \caption{\textit{A concrete regular hypergraph representation of the sparse-SYK Hamiltonian with $N=8$, $q=4$, and $k=1/2$ with hyperedges $=\{(2, 3, 6, 7)$, $(1, 4, 6, 8)$, $(3, 4, 5, 8)$, $(1, 2, 5, 7)\}$. Each vertex is contained in exactly $kq=2$ hyperedges.}}
    \label{fig:hyper_regular}
\end{figure}

A different approach to characterize the sparseness is to use {\it regular hypergraphs}, in which the Majorana fermions are identified as vertices of a hypergraph whose hyperedges contain $q$ vertices. The regularity condition imposes that each fermion appears exactly the same number of times in the list of hyperedges. The hypergraph is also uniform, meaning that all hyperedges contain the same number, $q$, of vertices. An important property of random-uniform-regular hypergraphs is that they are expected to be {\it expanders} \cite{Dumitriu2021}. That is, they are expected to have good connectivity even though they are sparse. Note  that the $q$-uniformity condition is related to the Hamiltonian being $q$-local but the regularity condition is imposed for convenience. A simple counting argument shows that the hypergraph should be $k q$-regular. Measures of connectivity of this type of hypergraph in the sparse-SYK context were studied in \cite{Caceres:2021nsa}. An example of a regular hypergraph is given in Figure \ref{fig:hyper_regular}.

\section{OTOCs in Sparse SYK}\label{sec:otocs_sparse_syk}

In \cite{Maldacena:2016hyu, Gu_2019}, it was shown that, in the all-to-all SYK, to obtain the leading order in $N$ and late time contributions to the OTOC, we should calculate ladder diagrams of dressed propagators. This is also true for the sparse model. In this section, we will present an analytic calculation of ladder diagrams in the sparse-SYK. 
We find that the  standard contribution obtained through the sum over ladder diagrams is corrected, in the sparse-SYK, by  higher order terms of order $1/(kq)^n$. In the all-to-all SYK  the ladder diagrams can be summed exactly, giving an expression for the OTOC from which a Lyapunov exponent can be computed \cite{Maldacena:2016hyu}.  The same ressumation does not go through in the case of the sparse model and an analytic expression for the ladder diagrams to all orders of $1/(kq)^n$ cannot be obtained. However, we found  an algorithm  to sum the diagrams up to any given order of $1/(kq)^n$ that we present in this section. The intricacies of the procedure are in Appendix \ref{app:algorithm}.

\subsection{Ladder Diagrams}\label{subsec:ladder}
        
As a first step, we will  analyze the combinatorics related to ladder diagrams in a sparse-SYK model with random pruning \cite{Xu:2020shn}. In this model, each interaction term in the Hamiltonian comes with a factor, $x_{abcd}$, which is either $0$ or $1$ with probability $p$.  This is telling us that each term in the Hamiltonian is deleted with probability $1-p$. In this model, we normalize the couplings to have a $p$ dependent variance,
\begin{align} \label{eq:variance}
    \expval{J_{abcd}^2} &= \frac{(q-1)! J^2}{pN^{q-1}},
\end{align}
where after averaging, the multiplication of two couplings is only non-zero if their indices are equal (up to permutations of one another). As in the standard SYK model, the ladder diagrams will be the diagrams that contribute to the OTOC \cite{Stanford:2015owe,Maldacena:2016hyu,Gu_2019},
\begin{align} \label{eq:OTOC}
    \mathcal{F}(t_1,t_2,t_3,t_4) = \frac{1}{N^2} \Tr(\rho^{1/4} \chi_i(t_1) \rho^{1/4} \chi_j(t_2) \rho^{1/4} \chi_i(t_3) \rho^{1/4} \chi_j(t_4)),
\end{align}
at leading order in late time $t$.  We will now proceed by showing how the ladder diagrams can be summed at zero temperature -- although the OTOC is clearly a finite temperature quantity, the zero temperature calculation will be sufficient for showing the general procedure of ladder diagram summation. The zero temperature calculation can be extended to finite temperatures in a fairly straightforward manner by writing retarded propagators along the real time portion of a Schwinger-Keldish contour and Wightman functions between the two real parts of the contour \cite{Stanford:2015owe,Maldacena:2016hyu,Fischler:2021rxy}.  

We find, as in \cite{Xu:2020shn}, a series in both $1/N$ and $1/(kq)$.  This means that at each order in $1/N$ we will find a series expression for a prefactor that is a proportionality constant for each diagram in $1/(kq)$. This series terminates at an order corresponding to the order of the ladder diagram being considered. At zeroth order in $1/(kq)$, we have the familiar diagrams of the dense model. But at higher orders, there are corrections organized in a $1/(kq)$ expansion.  To leading order in $N$, we find that a ladder diagram of $\mathcal{O}(J^{2n})$ (think a ladder with $n$ rungs) will contain a series in $1/(kq)$ up to order $1/(kq)^{n-1}$, i.e.
\begin{equation} \label{eq:prefactorGeneral}
    \begin{tikzpicture}[baseline=(current bounding box.center)]
        \begin{feynman}
            \vertex (ul);
            \vertex[below=1.3cm of ul] (ll);
            \vertex[right=1cm of ul] (um);
            \vertex[right=1cm of ll] (lm);
            \vertex[right=1.75cm of um] (ur);
            \vertex[right=1.75cm of lm] (lr);
            
            \diagram* [] {
                (ul) -- (um) -- (ur),
                (ll) -- (lm) -- (lr),
                (lm) -- [bend left] (um),
                (lm) -- [scalar, in=210, out=150] (um),
                (lm) -- [bend right,edge label'=\(\cdots\times n\)] (um),
            };
        \end{feynman}
    \end{tikzpicture}
    ~\equiv 
    \mathcal{F}_n \sim \frac{(-1)^n (q-1)^n J^{2n}}{N} \left(1 + \frac{c_1}{kq} + \frac{c_2}{(kq)^2} + \cdots + \frac{c_{n-1}}{(kq)^{n-1}}\right).
\end{equation}
The dashed line implies that the diagram is disorder averaged. For the rest of this section, dashed lines will be dropped from diagrams, and should be assumed. 

Let us now write down the first few terms of the ladder diagram sum
\begin{equation}
    \begin{tikzpicture}[baseline=(current bounding box.center)]
        \begin{feynman}
            \vertex (ul) {\(\tau_1, i\)};
            \vertex[below=1.3cm of ul] (ll) {\(\tau_2, i\)};
            \vertex[right=1cm of ul] (um);
            \vertex[right=1cm of ll] (lm);
            \vertex[right=1.75cm of um] (ur) {\(\tau_3, i\)};
            \vertex[right=1.75cm of lm] (lr) {\(\tau_4, i\)};
            
            \diagram* [] {
                (ul) -- (um) -- (ur),
                (ll) -- (lm) -- (lr),
            };
        \end{feynman}
    \end{tikzpicture}
+
    \begin{tikzpicture}[baseline=(current bounding box.center)]
        \begin{feynman}
            \vertex (ul) {\(i\)};
            \vertex[below=1.3cm of ul] (ll) {\(i\)};
            \vertex[right=1.2cm of ul] (um);
            \vertex[right=1.2cm of ll] (lm);
            \vertex[right=1.2cm of um] (ur) {\(j\)};
            \vertex[right=1.2cm of lm] (lr) {\(j\)};
            
            \diagram* [] {
                (ul) -- (um) -- (ur),
                (ll) -- (lm) -- (lr),
                (lm) -- [bend left, edge label = \(a\)] (um),
                (lm) -- [bend right, edge label' = \(b\)] (um),
            };
        \end{feynman}
    \end{tikzpicture}
    ~+
    \begin{tikzpicture}[baseline=(current bounding box.center)]
        \begin{feynman}
            \vertex (ul) {\(i\)};
            \vertex[below=1.3cm of ul] (ll)  {\(i\)};
            \vertex[right=1.2cm of ul] (um);
            \vertex[right=1.2cm of ll] (lm);
            \vertex[right=1.3cm of um] (um2);
            \vertex[right=1.3cm of lm] (lm2);
            \vertex[right=1.2cm of um2] (ur) {\(j\)};
            \vertex[right=1.2cm of lm2] (lr) {\(j\)};
            
            \diagram* [] {
                (ul) -- (um) -- [edge label = \(c\)] (um2) -- (ur),
                (ll) -- (lm) -- [edge label' = \(c'\)] (lm2) -- (lr),
                (lm) -- [bend left, edge label = \(a\)] (um),
                (lm) -- [bend right, edge label' = \(b\)] (um),
                (lm2) -- [bend left, edge label = \(a'\)] (um2),
                (lm2) -- [bend right, edge label' = \(b'\)] (um2),
            };
        \end{feynman}
    \end{tikzpicture}
+\cdots.
\end{equation}
The first two diagrams are given by
\begin{align}
    \mathcal{F}_0 &= \frac1N \left[-G(\tau_{13})G(\tau_{24}) + G(\tau_{14})G(\tau_{23})\right]\\
    \mathcal{F}_1 &= \frac{(q-1) J^2}{N} \int d\tau d\tau' \left[G(\tau_1 - \tau)G(\tau_2 - \tau')G^{q-2}(\tau - \tau')G(\tau-\tau_3) G(\tau' - \tau_4) - (\tau_3 \leftrightarrow \tau_4)\right], 
\end{align}
where $G(\tau)$ is the Euclidean-time propagator, i.e. the time-ordered product
\begin{align}
    G(\tau) &= \expval{\mathcal{T}(\chi(\tau)\chi(0))}.
\end{align}

Let us now focus on the combinatoric factor that appears in front of the integral, and is given generally by (\ref{eq:prefactorGeneral}).  First, let us look at the overall factor outside of the parenthesis in (\ref{eq:prefactorGeneral}).  The $(-1)^n$ comes from the addition of a fermion loop at each successive order.  The $(q-1)^n$ is a combinatoric factor coming from the number of ways Wick contractions can be done on the interaction vertices. Inside the parenthesis, we can find each term by considering the ways $\expval{J_{ia_1b_1c_1}J_{ia_1b_1c_2}\cdots}$ can be contracted at leading order in $N$. Focusing on $\mathcal{F}_1$, we have $\expval{J_{iabj}J_{iabj}}$ which gives us back \eqref{eq:variance} a priori, except now there is a sum over $i,j,a,b$ which will give a factor of $N^q$. We will also obtain a factor of $p$ since, $x_{iabj}x_{iabj} = x_{iabj}$ and $\expval{x_{iabj}} = p$. Ignoring the Wick contraction combinatorics, which amounts to the $(q-1)$ factor above, we have
\begin{align}
    \frac{1}{N^2}\frac{J^2}{p N^{q-1}} N^{q} p = \frac{J^2}{N}.
\end{align}
This is the $1$ inside the parenthesis in \eqref{eq:prefactorGeneral}. Every diagram in the sum will contribute the same $1$, however the terms of order $1/(kq)$ and up will change based on the order of perturbation theory we are working in.

The two rung diagram will give
\begin{align} 
    \expval{J_{iabc}J_{iabc'}J_{ja'b'c}J_{ja'b'c'}} &= \expval{J_{iabc}J_{iabc'}} \expval{J_{ja'b'c}J_{ja'b'c'}} + \expval{J_{iabc}J_{ja'b'c}} \expval{J_{iabc'}J_{ja'b'c'}} \nonumber \\
    & \qquad\qquad + \expval{J_{iabc}J_{ja'b'c'}} \expval{J_{iabc'}J_{ja'b'c}}. 
    \label{2ndordercalc}
\end{align}
There are now two possibilities to consider. The first is when $ja'b'$ is not a permutation of $iab$.  In this case, the second two terms in the cumulant expansion vanish while the first will give us a factor of $1$ as above.  When $ja'b'$ is a permutation of $iab$, then the combinatoric pre-factor will be
\begin{align}
    3 \frac{1}{N^2}\frac{J^4}{p^2 N^{2q-2}} N^{q}(q-1)! ~p = 3\frac{J^4 (q-1)!}{N N^{q-1} p} = \frac{J^4}{N}\frac{3}{kq}.
\end{align}
where the $3$ is from the three different terms in the sum. The last equality can be seen from the identity $p = kq!/N^{q-1}$.  So, we see that the one-rung diagram has two terms that contribute at leading order in $N$. One that is $k$ independent and one that is of order $1/(kq)$.  We can proceed in this way for higher order rung diagrams order by order and find a series in $1/(kq)$ at leading order in $N$.  At each order, care must be taken to consider which indices are permutations of the others and in how many ways they can be made permutations of each other.  The procedure is quite tedious; we have worked out the first six pre-factors and present them below for the reader.
\begin{align}\label{expansions}
    \mathcal{F}_1 &\sim 1\\
    \mathcal{F}_2 &\sim 1 + \frac{3}{kq}\\
    \mathcal{F}_3 &\sim 1 + \frac{9}{kq} + \frac{15}{(kq)^2}\\
    \mathcal{F}_4 &\sim 1 + \frac{18}{kq} + \frac{90}{(kq)^2} + \frac{105}{(kq)^3}\\
    \mathcal{F}_5 &\sim 1 + \frac{30}{kq} + \frac{300}{(kq)^2} + \frac{1035}{(kq)^3} + \frac{945}{(kq)^4}\\
    \mathcal{F}_6 &\sim 1 + \frac{45}{kq} + \frac{750}{(kq)^2} + \frac{5205}{(kq)^3} + \frac{13675}{(kq)^4} + \frac{10395}{(kq)^5}.
\end{align}
A general expression for the first two terms and the last term in the series is shown below
\begin{equation}
    \mathcal{F}_n \sim \frac{(q-1)^n J^{2n}}{N} \left(1 + \frac32 n (n-1) \frac{1}{kq} + \frac54 n(n-1)^2(n-2)\frac{1}{(kq)^2} + \cdots + \frac{(2 n)!}{2^n n!} \frac{1}{(kq)^{n-1}}\right).
\end{equation}
In Appendix \ref{app:algorithm}, we present  a straightforward and general algorithm for calculating a given coefficient to arbitrary order.

We would now like to make an attempt at summing the ladder diagrams.  Since we do not have an analytic expression for the ladder diagram to all orders in $J^2$ and $1/(kq)$, we choose to work at order $1/(kq)^2$. We can move between successive ladders (adding rungs to the zero-order diagram) through multiplication of a kernel $\mathcal{K}_n$,
\begin{align} \label{eq:kernelNextOrder}
    \mathcal{F}_{n}(\tau_1,\tau_2,\tau_3,\tau_4) &= \int d\tau d\tau'~ \mathcal{K}_n(\tau_1,\tau_2; \tau, \tau') \mathcal{F}_0(\tau, \tau'; \tau_3,\tau_4)\\
    &= \int d\tau d\tau'~ K^n \left(1 + \frac32 n (n-1) \frac{1}{kq} + \frac54 n(n-1)^2(n-2)\frac{1}{(kq)^2}\right) \mathcal{F}_0,
\end{align}
where 
\begin{align} 
    K(\tau_1,\tau_2; \tau_3,\tau_4) = -(q-1) J^{2} G(\tau_{13})\,G(\tau_{24})\,G^{q-2}(\tau_{34}).
\end{align}
We can now attempt to perform the sum over all ladders by viewing the integral operation in (\ref{eq:kernelNextOrder}) as matrix multiplication. We obtain
\begin{align}
    \mathcal{F} = \sum_{n=0}^\infty \mathcal{F}_n & = \sum_{n=0}^\infty K^n \left(1 + \frac32 n (n-1) \frac{1}{kq} + \frac54 n(n-1)^2(n-2)\frac{1}{(kq)^2}\right) \mathcal{F}_0 \nonumber \\
    &= \frac{1}{1 - K}\left(1 + \frac{3K^2}{kq (1 - K)^2} + \frac{15 K^3(1 + K)}{(kq)^2(1 - K)^4}\right) \mathcal{F}_0.
\end{align}
In all-to-all SYK, the sum of all ladder diagrams at early times can be shown to scale as $\mathcal{F} \sim e^{\lambda_L t}/N$. For large $kq$ the sum of ladder diagrams is the same as the all-to-all case, as expected. However, it is unclear how the corrections of the form $1/kq$ will contribute to the exponential behavior of $\mathcal{F}$.

\subsection{Diagram Counting Intuition and Hypergraphs}

The importance of ladder diagrams in all-to-all SYK is because they contain all $\mathcal{O}(\frac{1}{N})$ contributions to the four-point function. As we showed in the previous section, in the sparse model  there are additional contributions (compared to all-to-all) of the form $\frac{1}{kq}$. This implies that \textit{sparse-SYK has more diagrams at} $\mathcal{O}(\frac{1}{N})$. This may seem counterintuitive at first since sparse-SYK \textit{decreases} the number of couplings, however, an analysis of how the couplings themselves change elucidates this problem. Note that we have redefined the variance in the pruning model to be
\begin{equation}
    \expval{J_{ijkl}^2} = \frac{(q-1)! J^2}{pN^{q-1}} \approx \frac{J^2}{kq},
\end{equation}
where we have used $k = p/N \binom{N}{q} \approx p N^{q-1}/q!$. Therefore, the contributions of $\mathcal{O}(\frac{1}{N})$ no longer depend on the number of couplings $J_{ijkl}$, rather the $\mathcal{O}(\frac{1}{N})$ counting moves to the pruning coupling $x_{ijkl}$
\begin{equation}
    \expval{x_{ijkl}} = p \approx \frac{kq!}{N^{q-1}} = \expval{x_{ijkl}^n},
\end{equation}
where $n$ is any positive integer. Intuitively, the right equality follows because if $x_{ijkl}=1$, then the couplings with labels $\{i,j,k,l\}$ exist in the Hamiltonian. Hence, a diagram vertex representing this coupling is non-zero, so we can insert these vertices an arbitrary number of times and the diagram will still exist in the theory. In other words, if couplings with indices $\{i,j,k,l\}$ exist with probability $p$, then the probability of the same couplings contributing twice is not less likely (such as $p^2$), because said couplings already exist in the theory. Another version of this statement is that since $x_{ijkl}$ can only be 1 or 0, $1^n = 1$ and $0^n = 0$. Additionally, any non-zero term will have the same $1/kq$ contribution from the $J$'s, specifically the $n$th order ladder diagram will have $1/(kq)^n$ \textit{from the $J$'s}. Therefore, the order of any specific term in $\mathcal{O}(1/kq)$ is determined by how many $p$'s (from the pruning couplings) contribute in that term. Hence, the $1/kq$ counting for any given term is also determined by the sparsity couplings $x_{ijkl}$.

In summary, this means that the $\mathcal{O}(\frac{1}{N})$ counting moves to the $x_{ijkl}$ couplings, and since these behave differently from the $J_{ijkl}$ couplings, the contributions of $\mathcal{O}(\frac{1}{N})$ should change in the sparse pruning model.

An open question remains: we have performed this analysis using the random pruning method, yet the preferable numerical choice appears to be the hypergraph method, do we expect the ladder diagrams in the hypergraph method to yield the same expansion in $1/kq$? While we cannot comment on the expected behavior for a fixed hypergraph configuration as any given diagram would be dependent on which couplings are present, we also find in the subsequent section that a fixed hypergraph is insufficient to extract the Lyapunov exponent using the known numerical methods for the SYK model. It is only upon averaging over hypergraphs that the Lyapunov exponent can be extracted accurately, it is important to note that averaging over random pruning and averaging over hypergraphs (for sufficiently large averages) will lead to the same behavior. This is because when averaging over hypergraphs, the hypergraph configuration is chosen randomly, in which case (upon sufficient averaging) each coupling will occur will probability $p$ by construction. This is analogous to the random pruning model and hence we expect the same behavior, albeit the hypergraph should require less averaging due to its higher connectivity.

\section{Lyapunov numerics}\label{sec:numerics}
	
Numerical methods are needed to study the Lyapunov exponent at  finite $q$ and finite coupling in the all-to-all SYK because no closed form  is currently known in this regime\footnote{Perturbative expressions in $1/\beta J$ and in the $q \to \infty$ limit \cite{Maldacena:2016hyu} do exist.}.   The need for numerical methods to study the growth of OTOCs is even more pressing in the sparse-SYK model, where the general analytic relation between the Lyapunov exponent, $\lambda_L$, and the sparsity parameter $k$ is unknown. As we showed in the previous section,   the early time behavior of the OTOC contains  $1/kq$ contributions and it is unclear how to account for these contributions in models for extracting analytically the Lyapunov exponent. 

\begin{figure}
    \centering
    \includegraphics[scale=0.5]{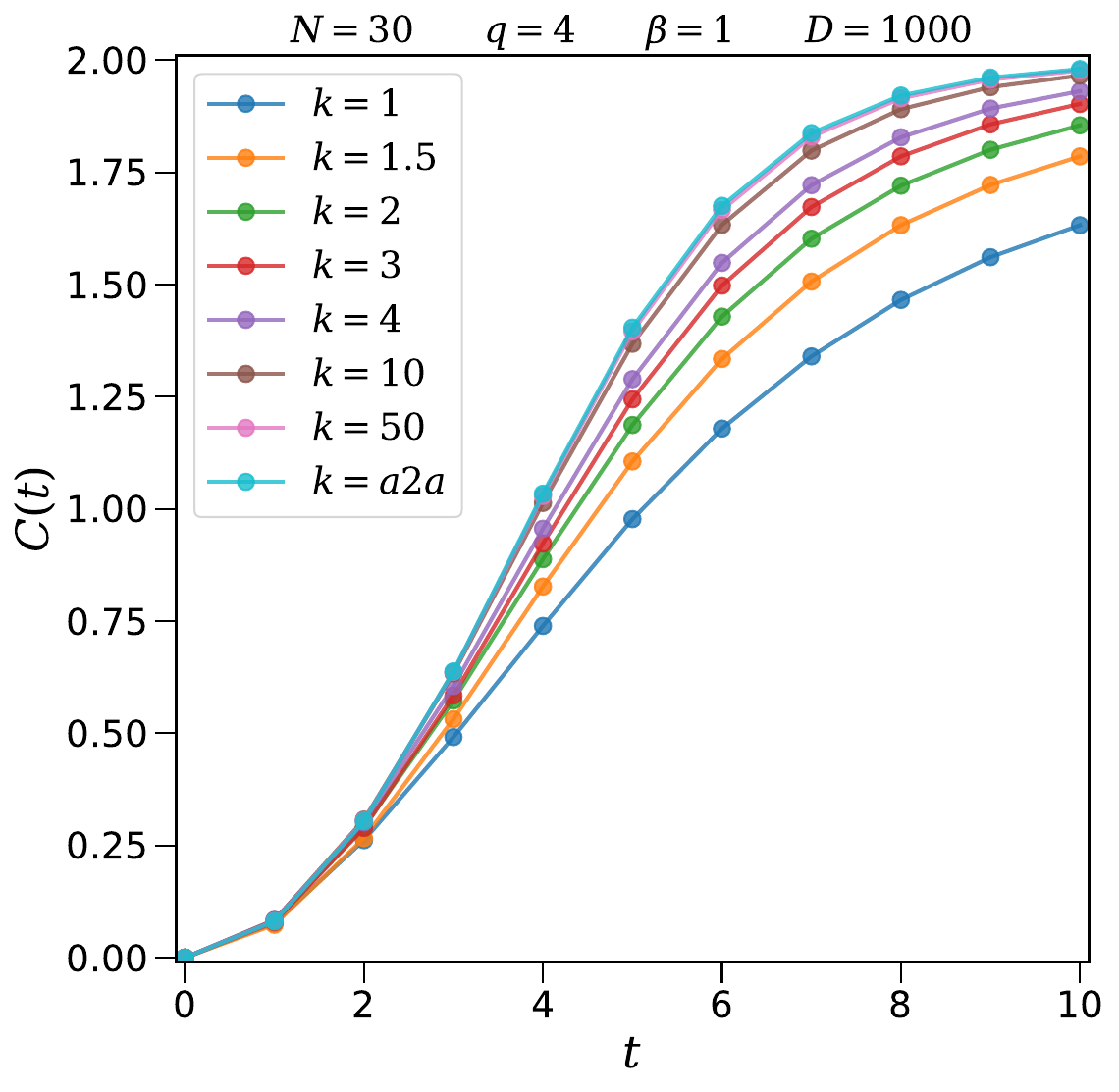}
    \caption{\textit{A comparison between the double commutators $C(t) = 2-2\mathcal{F}(t)$ of the all-to-all model (denoted 'a2a') and various sparsity parameters $k$ for the sparse model. All curves use the same number of fermions ($N=30$), number of fermions per interactions ($q=4$), temperature ($\beta = 1$), and number of disorder realizations ($D=1000$). For each $k$, hypergraphs are averaged over, a choice justified in the main body of the paper. Higher $k$'s closely resemble the curves of the all-to-all case, however as $k \to \mathcal{O}(1)$, i.e. the limit still expected to retain chaos, the exponential growth starkly decreases.}}
    \label{fig:sparsecomp}
\end{figure}

In this section we  numerically simulate the behavior of the regularized OTOCs \eqref{eq:OTOC} at finite $N$ in the sparse-SYK model. We employ the \textit{regular hypergraph} method of generating sparseness, as it ensures greater connectivity for any given realization of the Hamiltonian. The qualitative behavior from varying $k$ is shown in Figure \ref{fig:sparsecomp}, for which we see a decrease in the rate of exponential growth as $k$ decreases. This behavior is physically expected, as the sparse model reduces the number of connections between fermions, and hence it decreases the sensitivity to operator perturbations. At this stage, the computational benefit from sparse-SYK is undeniable.  Table \ref{table1} demonstrates that even for relatively low $N$, sparse simulations reap massive efficiency gains, an effect that further widens the higher $N$ one chooses to probe.

\begin{table}
    \begin{center}
    \begin{tabular}{ |p{3cm}||p{2cm}|p{2cm}|p{2cm}|p{2cm}|  }
 \hline
 \multicolumn{5}{|c|}{Time Comparisons Average, $q=4 \hspace{3mm}\beta = 1$ (112 Samples)} \\
 \hline
 Size ($N$) & All-to-All & $k=50$ & $k=10$ & $k=4$\\
 \hline
 \hline
26 & 00:02:14 & 00:00:33 & 00:00:10 & 00:00:04\\
\hline
28 & 00:06:22 & 00:01:19 & 00:00:23 & 00:00:10\\
\hline
30 & 00:17:44 & 00:03:09 & 00:00:52 & 00:00:22\\
\hline
32 & 00:47:42 & 00:07:15 & 00:01:55 & 00:00:48\\
\hline
34 & 02:06:27 & 00:16:26 & 00:04:14 & 00:01:48\\
\hline
36 & 05:27:22 & 00:37:26 & 00:09:28 & 00:04:02\\
 \hline
\end{tabular}
\end{center}
\caption{\textit{Time comparisons between all-to-all SYK and various sparseness ($k$'s) in sparse-SYK using the hypergraph method over a range of $N$'s from $N=26$ to $N=36$. All simulations were run on a single processor with $\beta =1$ and $q=4$, averaging over 112 samples to account for time differences in individual runs. Times are rounded to the nearest second.}}
 \label{table1}
\end{table}

If one wishes to capitalize on these efficiency benefits to quantify quantum chaos by means of a Lyapunov exponent, one needs a model for use in extraction. In the all-to-all model, the best known method for accurate extraction of the Lyapunov exponent \cite{Kobrin:2020xms} utilizes a symmetry of the OTOC expansion
\begin{align}\label{expansion}
    \text{OTOC} \approx C_0 + C_1 \frac{e^{\lambda_Lt}}{N} + C_2 \frac{e^{2\lambda_Lt}}{N^2} + \cdots\,,
\end{align}
which is invariant under
\begin{equation}\label{symmetry}
    t \to t + \frac{1}{\lambda_L}\log c, \quad  N \to c\,N.
\end{equation}
We know the sparse model will approach the all-to-all model in the large $k$ limit, and hence this symmetry is expected to be recovered in said limit. Our aim is to determine whether the aforementioned symmetry still holds in the sparse-SYK model, even if only approximately, in such a way that it allows a quantifiable extraction of the Lyapunov exponent for a given $k$.

\subsection{Extraction Methods}
    
Typical numerical methods for extracting the Lyapunov exponent via the OTOC involve performing an exponential fit in the exponential regime of the OTOC \cite{Yao:2016ayk, Lantagne-Hurtubise:2019svg, Shen:2017kez, Keles:2018akp}. However, this method was found to be insufficient at finite $N$ and $q=4$ for the all-to-all model, where the OTOC at finite-size could only be matched to a simple exponential for a very short time interval. It has been shown \cite{Kobrin:2020xms} that a numerical derivative method exploiting the symmetry \eqref{symmetry} is most effective in extracting the Lyapunov exponent. For each value of $N$, a fitted Lyapunov exponent $\lambda_\text{fit}(N)$ was found as follows. Since a shift under the symmetry \eqref{symmetry} leaves the OTOC invariant, we can pick a constant value of the OTOC, say $\mathcal{F}^*$, between all $N$'s simulated and use interpolation to find the time $t^*$ associated to each $\mathcal{F}^*$. It follows that the fitted Lyapunov exponent between two curves (assuming the symmetry holds) is
\begin{align}\label{numdev}
    \frac{1}{\lambda_\text{fit}} = \frac{\text{Log}(N_2) - \text{Log}(N_1)}{t_2^* - t_1^*},
\end{align}
which could then be plotted vs $1/N$ to see the effect as the system approached large $N$. Taking a second order fit in $1/N$ for $\lambda_\text{fit}$, $\lambda_L$ is defined as the value when $1/N \to 0$

\begin{figure}[h]
\centering
\begin{subfigure}{0.475\textwidth}
    \includegraphics[width=\textwidth]{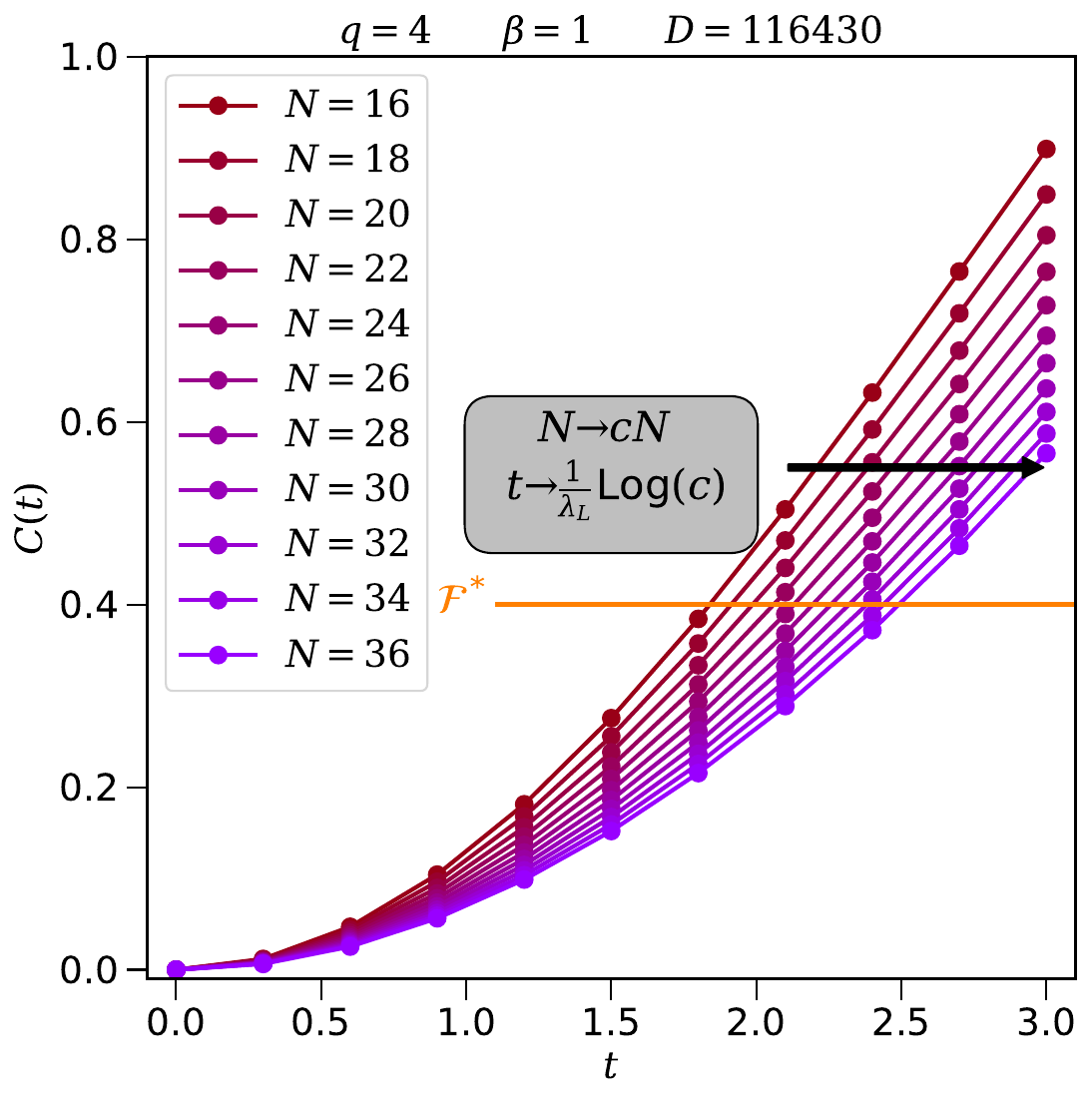}
    \caption{}
    \label{fig:aplt1}
\end{subfigure}
\hfill
\begin{subfigure}{0.475\textwidth}
    \includegraphics[width=\textwidth,height=7cm]{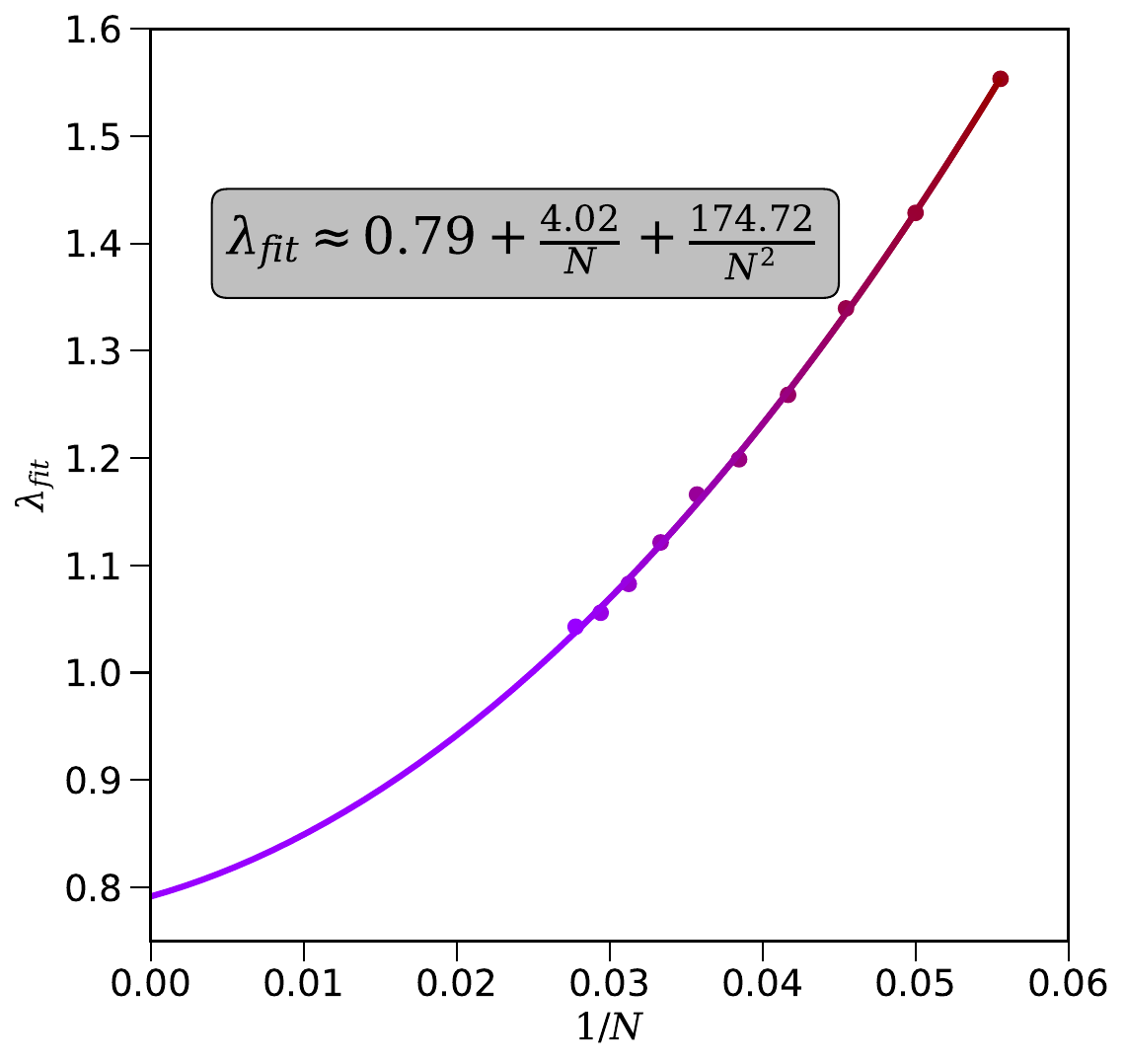}
    \caption{}
    \label{fig:aplt2}
\end{subfigure}
        
\caption{\textit{Plot visualizing the extraction of $\lambda_{fit}$ in all-to-all SYK ($q=4, \beta=1$, $116430$ disorder realizations) from the double commutator $C(t)$ via a numerical derivative. (a) Plots of the $C(t)$ versus time for $N=16$ to $N=36$. To calculate the numerical derivative, a horizontal cross section $\mathcal{F}^*$ must be chosen, here $\mathcal{F}^*=0.4$. The symmetry (labelled within the grey box) describes at what times subsequent curves should intersect $\mathcal{F}^*$. (b) After the numerical derivative between subsequent curves is calculated via \ref{numdev}, $\lambda_{fit}$ is plotted versus $1/N$ and fitted to a 2nd order polynomial (in grey box) to extract the large N Lyapunov exponent, here $\lambda_L=0.79$.}}
\label{fig:otocextract}
\end{figure}

\begin{align}\label{lambdafit}
    \lambda_\text{fit}(N) \approx \lambda_L + \frac{\lambda_1}{N} + \frac{\lambda_2}{N^2}.
\end{align}

In our simulations, we use $\mathcal{F}^*=0.4$. See Figure \ref{fig:otocextract} for a concrete example of this procedure. We test the robustness of this choice by using different values in Appendix \ref{app:robustness}. Due to the success of the numerical derivative method exploiting the symmetry \eqref{symmetry}, we choose to closely match the numerical tools employed by \cite{Kobrin:2020xms} in their extraction of the Lyapunov exponent. This includes utilizing \textit{dynamite} \cite{dynamite}, a Python package allowing fast simulation of quantum dynamics with Krylov subspace implementation \cite{park1986unitary}. As stated in Section \ref{subsec:chaos}, a \textit{regularized} OTOC 
\begin{equation}
    \mathcal{F}(t) = \langle W(t)\rho^{1/4}V(0)\rho^{1/4}W(t)\rho^{1/4}V(0)\rho^{1/4}\rangle_\beta
\end{equation}
has been shown to reduce finite-size effects, and here $V$ and $W$ are chosen as distinct single Majorana operators. Additionally, it is known for large $N$ that the thermal expectation value involving the trace can be closely approximated by calculating the expectation value via Haar random states \cite{Goldstein:2005aib, Luitz:2016kqa, Kobrin:2020xms}, i.e. for a Haar random state $|\psi\rangle$
\begin{align}
    \text{tr}\left[e^{-\beta H}\hat{O}\right] \approx \overline{ \langle \psi | e^{-\beta H}\hat{O} | \psi \rangle}.
\end{align}
The finite $N$ error in this approximation can be reduced through sufficient averaging over different initial random states, represented by the overline. It has been shown that averaging these states separately or simultaneously with the couplings $J_{ijkl}$ in the all-to-all case had no significant difference in error. Hence, in this work, we also employ simultaneous averaging of the couplings and initial random states.

We find another source of averaging that is important to the extraction of the Lyapunov exponent not present in the all-to-all model, namely averaging over many hypergraphs (by generating regular hypergraphs using different seeds). As we will explain below, the fixed hypergraph method does not produce the expected symmetry. Only after averaging over many hypergraphs does the desired symmetry becomes apparent (see Appendix \ref{app:efficiency}). Thus, only for an aggregate of hypergraphs the symmetry-based method could be used to extract the Lyapunov exponent.

In order to quantify the error of $\lambda_L$, we  simulate a large number of realizations $\gtrsim 100,000$ and subdivide these realizations into subgroups, from which we extract the Lyapunov exponent for each subgroup. By performing statistics on many of these subgroups, we can derive a mean Lyapunov exponent $\bar{\lambda}$, a standard deviation $\sigma$, and a coefficient of determination $R^2$. We explain this procedure in more detail in what follows.

\afterpage{
\begin{figure}[H]
\centering
\begin{subfigure}[b]{0.45\textwidth}
    \includegraphics[width=\textwidth]{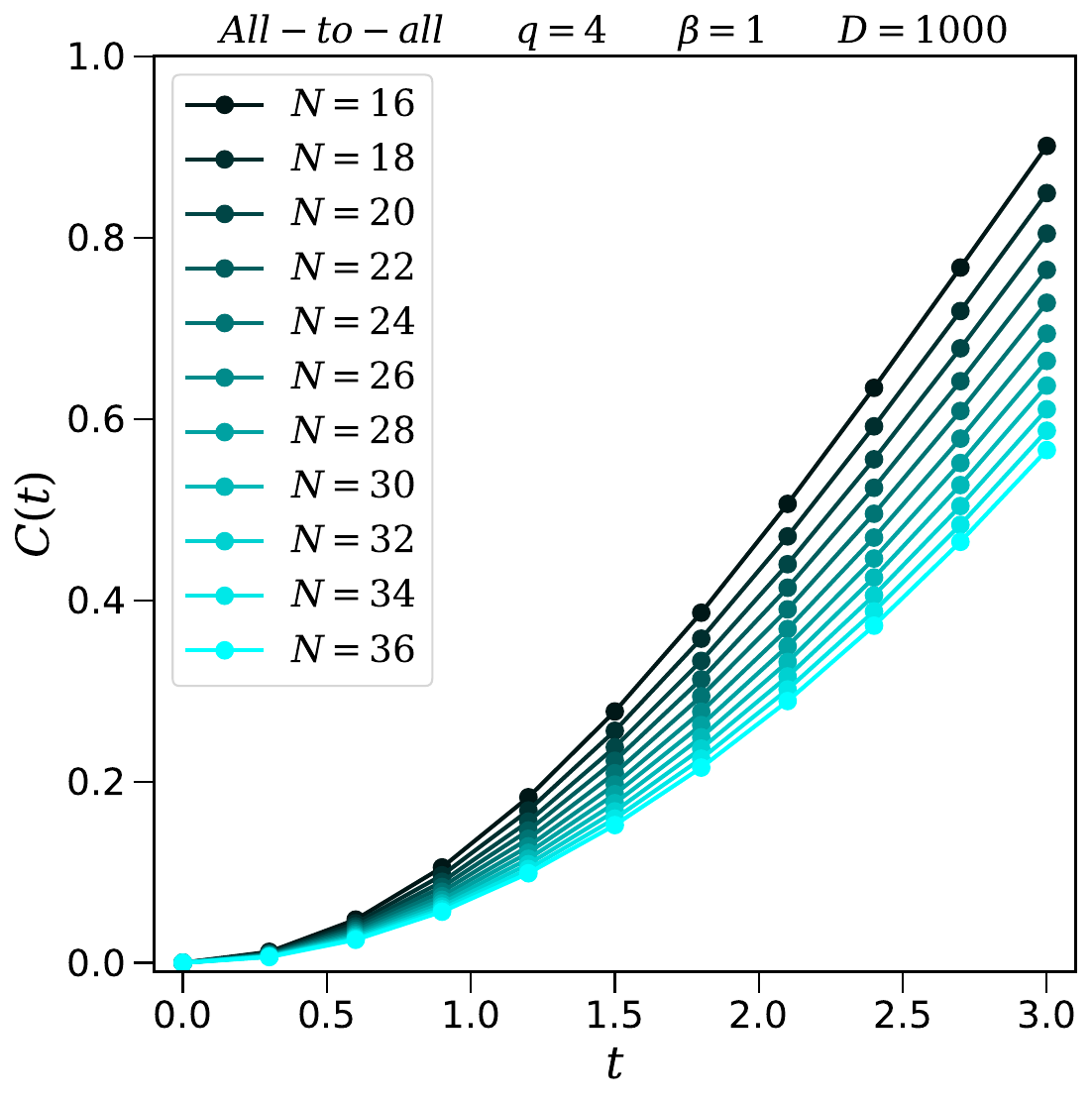}
    \caption{}
    \label{fig:first}
\end{subfigure}
\hfill
\begin{subfigure}[b]{0.45\textwidth}
    \includegraphics[width=\textwidth]{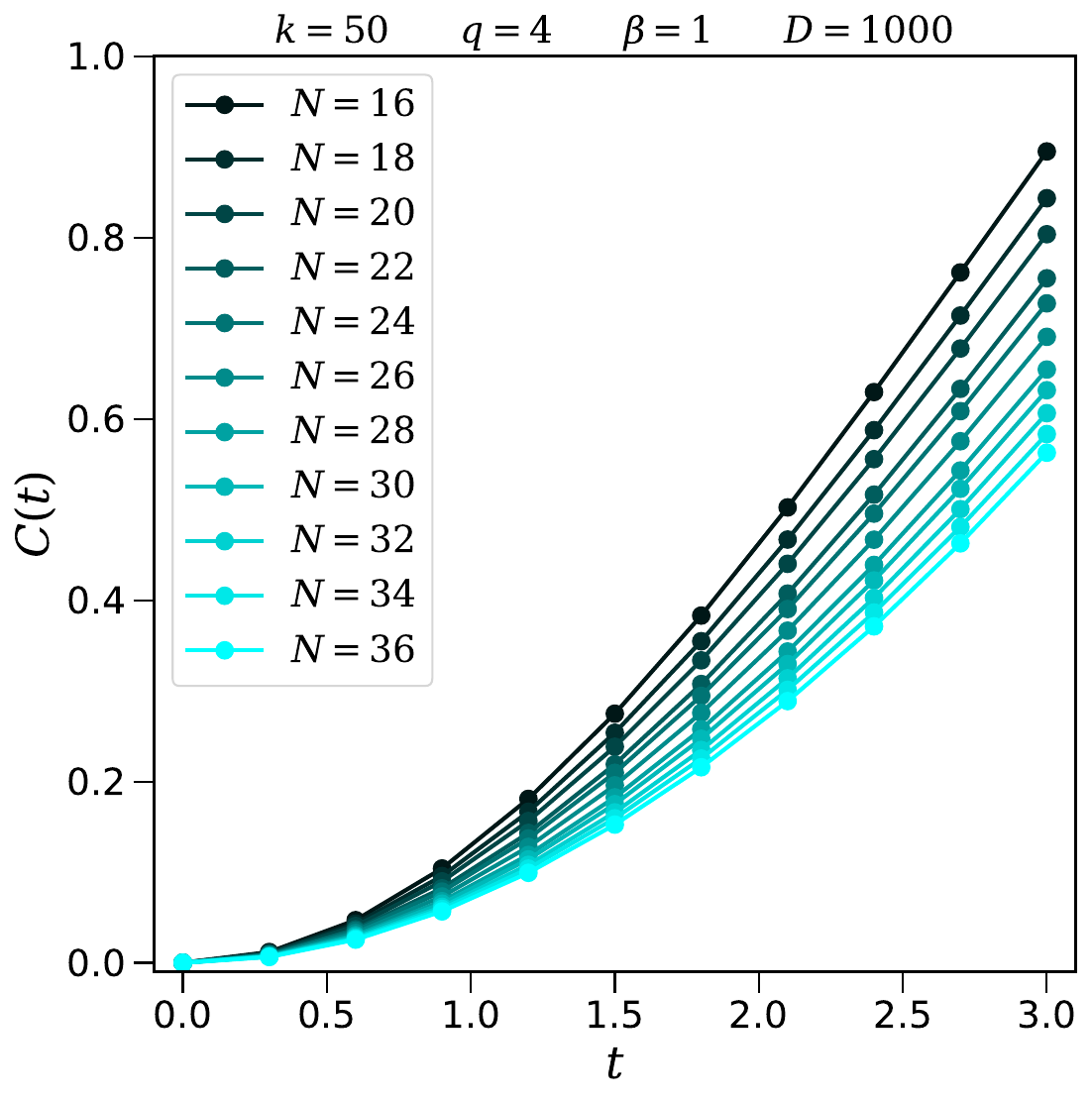}
    \caption{}
    \label{fig:second}
\end{subfigure}

\begin{subfigure}[b]{0.45\textwidth}
    \includegraphics[width=\textwidth]{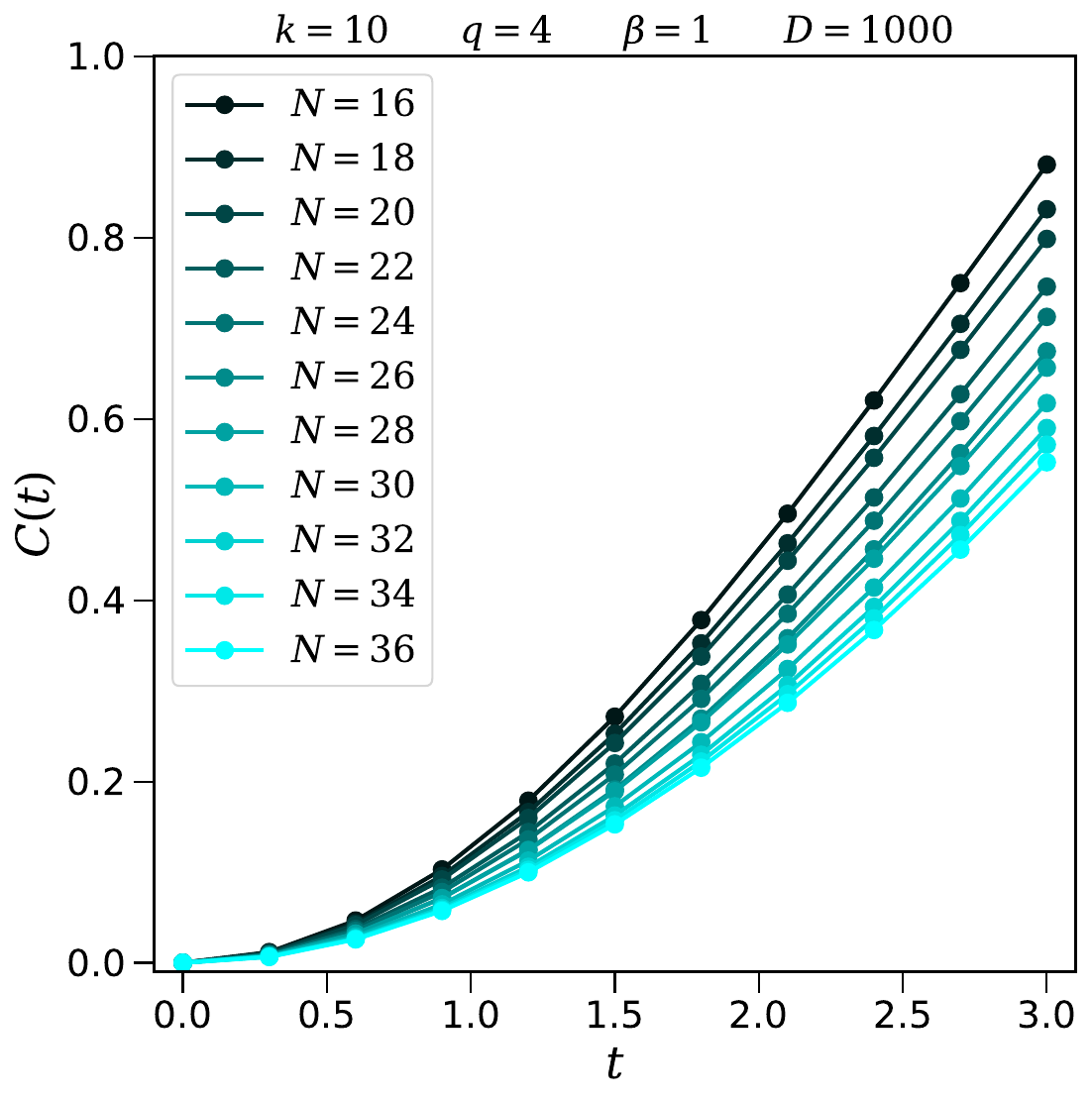}
    \caption{}
    \label{fig:third}
\end{subfigure}
\hfill
\begin{subfigure}[b]{0.45\textwidth}
    \includegraphics[width=\textwidth]{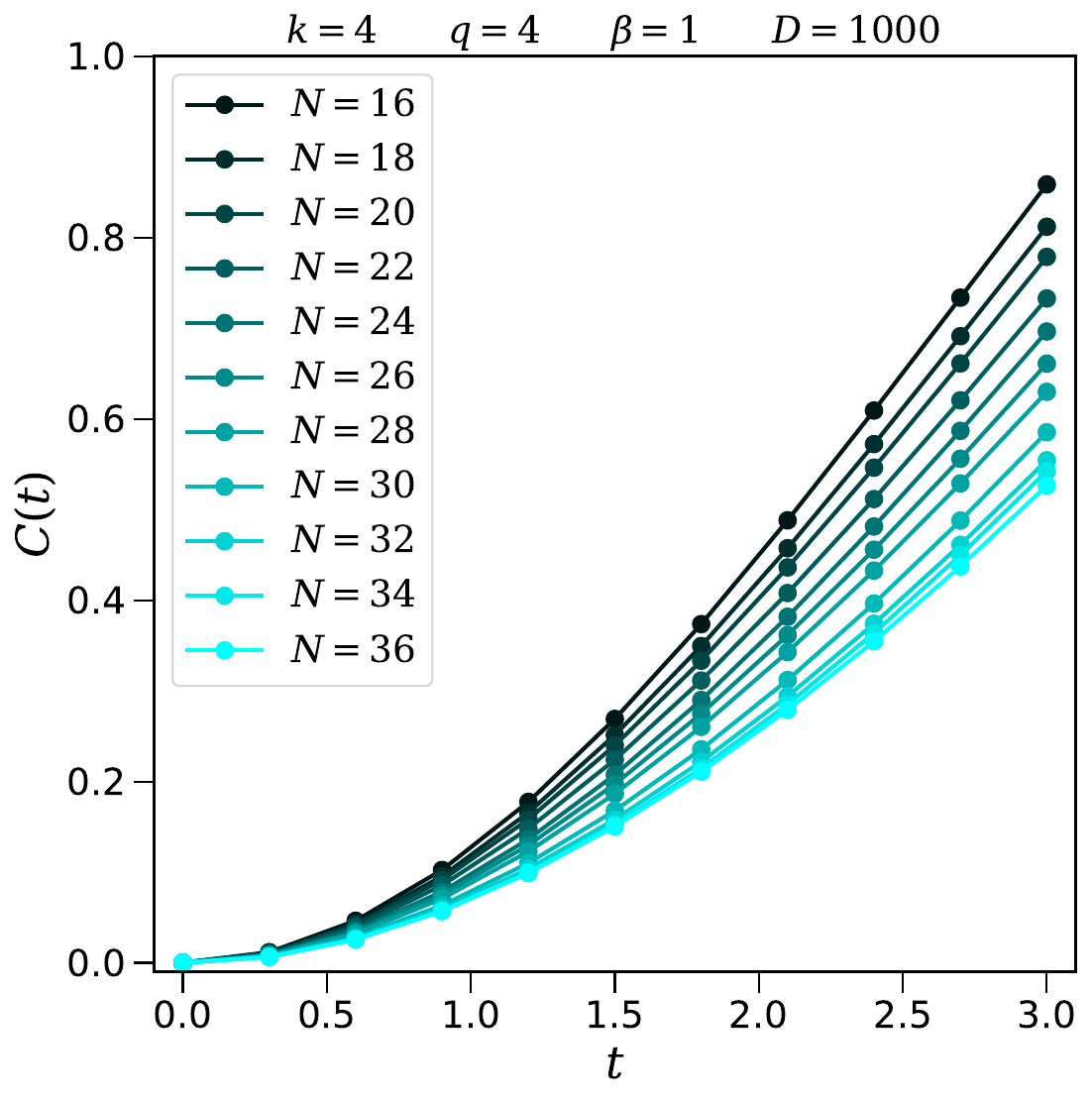}
    \caption{}
    \label{fig:fourth}
\end{subfigure}
        
\caption{\textit{Comparison of $C(t) = 2 - 2\mathcal{F}(t)$ between All-to-all and various Sparse models with $D=1000$ disorder realizations. Presented data is averaged over many different hypergraphs. $(a)$ In the All-to-all model, the symmetry visually holds as $N$ increases; subsequent curves get successively closer together. $(b,c,d)$ Plots for $k=50,10,4$ respectively. The symmetry visually appears to break down in the Sparse regime at this level of realizations, i.e. the spacing between subsequent curves is not consistent with the symmetry.}}
\label{fig:comparisons}
\pagebreak
\end{figure}
\pagebreak
}

\subsection{Averaging over Hypergraphs}\label{subsec:averagingHyper}

\afterpage{
\begin{figure}[H]
\centering
\begin{subfigure}{0.49\textwidth}
    \includegraphics[width=\textwidth]{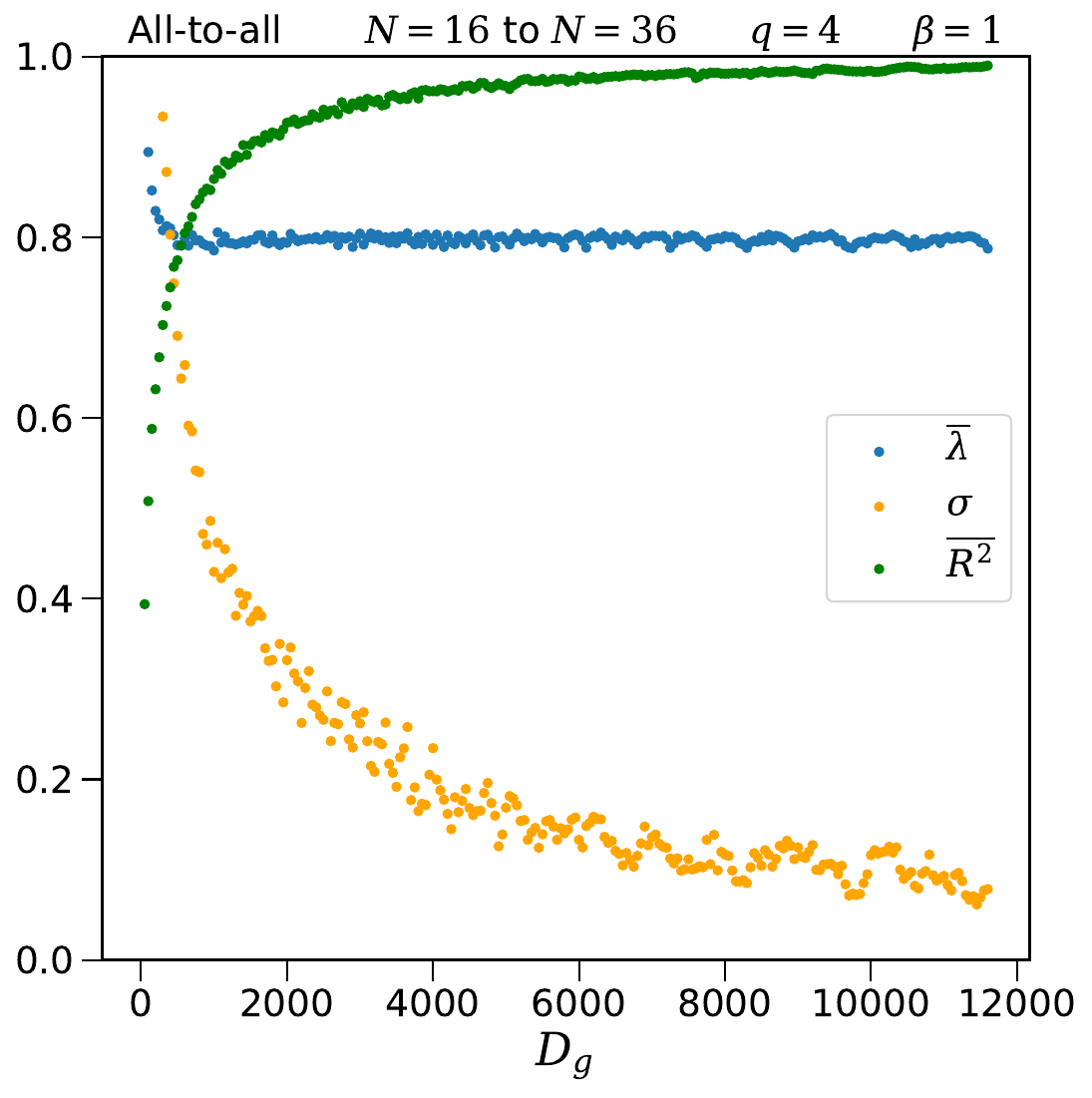}
    \caption{}
    \label{fig:1Dmax}
\end{subfigure}
\hfill
\begin{subfigure}{0.49\textwidth}
    \includegraphics[width=\textwidth]{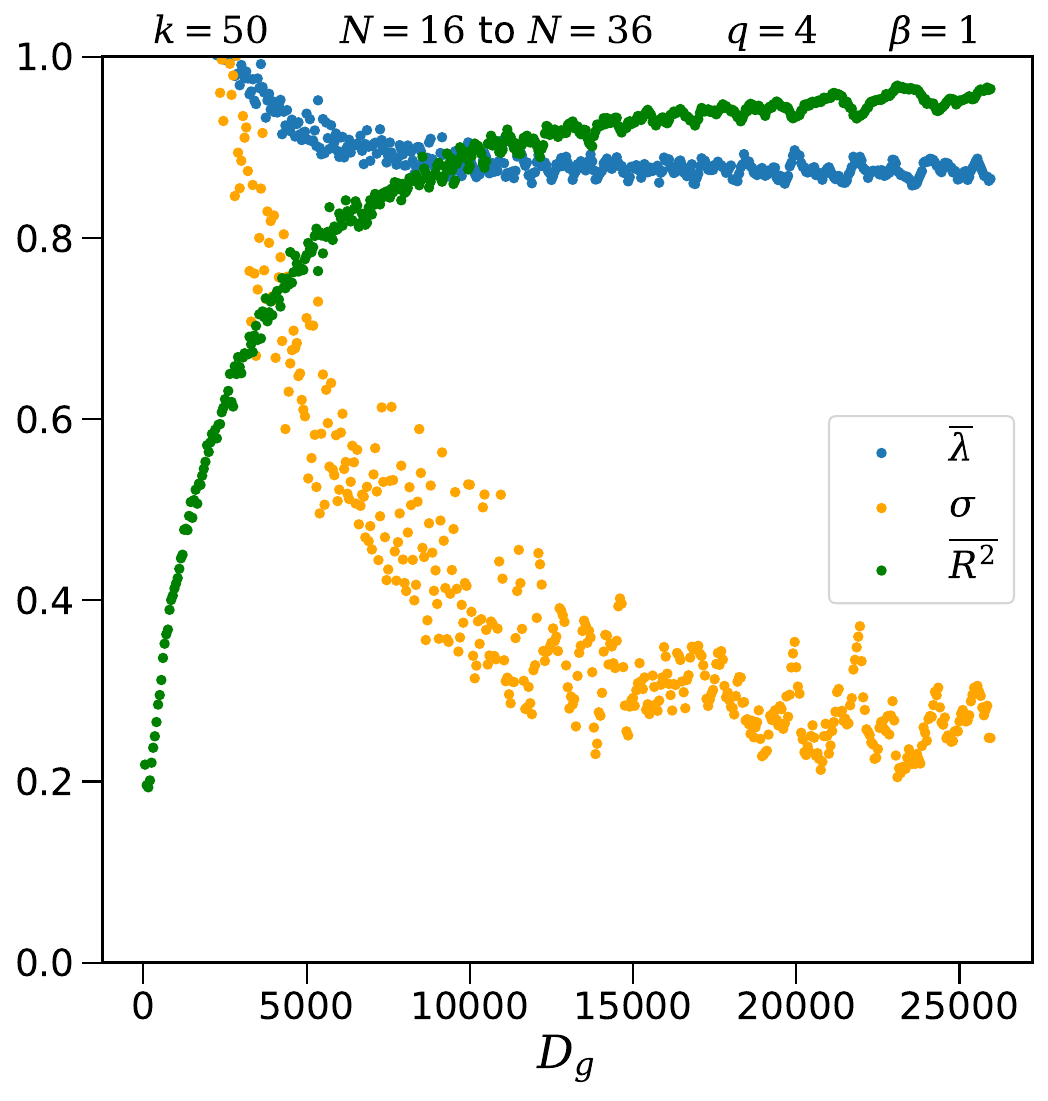}
    \caption{}
    \label{fig:2Dmax}
\end{subfigure}

\begin{subfigure}{0.49\textwidth}
    \includegraphics[width=\textwidth]{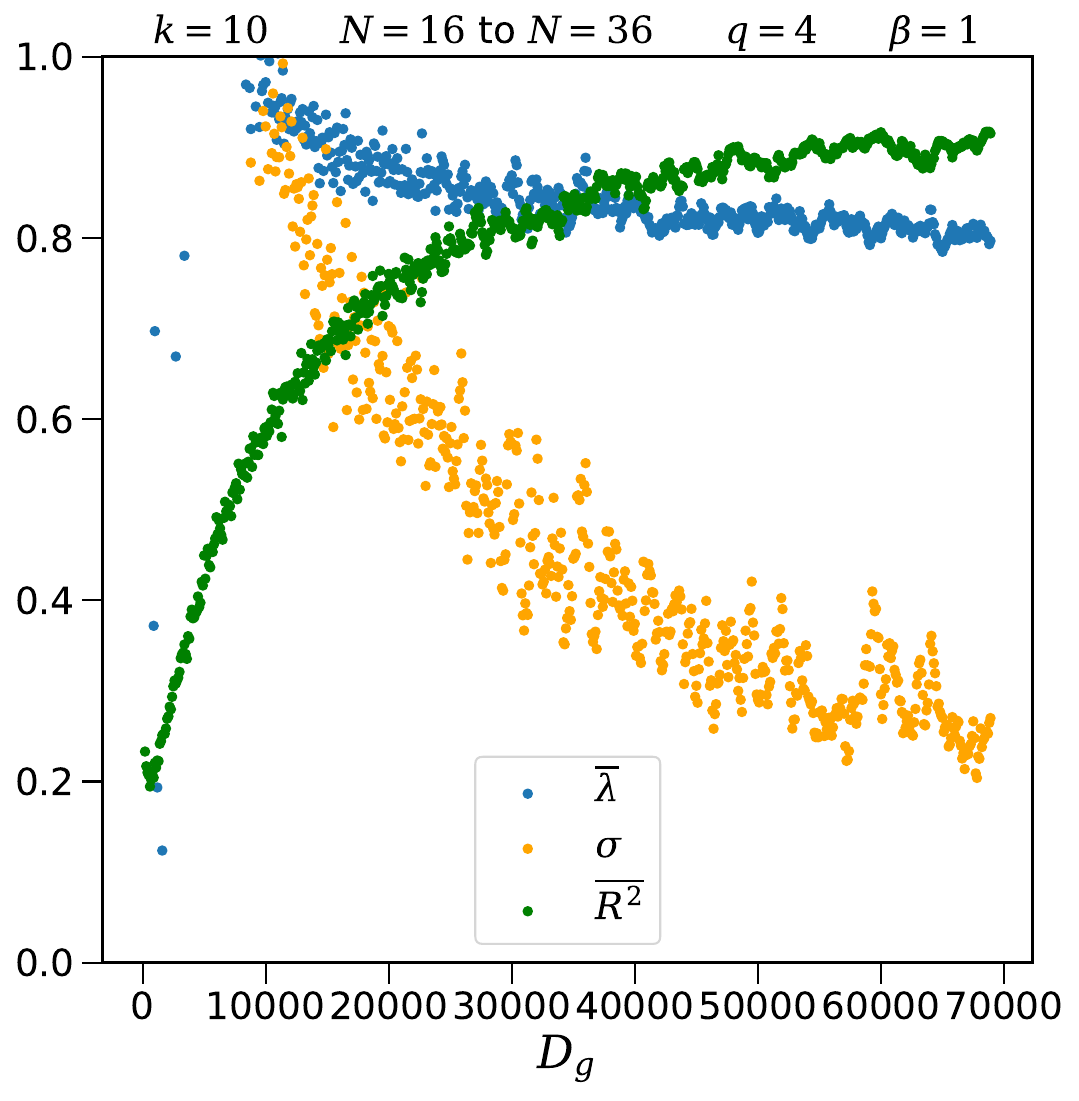}
    \caption{}
    \label{fig:3Dmax}
\end{subfigure}
\hfill
\begin{subfigure}{0.49\textwidth}
    \includegraphics[width=\textwidth]{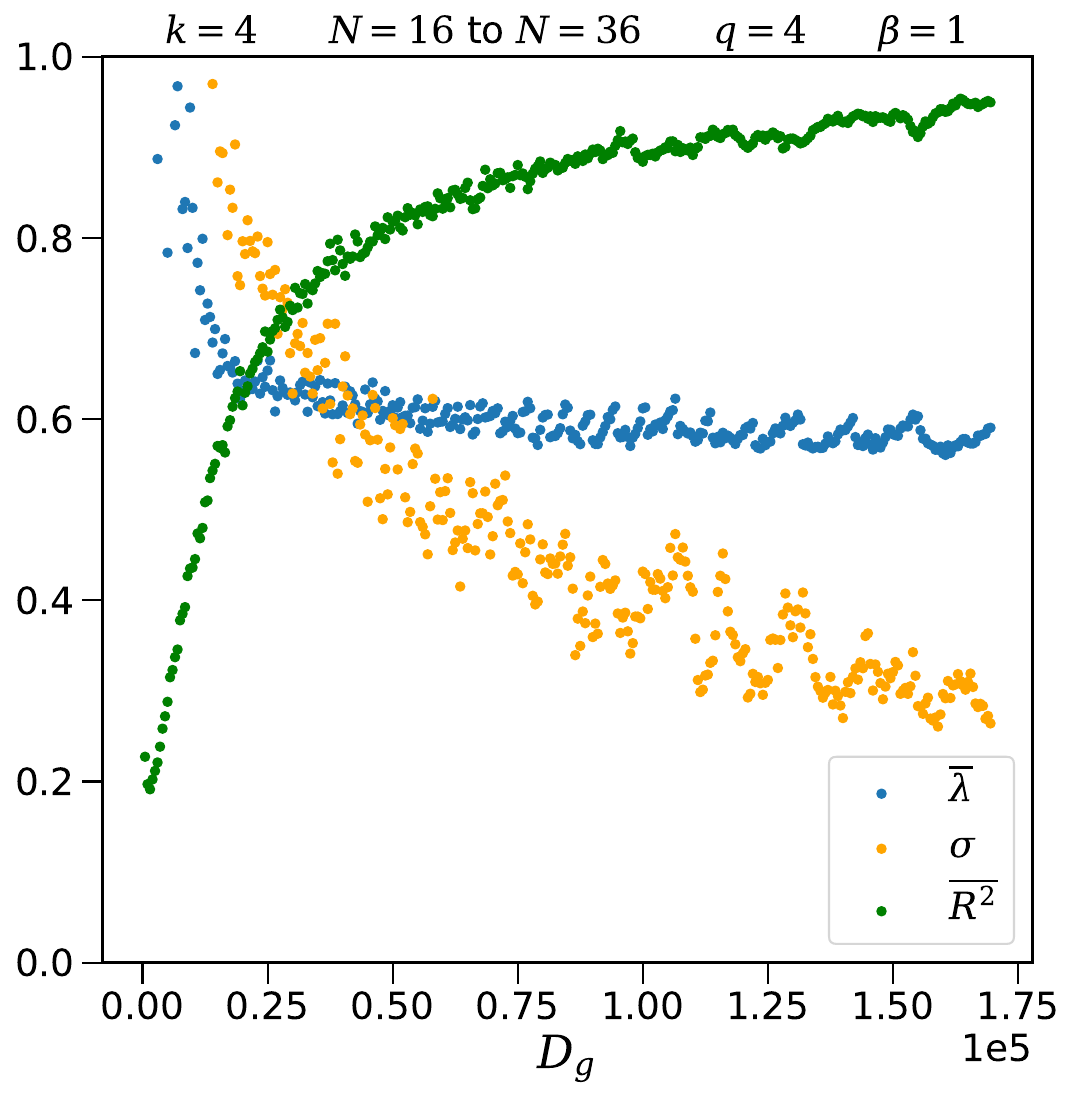}
    \caption{}
    \label{fig:4Dmax}
\end{subfigure}
        
\caption{\textit{Many realizations are subdivided into groups sized $D_g$, from which a Lyapunov exponent and $R^2$ is extracted. This process is repeated $D_{tot}/D_g$ times, the average values of $\lambda_L$ (\underline{blue}) and $R^2$ (\underline{green}) are plotted, as well as the standard deviation of $\lambda_L$, $\sigma$ (\underline{orange}). Resolution of plots all have $\Delta D_g = 50$. \textbf{(a)} all-to-all model. Exhibits convergence for plotted quantities; $R^2 \to 1$ shows agreement with the symmetry. \textbf{(b,c,d)} $k=50,10,4$ respectively. For similar values of $R^2$, display similar values of error (even compared to (a)), however fluctuations increase as compared to all-to-all.}}
\label{fig:Dmaxmain}
\pagebreak
\end{figure}
\pagebreak
}
   
The ultimate goal of this section is to glean whether the symmetry as described by \eqref{symmetry} holds in the sparse-SYK model, and if so to extract Lyapunov exponents and determine their dependence on the sparsity parameter $k$. We can get a qualitative idea of whether the symmetry holds from Figure \ref{fig:comparisons}, where we plotted the Double Commutators of the sparse model in comparison to the all-to-all model over $D=1000$ realizations. These results employ simultaneous averaging over hypergraphs, couplings, and random initial states. At this level of averaging, the inconsistent behavior in spacing between subsequent curves suggest that the symmetry breaks down as sparsity increases. This behavior admits at least three possible explanations. Either the symmetry itself does not hold in the sparse model, the sparse model is more susceptible to finite-sized effects, or the averaging over hypergraphs produces random fluctuations requiring a higher degree of averaging. 

A quantitative approach must be employed to separate these possibilities. To test the impact of random fluctuations, we note that an average over a sufficiently large number of realizations must converge to a certain value, and therefore repeating the experiment using the same number of realizations should not change that value. We measure the dependence on the number of realizations by repeating the extraction of the Lyapunov exponent many times, where each extraction utilizes $D_g$ realizations. From these repetitions, we obtain a mean value for the Lyapunov exponent $\overline{\lambda}$ and a standard deviation $\sigma$ for each $D_g$. By varying $D_g$, we can analyze how the error on $\overline{\lambda}$ is modified. The choice of how many repetitions to average over is completely arbitrary, we find that the choice yielding the smoothest results for Lyapunov extraction is to average the ``maximum" number of times, i.e. if there are $D_\text{tot}$ total number of realizations, then we can make $D_\text{tot}/D_g$ unique groupings of $D_g$ realizations and hence we average over all of these groups. When $D_\text{tot}/D_g<10$ we stop increasing $D_g$, so as to allow sufficient samples to be averaged over. Another possible choice is to choose a set number of samples to average over regardless of $D_g$, we do this with 10 and 20 samples in Appendix \ref{app:robustness} as a test of robustness.

The other quantitative tool utilized is the coefficient of determination $R^2$, which represents in our case how well the numerical derivative data of a given group fits \eqref{lambdafit}. It has the formula
\begin{align}
    R^2 = 1- \frac{SS_\text{res}}{SS_\text{tot}}.
\end{align}
The term $SS_\text{res}$ refers to the sum of squares of residuals, $SS_\text{res}=\sum_i (x_i-\hat{x_i})^2$, $x_i$ refers to our numerical data and $\hat{x_i}$ refers to the prediction based upon our model from \eqref{lambdafit}. This represents the total sum of variances between our model's prediction and the numerical data. $SS_\text{tot}$ on the other hand is the sum of the variance of the data, $SS_{tot}=\sum_i (x_i - \overline{x})^2$. The ratio $SS_\text{res}/SS_\text{tot}$, also called the fraction of variance unexplained, represents the ratio between the ``unexplained" variance (from the discrepancy between model and data) and the total variance. If a set of data implements the symmetry given by \eqref{symmetry}, $\lambda_\text{fit}$ will closely match the data, implying $SS_\text{res}/SS_\text{tot}\to 0$ and therefore $R^2 \approx 1$. Hence, by analyzing how $R^2$ changes with $D_g$, we can see whether simply greater averaging is sufficient to recover the symmetry. 

Due to the outputs of the mean Lyapunov exponent, associated standard deviation, and $R^2$ all roughly lying in the range $[0,1]$, we can plot these values together in the same plot, which can be seen in Figure \ref{fig:Dmaxmain}. 
\begin{figure}[t]
    \centering
\includegraphics[width=.49\textwidth]{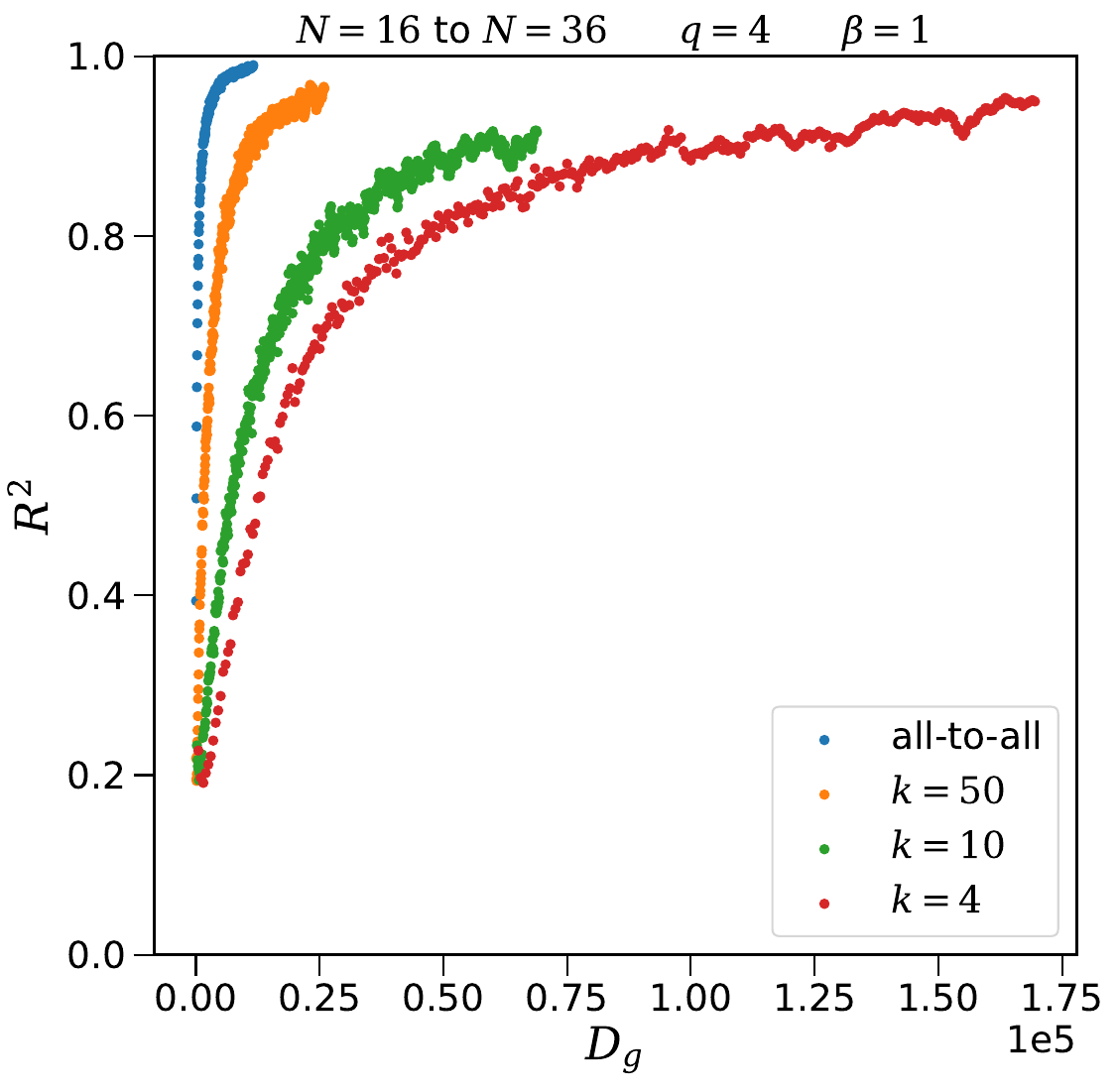}
    \caption{\textit{$R^2$ for the plots in Figure \ref{fig:Dmaxmain}. As the model becomes more sparse, the number of realizations has to increase considerably to obtain reliable results. As in previous plots, $N=16$ to $36$, $q=4$, $\beta=1.$}}
    \label{fig:Rsquare}
\end{figure}
These plots indicate that random fluctuations, caused in large part because hypergraphs must be averaged over, have a much larger effect for sparse models. It is not until far larger numbers of disorder realizations in the Sparse case that similar values for error and $R^2$ as compared to the all-to-all model are achieved. At this level of disorder groupings, the all-to-all model demonstrates convergence in the mean, standard deviation, and $R^2$ values with groups of roughly $D_g \sim 10,000$, with $R^2 \to 1$ suggesting the symmetry based model closely matches the data. As $k$ decreases, the stability of this convergence slightly decreases, with somewhat higher fluctuations. However, given a large enough $D_g$, such as $D_g \sim 175,000$ for $k=4$, we can also recover $R^2 \to 1$ and some convergence in error and mean Lyapunov exponent. We illustrate this point in Figure \ref{fig:Rsquare}.

Overall, we find that as $k$ decreases, the number of disorder realizations goes up to (and exceeds) an order of magnitude greater than in the all-to-all case. This suggests the following interpretation: as $k$ decreases, the structure/connections of the sparse Hamiltonian further and further diverge from that of the all-to-all model. To compensate for a given hypergraph being a poor representation of the all-to-model from which it is constructed, hypergraphs (and hence the connections/structure) must be averaged over if the symmetry is to be recovered. The subsequent section provides further evidence for this interpretation.


One important observation from Figure \ref{fig:Dmaxmain} is the appearance of the $k=50$ average Lyapunov exponent being higher than that of the all-to-all model, this is surprising as increasing sparseness should decrease the chaotic properties. This is however an artifact of numerical error, specifically that the Lyapunov exponent approaches its asymptotic value from above. At our level of disorder realizations, the standard deviation and $R^2$ (for all $k$) make it clear that full convergence has not yet been achieved, the $k=10$ and $k=4$ cases demonstrate that as one approaches convergence by increasing $D_g$, the average Lyapunov exponent continues to decrease. They also demonstrate that for lower $k$, convergence is not achieved until higher levels of $R^2$ are achieved as compared to the all-to-all model. We expect that given enough disorder realizations, the extracted Lyapunov exponent for $k=50$ would drop below that of the all-to-all model. A reasonable cause for why the standard deviation may asymptote above zero is due to finite $N$ effects, and it appears for the $k=50$ case this asymptote may be above that of the all-to-all case. It is also possible that some disagreement with symmetry-based models is due to finite $N$ behavior acting more strongly in the sparse model. Hence, our next goal is to investigate whether this is the case, and if simulating larger $N$ can lead to more accurate extraction of the Lyapunov exponent.

\subsection{Is accurate extraction possible at larger $N$?}\label{subsec:largerN}

\afterpage{
\begin{figure}[H]
\centering
\begin{subfigure}{0.49\textwidth}
    \includegraphics[width=\textwidth]{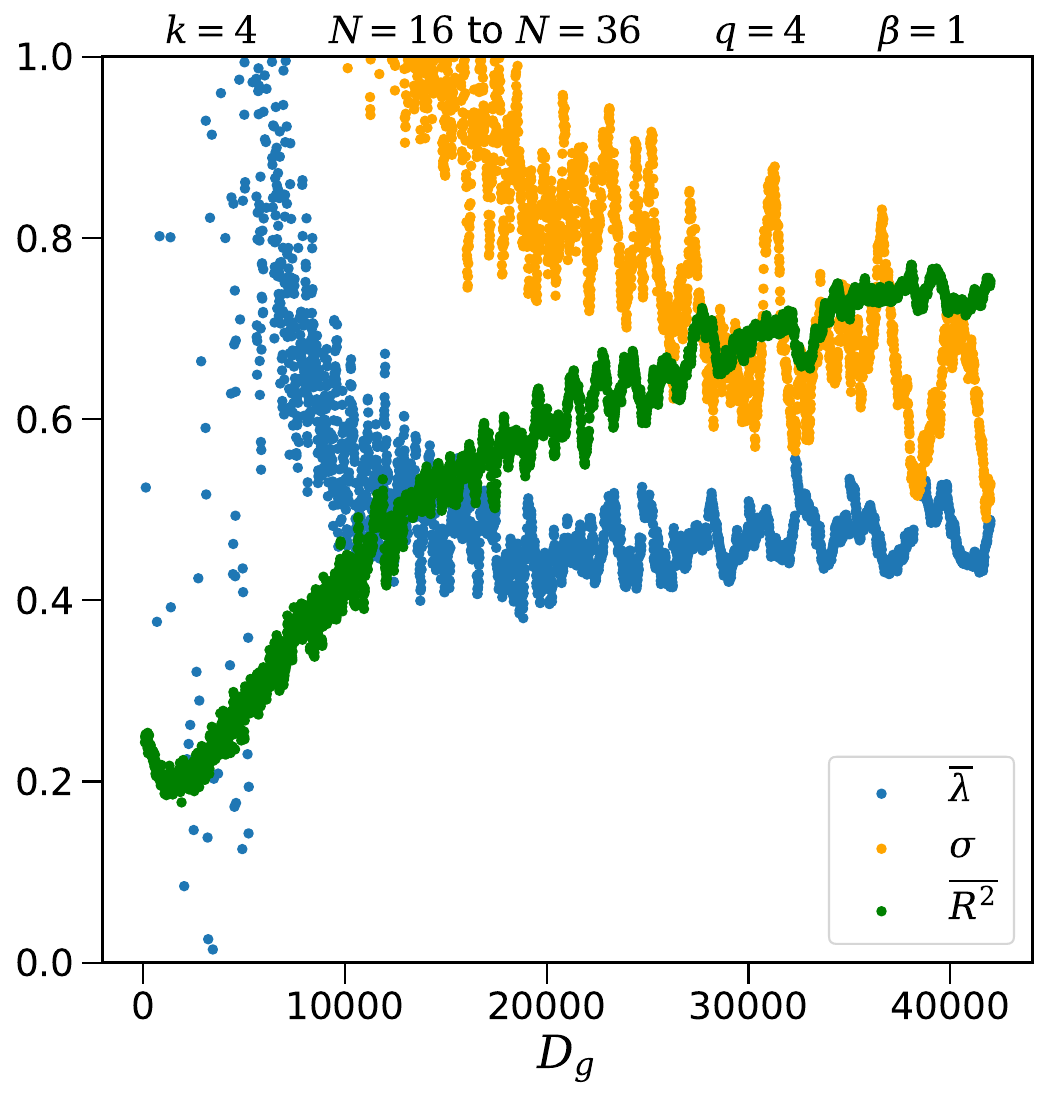}
    \caption{}
    \label{fig:firstmax}
\end{subfigure}
\hfill
\begin{subfigure}{0.49\textwidth}
    \includegraphics[width=\textwidth]{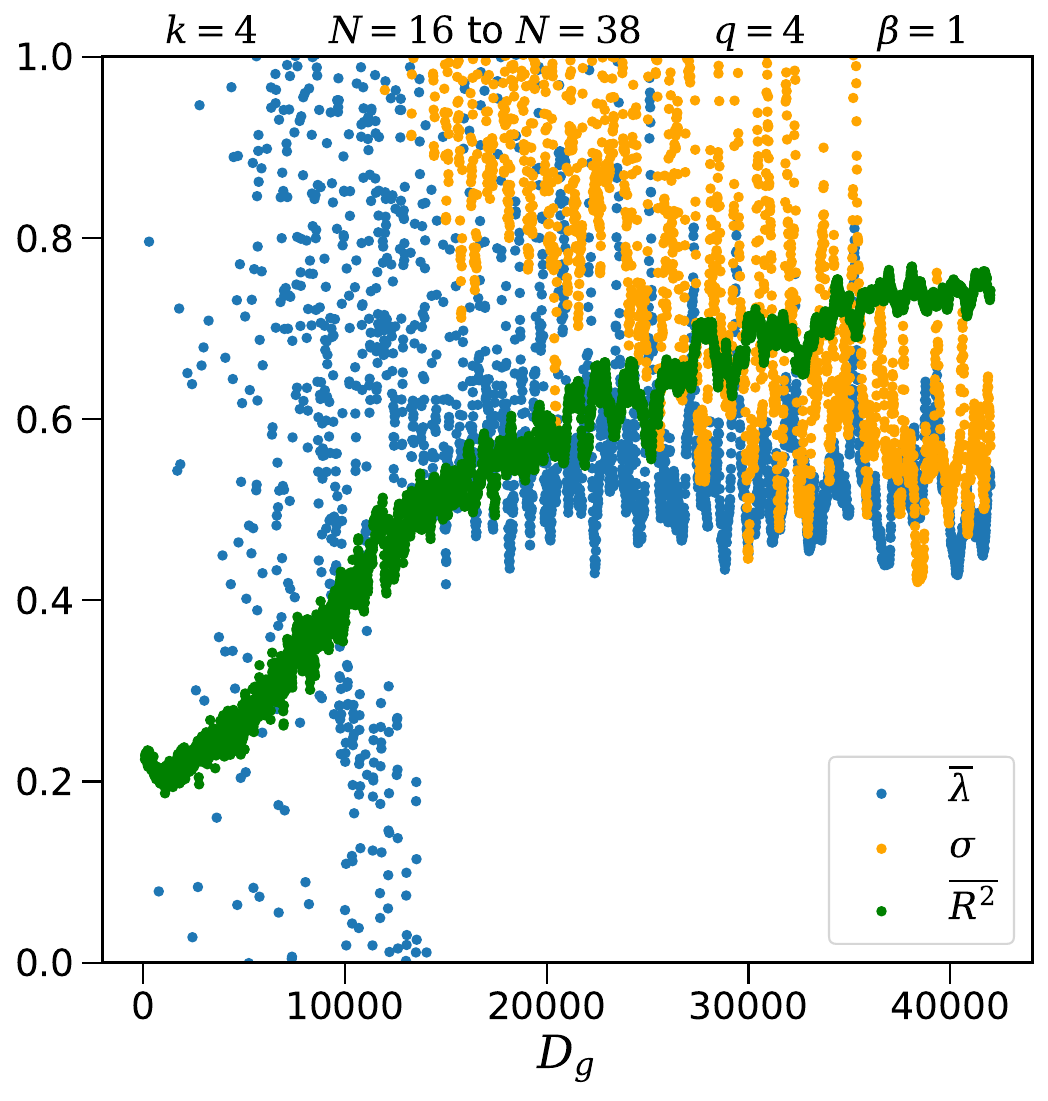}
    \caption{}
    \label{fig:secondmax}
\end{subfigure}

\begin{subfigure}{0.49\textwidth}
    \includegraphics[width=\textwidth]{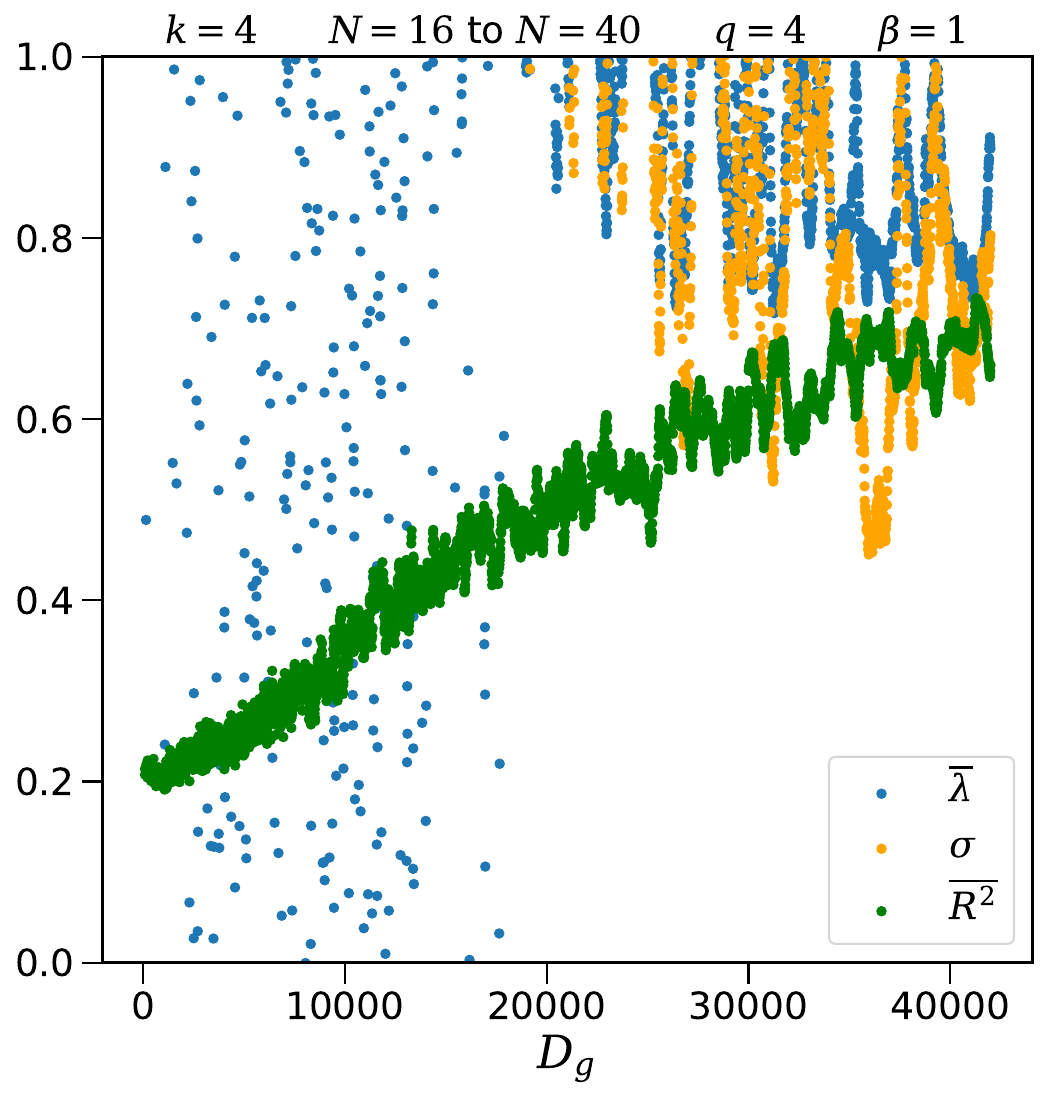}
    \caption{}
    \label{fig:thirdmax}
\end{subfigure}
\hfill
\begin{subfigure}{0.49\textwidth}
    \includegraphics[width=\textwidth]{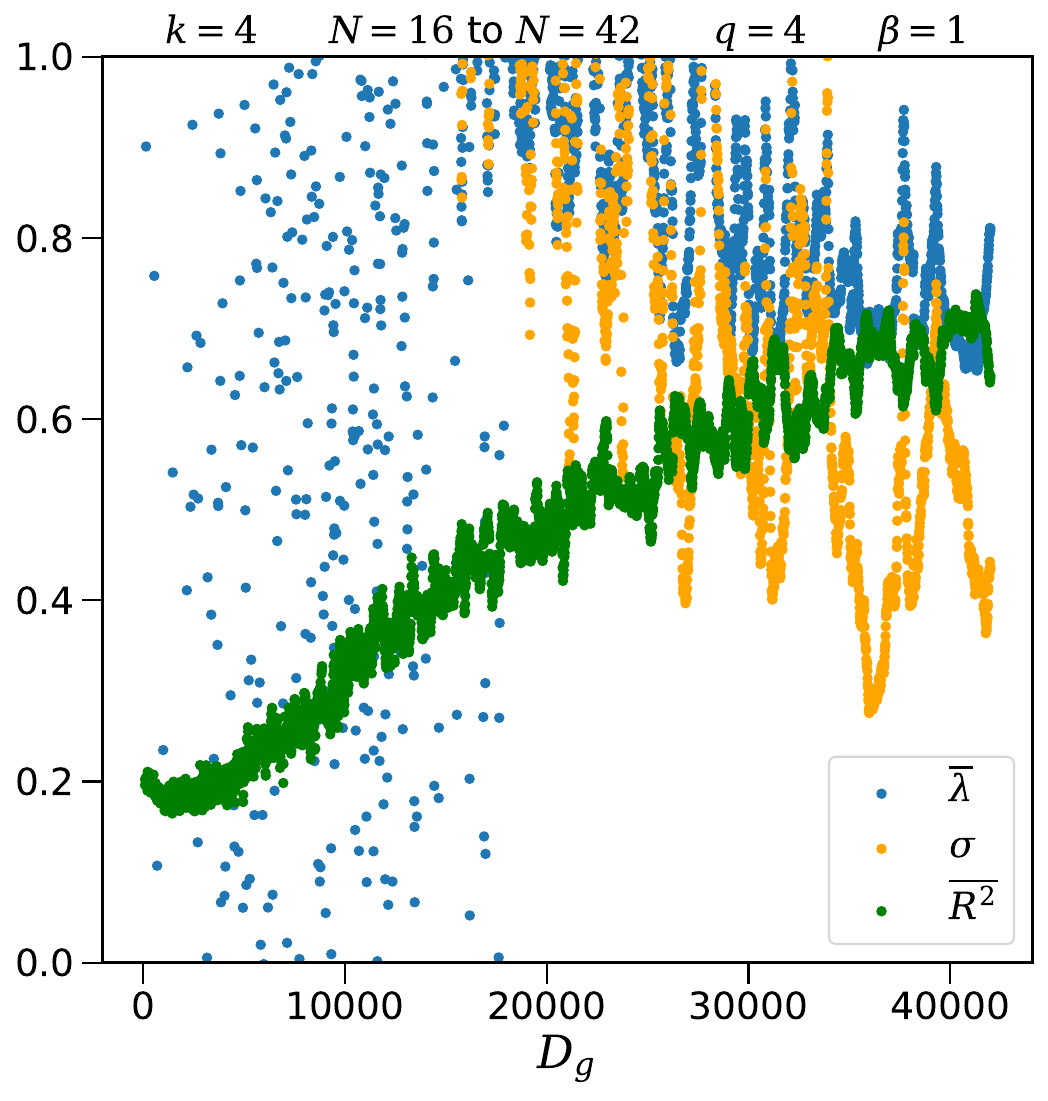}
    \caption{}
    \label{fig:fourthmax}
\end{subfigure}
        
\caption{\textit{Many realizations are subdivided into groups sized $D_g$, from a Lyapunov exponent and $R^2$ is extracted. This process is repeated $D_{tot}/D_g$ times, the average values of the $\lambda_L$ (\underline{blue}) and $R^2$ (\underline{green}) are plotted, as well as the standard deviation of $\lambda_L$ (\underline{orange}). Resolution of plots all have $\Delta D_g = 10$. All plots are of k=4 sparse model, and have the same max $D_g$. \textbf{(a)} Max value $N=36$, scaled down version of Figure \ref{fig:Dmaxmain} (d), baseline of comparison. \textbf{(b,c,d)} Max value $N=38,40,42$ respectively. All exhibit greater fluctuations and no significant change to $R^2$ for this range of $D_g$.}}
\label{fig:Nmax}
\pagebreak
\end{figure}
\pagebreak
}

It is natural to ask if going to larger $N$ will allow us to extract the Lyapunov exponent more accurately, as it does for the all-to-all model. Unfortunately, we find that at fixed $k$, the answer seems to be no. In this section, we probe higher $N$'s in the sparse model to test if finite-sized effects can explain some of the disagreement with symmetry-based models. Our methods match those used in the previous section, except rather than comparing different $k$'s we compare different ranges of $N$'s for the $k=4$ model, which approaches the minimum limit of sparsity while still retaining chaos, and is correspondingly the fastest of our choices of $k$ to simulate. We chose to retain the lower bound of $N=16$ when extracting the Lyapunov exponent for all higher $N$'s as it led to better agreement with symmetry methods. In Appendix \ref{app:robustness} we have also included an alternative analysis with a constant range of $N$'s (e.g. for $N=42$ the range would be $N=22$ to $N=42$). 

The corresponding plots for these simulations can be found in Figure \ref{fig:Nmax}. We find that not only do random fluctuations actually increase in size with higher $N$, but also the $R^2$ value has no better agreement as $N$ increases. There is a likely explanation for these trends.
Recall that the number of interactions scales as $\sim N^q$ for the all-to-all model and as $\sim k N$ in the sparse models. Therefore, as $N$ increases, the ratio of interaction terms between the all-to-all and sparse models grow as $N^{q-1}$. Our results in Section \ref{subsec:averagingHyper} show that the further one strays from the all-to-all configuration, the more averaging one must do to recover the symmetry. Thus, as $N$ increases and the difference in interaction terms grows larger, the sparse model requires higher levels of averaging to recover the symmetry.

Hence we have demonstrated that for low $k$, there does not appear to be a simple method to consistently extract the Lyapunov exponent at finite $N$ using currently known methods without averaging over massive number of realizations. Another approach one could employ is varying $\beta$, however it is unlikely this would lead to better behavior, certainly for $\beta >1$ as fluctuations rise in this case, in part due to fewer values of $N$ being suitable for extraction since the OTOC expansion being valid requires $N \gtrsim \beta J$.

\subsection{Efficiency and Lyapunov Exponents}\label{subsec:eff}

The previous sections have made evident that higher degrees of sparsity (lower $k$) require greater averaging (and hence greater computation times) in order to accurately extract the Lyapunov exponent. We now seek to determine whether the computational benefit inherit in sparse-SYK is retained after the large amounts of disorder averaging that must be performed. In this section, we present a rudimentary method to determine the relative efficiencies of sparse models as compared to all-to-all, granting that more sophisticated methodology could be employed to perhaps gain greater insight as to how they compare. Based upon our methodology, we find that the computational trade-off between inherent speed increase in sparse models and time increase due to greater disorder averaging roughly cancel out, meaning accurate Lyapunov extraction is no more efficient for low $k$ models as it is for all-to-all models. We also present the numerically calculated Lyapunov exponents for our given range of $k$'s.

Our rudimentary choice of calculating efficiency starts with choosing what qualifies as ``accurate" Lyapunov extraction. Given our ranges of simulated $D_g$, a goodness of fit of $R^2=0.9$ was our choice of where to begin considering the extraction as accurate enough. Since we do not have exact curves corresponding to how $R^2$ varies with $D_g$, we instead chose to analyze the corresponding values of $D_g$ where the numerically calculated values of $R^2$ were within 0.005 of $R^2=0.9$. This gives us a range of $D_g$ values, interpreted to mean that on average, samples extracted with those values of $D_g$ will produce curves for $\lambda_\text{fit}$ with a goodness of fit of $ R^2\approx 0.9$. For data that passes through $R^2=0.9$ multiple times, the largest continuous cluster is chosen.

We also need to quantify the time advantage sparse-models have over all-to-all models, which can be straightforwardly done by finding the time (on average) it takes to simulate a single disorder realization for our range of $N$'s ($N=16$ to $N=36$). The full list of times per $N$ is provided in Appendix \ref{app:compu_specs}, we present in this section just the total time for all values of $N$. By dividing the total time (in seconds) by the total time of the all-to-all model, one achieves a ``time factor", representing the ratio of how much faster a given sparse model is per realization as compared to the all-to-all model. Multiplying the previously described ranges of $D_g$ by these time factors will yield the relative number of disorder realizations needed to achieve accurate (as deemed by $R^2=0.9$) Lyapunov extraction. For the calculated values of the Lyapunov exponent and associated standard deviation, we average over the data with $R^2>0.9$ (rounded within 0.01), which all achieve convergence within those ranges. Results of this analysis can be found in Table \ref{tab:eff}. For discussion regarding the appearance of the $k=50$ extracted Lyapunov exponent being higher than that of the all-to-all model, refer to the end of section 4.2.

This analysis demonstrates that the necessity of greater disorder averaging required for accurate Lyapunov extraction within sparse-SYK roughly negates the computational benefit provided per realization by sparse models, since the relative efficiency is roughly comparable between all models.

\begin{table}[H]
    \begin{center}
    \begin{tabular}{ |p{4.5cm}||p{2cm}|p{2.2cm}|p{2.2cm}|p{2.3cm}|  }
 \hline
 \multicolumn{5}{|c|}{Efficiency Table (112 Samples for Time)} \\
 \hline
  & All-to-All & $k=50$ & $k=10$ & $k=4$\\
 \hline
 \hline
Total Time ($N=16$ to 36) & 08:49:51 & 01:06:44 & 00:17:11 & 00:07:18\\
     \hline
     Time Factor & 1 & 0.126 & 0.0324 & 0.0138\\
     \hline
     $D_g$ Range, $R^2 \approx 0.9$ & 1400-1500  & 9700-12250 & 53500-68000 & 90000-111000\\
     \hline
     \hline
     Relative Efficiency & 1400-1500 & 1222-1543 & 1733-2203 & 1242-1531\\
     \hline
     Lyapunov Exponent & 0.80 & 0.88 &  0.81 & 0.58\\
     \hline
     Standard Deviation & 0.16 & 0.30 & 0.28 & 0.34 \\
     \hline
\end{tabular}
\end{center}
\caption{\textit{Efficiency comparisons between all-to-all SYK and various sparse models. The total time is the average total time (over 112 samples) to simulate a single disorder realization. The time factor is the fraction of time (compared to all-to-all) spend per disorder realization. The $D_g$ range is the range of disorder groupings which yield a value of $R^2=0.9$ within 0.005. The relative efficiency is the range of $D_g$ multiplied by the corresponding time factor. The Lyapunov exponent and standard deviation is the average of the corresponding quantities over the range of $D_g$'s with $R^2>0.9$ from Figure \ref{fig:Dmaxmain}.}}
 \label{tab:eff}
\end{table}

\section{Conclusions}

We studied the quantum chaotic behavior in the sparse-SYK model as diagnosed by out-of-time ordered correlators (OTOCs). We approached the problem through both analytical and numerical methods. In the large $N$ limit, where the OTOCs are given by a sum of ladder diagrams, we found an algorithm for calculating the $1/kq$ corrections to the ladder diagrams. This allows us to obtain analytic expressions for ladder diagrams of arbitrary order expanded to arbitrary orders in $1/kq$ -- see Appendix \ref{app:algorithm} for details. Importantly, the kernel in \eqref{eq:kernelNextOrder} obtained in this procedure is dependent on the ladder diagram order, a feature not shared by the all-to-all SYK. Thus, the summation over ladder diagrams needs to be truncated at some order in $1/kq$, and the summation to extract the corrections to the early time exponential behavior of the OTOC cannot be performed with the same tools as in the all-to-all model. 

The numerical approach employed in this work followed the symmetry-based method in \cite{Kobrin:2020xms},  previously applied to the all-to-all SYK, which is the only known accurate method for extraction of the Lyapunov exponent from finite $N$ simulations. We found that sparse models with $k \lesssim 50$ require far larger averaging of disorder realizations for accurate extraction as compared to the all-to-all model. In a fixed regular hypergraph sparse model, the extraction of the Lyapunov exponent led to inconsistent results, i.e., two fixed regular hypergraphs with the same level of sparseness resulted in different Lyapunov exponents even when we significantly increase the number of disorder realizations. This indicates that the symmetry \eqref{symmetry} does not hold for a fixed hypergraph configuration. In order to recover the symmetry, we performed an additional average over regular hypergraphs with the same sparseness parameter $k$. We find that even with the extra average, the fluctuations in the Lyapunov exponent are still larger compared to the all-to-all SYK, and these fluctuations increase even more as $k$ decreases. 

As shown in Section \ref{sec:numerics}, a reliable extraction of the Lyapunov exponent is still possible for low values of $k$  but a very large number of  realizations is needed. 
Therefore, there is a trade-off between efficiency and variance in the extracted Lyapunov exponent in the sparse-SYK. One could argue that these fluctuations can be reduced through probing larger values of $N$. However, as we argued in Section \ref{subsec:largerN}, the opposite effect takes place for Lyapunov extraction in the sparse-SYK model. The reduction in computation time of a single realization is offset by the larger number of simulations needed to be performed. In our numerical tests, the offset of the two effects is roughly comparable, i.e., the computational resources required to achieve similar levels of accuracy for Lyapunov extraction are roughly the same between all-to-all and sparse models.

We remark that our numerical results are consistent with the previous statement in \cite{Garcia-Garcia:2020cdo} that quantum chaos is present in the sparse-SYK model for $k\sim\mathcal{O}(1)$, derived from level statistics \cite{Garcia-Garcia:2016mno}. Interestingly, the analysis of \cite{Garcia-Garcia:2020cdo} also revealed the existence of emergent symmetries in the sparse-SYK model. It is possible that these emergent symmetries are correlated with the large fluctuations observed in the Lyapunov exponent.

Our results suggest several directions to explore in the future.
\begin{itemize}
    \item In \cite{Shankar:2023pax} the authors calculate the Lyapunov exponent in a variant of the SYK model by solving the Schwinger-Dyson equations numerically. It would be interesting to investigate if this method can be an alternative way to extract the Lyapunov exponent of the sparse-SYK model.
    \item Our numerical results  indicate  that the early time behavior of the OTOCs is affected by the sparseness of the model. It would be interesting to further explore the summation of the ladder diagrams to determine the leading corrections to the early time exponential ansatz $\mathcal{F}\sim e^{\lambda t}/N$ in the sparse-SYK model.
    \item  A better understanding of the emergent symmetries found in \cite{Garcia-Garcia:2020cdo}, especially their effect at larger $N$, could also provide insights on how the sparseness affect the behavior of OTOC expansion \eqref{eq:otoc_expansion}.
\end{itemize}

\acknowledgments

The authors would like to thank Javier Mas, Alfonso Ramallo, Brian Swingle and Shenglong Xu for useful discussions.  The work of EC, BK, and AM was supported by the National Science Foundation under Grant Number PHY-2112725. BK is also supported by NSF PHY-2210562 and TG by PHY-1914679. The work of TG was performed under the auspices of the U.S. Department of Energy by Lawrence Livermore National Laboratory under contract DE-AC52-07NA27344. EC thanks the Instituto de F\'isica Te\'orica (IFT) at UAM, Madrid, for hospitality during the last stages of this work.  The authors also acknowledge the Texas Advanced Computing Center (TACC) at The University of Texas at Austin for providing HPC resources that have contributed to the research results reported within this paper. URL: http://www.tacc.utexas.edu.

\begin{appendices}
\section{Sparse SYK Ladder Diagram General Algorithm}\label{app:algorithm}

\subsection{Simplifying Diagram Calculations}

As seen from the calculations in Section \ref{sec:otocs_sparse_syk}, general ladder diagram calculations will heavily involve combinatorics, especially as their order increases. Additionally, since the kernel is dependent on the order of the diagram, finding a given ladder diagram to arbitrary order is a non-trivial procedure. However, as we will see from calculating the first several ladder diagrams, a pattern emerges. In this section, we will develop the notation and motivation behind this procedure.

Let us return to the example of the two rung ladder diagram
\begin{equation} \label{eq:ExactLabel}
    \begin{tikzpicture}[baseline=(current bounding box.center)]
        \begin{feynman}
            \vertex (ul) {\(i\)};
            \vertex[below=2cm of ul] (ll)  {\(i\)};
            \vertex[right=2cm of ul] (um);
            \vertex[right=2cm of ll] (lm);
            \vertex[right=2cm of um] (um2);
            \vertex[right=2cm of lm] (lm2);
            \vertex[right=2cm of um2] (ur) {\(j\)};
            \vertex[right=2cm of lm2] (lr) {\(j\)};
            
            \diagram* [] {
                (ul) -- (um) -- [edge label = \(c\)] (um2) -- (ur),
                (ll) -- (lm) -- [edge label' = \(c'\)] (lm2) -- (lr),
                (lm) -- [bend left, edge label = \(a\)] (um),
                (lm) -- [bend right, edge label' = \(b\)] (um),
                (lm2) -- [bend left, edge label = \(a'\)] (um2),
                (lm2) -- [bend right, edge label' = \(b'\)] (um2),
            };
        \end{feynman}
    \end{tikzpicture}
\end{equation}
General ladder diagrams require $c=c'$ to be $\mathcal{O}(\frac{1}{N})$. In this case, we notice the Gaussian moment expansion \eqref{2ndordercalc} simplifies to
\begin{align}
    \expval{J_{iabc}J_{iabc}J_{ja'b'c}J_{ja'b'c}} &= \expval{J_{iabc}J_{iabc}} \expval{J_{ja'b'c}J_{ja'b'c}} + 2\expval{J_{iabc}J_{ja'b'c}} \expval{J_{iabc}J_{ja'b'c}}.
\end{align}
Recall from \eqref{eq:variance} that only if two couplings' indices are equal (up to permutation) may the multiplication of the two be non-zero after averaging. Let us adopt a convention of labelling vertices from left to right in the ladder diagram. At the order of $1/N$ the vertices connected by rungs have the same indices, so we can label them by the same vertex number. Our convention will involve replacing couplings by their corresponding vertex numbers, for instance $J_{iabc} \to 1$. In this new labeling scheme, (\ref{eq:ExactLabel}) corresponds to

\begin{equation}
    \begin{tikzpicture}[baseline=(current bounding box.center)]
        \begin{feynman}
            \vertex (ul);
            \vertex[below=2cm of ul] (ll);
            \vertex[right=2cm of ul] (um) {\(1\)};
            \vertex[right=2cm of ll] (lm) {\(1\)};
            \vertex[right=2cm of um] (um2) {\(2\)};
            \vertex[right=2cm of lm] (lm2) {\(2\)};
            \vertex[right=2cm of um2] (ur);
            \vertex[right=2cm of lm2] (lr);

            \diagram* [] {
                (ul) -- (um) -- [edge label = \(\)] (um2) -- (ur),
                (ll) -- (lm) -- [edge label' = \(\)] (lm2) -- (lr),
                (lm) -- [bend left, edge label = \(\)] (um),
                (lm) -- [bend right, edge label' = \(\)] (um),
                (lm2) -- [bend left, edge label = \(\)] (um2),
                (lm2) -- [bend right, edge label' = \(\)] (um2),
            };
        \end{feynman}
    \end{tikzpicture}
\end{equation}
Similarly, we can replace \eqref{2ndordercalc} with
\begin{align}
    \expval{1122}_J &= \expval{11}_J \expval{22}_J + 2\expval{12}_J \expval{12}_J.
\end{align}

In our new convention, the term $\expval{12}_J$ should be interpreted to mean that after averaging, $\expval{12}_J = \frac{J^2}{kq}$, only if vertices 1 and 2 have indices which are equal up to permutations of each other. Recall that the two relevant cases here are if vertex 1 and 2 are or are not permutations of each other. In the latter case (not permutations), $\expval{12}_J=0$ (not permutations), so only the term $\expval{11}_J\expval{22}_J$ contributes. Let us represent the sparsity couplings $\expval{x_{abcd}x_{abcd}} = \expval{x_{abcd}}$ as $\expval{11}_x = \expval{1}_x$, we found the pruning couplings contributed a factor of $\expval{1}_x \expval{2}_x = p^2$.

In the other case, vertices 1 and 2 do have indices equal up to permutations. However since all vertices are equal, the averaging over $x$ reduces to $\expval{1122}_x = \expval{1}_x = p$, so overall only a single factor of $p$ is contributed from the $x$'s. Recall that the total expression (in our new notation) is $\expval{1122}_J\expval{1122}_J$, so the factors from averaging over $J$'s must also be considered. Overall, our goal is to find the numerical/combinatorial terms multiplying these $p$'s since these are what determine the order in $1/kq$, which are embedded in averaging over the $J$'s. By using \eqref{2ndordercalc} and dropping the factors of $J^2$ (these can be pulled out across the entire expression), let us create a table which counts all relevant contributions: 

\begin{center}
\begin{tabular}{|| c | c | c ||}
\hline
$\expval{1122}_J$ = & $\expval{11}_J \expval{22}_J + 2\expval{12}_J \expval{12}_J$ & Total \\
\hline
\hline
$p^2$ & 1 & 1 \\ 
\hline
$p$ & 1 \hspace{3mm} + \hspace{3mm} 2 & 3 \\
\hline
\end{tabular}
\end{center}

Let us repeat this process for the ladder diagram with three rungs in full detail. For this case we have
\begin{align}
    \expval{112233}_J & = \expval{11}_J \expval{2233}_J + 2\expval{12}_J \expval{1233}_J + 2\expval{13}_J \expval{1223}_J 
    \label{3rdorder} \\
    & = \expval{11}_J \expval{22}_J \expval{33}_J + 2\expval{11}_J\expval{23}_J\expval{23}_J + 2\expval{12}_J \expval{12}_J\expval{33}_J + \nonumber \\
    & \qquad \qquad \qquad \qquad \qquad + 2\expval{13}_J \expval{13}_J\expval{22}_J+8\expval{13}_J\expval{12}_J\expval{23}_J.
    \label{3rdorderexpand}
\end{align}
Let us organize again by order in $p$, the case $p^3$ is given by no vertices being equal up to permutations. In that case, as was the case for $n=2$, only the first term contributes. For order $p^2$, the counting gets trickier. There are three cases which contribute, if vertices 1 and 2 are equal, vertices 1 and 3 equal, and vertices 2 and 3 are equal. For 1 and 2 equal, we have $\expval{13}_J=0=\expval{23}_J$, leaving a coefficient of $1+2=3$. Each other case of order $p^2$ has this same numerical coefficient, making the total numerical coefficient 9. For order $p$, all terms contribute giving a numerical constant of 15.

Since all we care about is the order of $p$, not the specific case of which vertex is a permutation of another, we notice that every term with coefficient 2 in \eqref{3rdorderexpand} is the same up to relabelling of vertices, therefore they contribute at the same order of $p$. We therefore wish to label these terms by a factor which suggests that they all contribute equally at the same order of $p$. If the highest order of $p$ a term contributes is $p^k$, we will relabel these terms by the factor $\mathcal{P}^k$. In this way may rewrite \eqref{3rdorderexpand} as
\begin{align}\label{chiintro}
   \expval{112233}_J \sim \mathcal{P}^3 + 6\mathcal{P}^2 + 8\mathcal{P}.
\end{align}
Our current notation is still burdensome, as the expansion contains terms with both $\expval{\cdot}_J$ and $\expval{\cdot}_x$, our goal will be to combine them. We can do this by notationally decomposing our average over $x$ into groups of 2 vertices (like that which is naturally done from the Gaussian moment expansion for the $J$'s). To see this concretely, let $\expval{\cdot}_d$ be a term in the decomposed average, such that $\expval{aabb\hdots}_x \equiv \expval{aa}_{d} \expval{bb}_d \expval{\hdots}_d$. Note that these decomposed terms are \textit{not} a true average nor should they be interpretted as one, they can recombine/decompose freely (to conserve the relevant structure) but only after a total recombination of the $\expval{\cdot}_d$'s do they represent something physically meaningful, namely it recovers the original average over $x$. Their purpose is simply for notational convenience. The use of this becomes apparent when we let $\expval{ab} \equiv \expval{ab}_J \expval{ab}_d$, meaning we have the following identities 
\begin{gather}
    \expval{aa}_J \expval{aa}_x = \expval{aa} \sim \mathcal{P} \\
    \langle ab \rangle_J \langle ab \rangle_J \expval{abab}_x = \Big( \expval{ab}_J \expval{ab}_d \Big) \Big( \expval{ab}_J \expval{ab}_d \Big) =\expval{ab} \expval{ab} \sim \mathcal{P} \label{recrules} \\
    \langle ab \rangle_J \langle ac \rangle_J \expval{abac}_x = \expval{ab} \expval{ac} \sim \langle bc \rangle.
\end{gather}
The second identity is true because the term is only non-zero when $\expval{ab}_J \neq 0$, meaning vertices $a$ and $b$ are permutations of one another. In that case, $\expval{abab}_x = \expval{a}_x = p$, meaning the highest order the term can contribute is $p^1$, hence the term corresponding to $\mathcal{P}$. The third identity can not be further simplified (ie, we cant set it equal to $\mathcal{P}$) because if $\expval{bc} \sim \mathcal{P}$, that would imply $\expval{bc} \expval{bc} \sim \mathcal{P}^2$, which contradicts the second identity. The origin of this comes from equivalent vertices appearing in pairs in the ladder diagram.

Another way to see why we make this replacement is to note as stated earlier that a relabelling of indices will not change the way in which the counting at each order of $p$ is done, therefore we can always choose to relabel terms such that they combine into the form \eqref{chiintro}, for example
\begin{align}
    \expval{112233} & = \expval{11} \expval{2233} + 2\expval{12} \expval{1233} + 2\expval{13} \expval{1223} \nonumber \\
    & = \expval{11} \expval{2233} + 4\expval{12} \expval{1233} \nonumber \\
    & = \expval{11} \expval{22} \expval{33} + 2\expval{11}\expval{23}\expval{23} + 4\expval{12} \expval{12}\expval{33}+8\expval{13}\expval{12}\expval{23} \nonumber \\
    & = \expval{11} \expval{22} \expval{33} + 6\expval{11}\expval{23}\expval{23} +8\expval{13}\expval{12}\expval{23} \nonumber \\
    & \sim \mathcal{P}^3 + 6\mathcal{P}^2 + 8\mathcal{P}.
\end{align}
Note that the process of finding the coefficient at $\mathcal{O}(p^k)$ is not as simple as just reading off the coefficient of $\mathcal{P}^k$ as written above, recall for example the term $\expval{11} \expval{22} \expval{33} \sim \mathcal{P}^3$ contributed for every possible case of order $p^2$. Therefore we need to multiply each coefficient by all the cases where it contributes. Let us define two vertices as ``connected" (forming a ``connection") if their indices are equal up to permutations. Additionally, we will define the number of ``free" vertices as the number of vertices whose indices may take on independent values while still ensuring the term is non-zero. For instance, $\expval{aa}\expval{bb}$ has two free vertices, since vertices $a$ and $b$ can both take on independent indices. On the other hand, $\expval{ab}\expval{ab}$ is only non-zero if $a$ and $b$ are connected, meaning there is only one free vertex (say $a$, since $b$ must have its vertices equal up to permutation of $a$).

For the third-order ladder diagram, there must therefore be one additional connection (involving two vertices) for the leading term to contribute at order $p^2$. The term $\expval{11} \expval{22} \expval{33}$ has three free vertices (1,2, and 3), of which we choose two to be equal at order $p^2$, hence its contribution will be multiplied by ${3 \choose 2}$. This is not the case for the term $\expval{11}\expval{23}\expval{23}$; vertices 2 and 3 already must be equal for the term to be non-zero.

The general counting procedure will therefore be as follows: create a table with the left column denoting the order of $p$, and the top row corresponding to the expansion of the average of the couplings in terms of $\mathcal{P}$. For order $p^k$, take the coefficient of $\mathcal{P}^k$ and all higher powers of $\mathcal{P}$ and multiply them by the number of combinations required to form enough connections to be of order $p^k$. For instance at order $p$, $\expval{11} \expval{22} \expval{33}$ must undergo 2 connections involving 3 free indices, so that means out of 3 free vertices choose 3 to be connected, which is ${3 \choose 3}=1$. Similarly for $\expval{11} \expval{23} \expval{23}$ there are 2 free vertices (2 or 3 must be equal to the other to be non-zero, removing a free vertex) with one connection involving 2 vertices, implying a factor of ${2 \choose 2}=1$. In general this pattern will hold. Another reason to see why their combinatoric factors should be 1 is that at order $p$, all vertices are equal so there is only one way to combine them all. So our table for the third order ladder diagram becomes
\begin{center}
\begin{tabular}{|| c | c | c ||}
\hline
$\expval{112233}$ = & $\mathcal{P}^3+6\mathcal{P}^2+8\mathcal{P}$ & Total \\
\hline
\hline
$p^3$ & 1 & 1 \\
\hline
$p^2$ & 1 ${3 \choose 2}$ +  6 & 9 \\ 
\hline
$p$ & 1 \hspace{1mm} + \hspace{1mm} 6 \hspace{1mm} + \hspace{1mm} 8 & 15 \\ 
\hline
\end{tabular}
\end{center}

We will continue to 4th order to further demonstrate how these combinatoric factors work. At order $p^2$, the factor $\mathcal{P}^4$ undergoes two connections, however this time there are two cases. With four free vertices, three vertices could be all connected (making 2 total connections) or two sets of two connections could be formed. The latter of which has ${4 \choose 2}$ for the first choice and ${2 \choose 2}=1$ for the second choice. Therefore the table becomes
\begin{center}
\begin{tabular}{|| c | c | c ||} 
\hline
$\expval{112233}$ = & $\mathcal{P}^4+12\mathcal{P}^3+44\mathcal{P}^2+48\mathcal{P}$ & Total \\
\hline
\hline
$p^4$ & 1 & 1 \\ 
\hline
$p^3$ & ${4 \choose 2}$+ 12 & 18 \\ 
\hline
$p^2$ & ${4 \choose 3}$ +${4 \choose 2}$${2 \choose 2}$  + 12${3 \choose 2}$ + 44 & 90 \\ 
\hline
$p$ & 1\hspace{2mm}+\hspace{2mm}12\hspace{2mm}+\hspace{2mm}44\hspace{2mm}+\hspace{2mm}48 & 105 \\ 
\hline
\end{tabular}
\end{center}
For this procedure to work for arbitrary order of ladder diagram, we must find a closed form for the expansion of averaged couplings into $\mathcal{P}$'s as done above, and find a general method for calculating the combinatoric factors. These will both be addressed in the next section.

\subsection{General Procedure}

Let us first address the combinatoric factors. Recall based on our definition of free vertices that the monomial $\mathcal{P}^k$ has $k$ free vertices. At order $p^k$ no connections need be formed, so the term does not gain a combinatoric factor (although still has a coefficient multiplying $\mathcal{P}^k$). The term $\mathcal{P}^{k+1}$ has $k+1$ free vertices, and must undergo 1 connection to be of order $p^k$. Out of the $k+1$ free vertices, two must be chosen to form a connection, hence the combinatoric factor becomes ${k+1 \choose 2}$. For $\mathcal{P}^{k+2}$, there must be two connections between free vertices at order $p^k$. However we must choose how to partition these connections, they can either all be grouped together, meaning three free vertices are all permutations of one another (2 connections) or be separated into two different connections (1+1 connections). For any grouping of connections, there is always one more vertex than connection, which is what the combinatorics are performed on. So for the two connections grouped together, the combinatoric factor will be ${k+2 \choose 2+1}={k+2 \choose 3}$. For two separate connections, we do the combinatorics out of all the free vertices with one connection ${k+2 \choose 1+1}$ and then do combinatorics on the remaining free vertices for the next connection ${k \choose 1+1}$ making a total combinatoric factor of ${k+2 \choose 2}{k \choose 2}$.

As an example, let us calculate the order $p^2$ contribution at $n=6$ order in ladder diagrams. Additionally, we will denote the $n$th order diagram by $a_n(\mathcal{P})$. It turns out at $n=6$ this becomes
\begin{align}\label{chi6}
   a_6(\mathcal{P}) \equiv \expval{112233445566} \sim  \mathcal{P}^6+30\mathcal{P}^5+340\mathcal{P}^4+1800\mathcal{P}^3+4384\mathcal{P}^2+3840\mathcal{P}.
\end{align}
So at order $p^2$, we start with the combinatoric factors of $\mathcal{P}^6$. $6-2=4$ connections must be made. These connections can be partitioned into the following sets: $\{4\}$, $\{3,1\}$, $\{2,2\}$, $\{2,1,1\}$, $\{1,1,1,1\}$. We add one to each of these sets to represent the possible combinatorics on the free vertices, so $\{5\}$, $\{4,2\}$, $\{3,3\}$, $\{3,2,2\}$, $\{2,2,2,2\}$. We then need to find the combinatoric factors by taking the $j=6$ free vertices and doing successive choose functions on each member of the set, this will give a factor of the form
\begin{align}
    \mathcal{P}^6 \hspace{1mm} \text{at order}\hspace{1mm} p^2:\hspace{3mm} {6 \choose 5}+{6 \choose 4}{2 \choose 2}+{6 \choose 3}{3 \choose 3}+{6 \choose 3}{3 \choose 2}{1 \choose 2}+{6 \choose 2}{4 \choose 2}{2 \choose 2}{0 \choose 2}
\end{align}
Notice that terms like ${0 \choose 2} = 0 ={1 \choose 2}$ make those coefficients vanish, so those terms do not contribute. These factors effectively mean it is not possible to construct the sufficient connections with that partition set. Repeating this process for all $\mathcal{P}$'s along with using the coefficients from \eqref{chi6} yields

\begin{align}
    {6 \choose 5}+{6 \choose 4}{2 \choose 2}+{6 \choose 3}+30\left( {5 \choose 4} +  {5 \choose 3}\right)+340\left( {4 \choose 3} +  {4 \choose 2}\right)+1800{3 \choose 2} + 4384 =13675
\end{align}
Since $\expval{112233445566}$ contains 6 powers of $\expval{J_{abcd}^2}\sim 1/p$, at order $p^2$ to total contribution from $p$'s is $1/p^4 \sim 1/(kq)^4$, which matches our coefficient in the expansion from \eqref{expansions}.

The general procedure for combinatoric factors should become clear from this. At order $p^k$ and ladder diagram order $n$, the combinatoric factor for $\mathcal{P}^j$ is as follows: if $k>j$, then the factor is zero. If $k=j$, the factor is 1 (effectively zero choices of free vertices need to be made). If $j>k$, then $j-k \equiv m$ number of connections need to be made. The number of distinct combinatoric factors is equal to the number of partition numbers of $m$, since $m$ different connections must be partitioned into groups. For any given group of connections, the number of vertices is one more than the number of connections in said group. Create a list corresponding to each possible partition of $m$ (so the number of different lists will be the number of partitions of $m$), and add 1 to each member of the list to represent the number of free vertices. For a given list $m_l$, place each member $m_{ls}\in m_l$ into the lower value of a choose function, ${\cdot \choose m_{ls}}$, and multiply all of these choose functions together. The upper value will start with $j$ (coming from the free vertices available from $\mathcal{P}^j$), and then decrease by the number of free vertices connected $m_{l1}$, which will be of the form ${j \choose m_{l1}}{j-m_{l1} \choose m_{l2}}{j-m_{l1}-m_{l2} \choose m_{l3}}\cdots$. This process will repeat for all sets of partitions $m_l$, and then for all values $j$ from $\mathcal{P}^j$ such that $j>k$. After multiplication by the corresponding coefficient of $\mathcal{P}^j$ and summing all terms together, this will form the total factor contributing to $p^k$.

The last step of our procedure is to find the expansion of $a_n(\mathcal{P})$ in terms of $\mathcal{P}$'s for arbitrary $n$. The first step is to form a recursion relation for $\langle 11 \cdots nn \rangle$ in terms of $\mathcal{P}$'s. There are 2 ways for vertices 1 and 2 to be paired, and $n-1$ different vertices that may be paired with vertex 1. Since we have the ability to relabel vertices without impacting the expansion of $\mathcal{P}$'s, we arrive at
\begin{align}
    a_n(\mathcal{P}) = \langle 11\cdots nn \rangle = \expval{11} \langle \underbrace{2233\cdots nn}_{n-1} \rangle + 2(n-1)\langle 12 \rangle \langle 12 \underbrace{33\cdots nn}_{n-2} \rangle,
\end{align}
where the underbrace refers to the number of unique vertices (before considering permutations). Since $\expval{22\cdots nn}$ involves $n-1$ unique vertices which we are free to relabel, it is therefore equivalent to $a_{n-1}(\mathcal{P})$. Additionally $\expval{11}=\mathcal{P}$, which leaves the second term. We can choose to define the second term as $b_k(\mathcal{P})$ and find its recursion relation by again expanding and relabelling
\begin{align}
    b_k(\mathcal{P}) \equiv \langle ab \rangle \langle ab \underbrace{11\cdots kk}_k \rangle = \expval{ab}\expval{ab} \langle 11\cdots kk \rangle +2k\langle ab \rangle \expval{a1} \langle b1 \underbrace{22\cdots kk}_{k-1}\rangle.
\end{align}
Using \eqref{recrules} and our above expressions we can rewrite our recursion equations as a set of two coupled recursion relations
\begin{align}
    a_n(\mathcal{P})  & = \mathcal{P} a_{n-1}(\mathcal{P}) +2(n-1)b_{n-2} \\
    b_k(\mathcal{P})  & = \mathcal{P} a_{k}(\mathcal{P}) +2kb_{k-1}(\mathcal{P}).
\end{align}
These sets of equations can be solved to find
\begin{align}
    a_n(\mathcal{P})  = 2^{n-1}\mathcal{P} \frac{\Gamma(\frac{\mathcal{P}}{2}+n)}{\Gamma(1+\frac{\mathcal{P}}{2})}.
\end{align}
While this solution may at first glance seem complicated, an expansion for the first few $n$'s makes the pattern apparent:
\begin{align}
    a_1(\mathcal{P}) & = \expval{11} = \mathcal{P} \\
    a_2(\mathcal{P}) & = \expval{1122} = \mathcal{P}(\mathcal{P} +2) \\
    a_3(\mathcal{P}) & = \expval{112233} = \mathcal{P}(\mathcal{P} +2)(\mathcal{P}+4) \\
    a_4(\mathcal{P}) & = \expval{11223344} = \mathcal{P}(\mathcal{P} +2)(\mathcal{P}+4)(\mathcal{P} + 6).
\end{align}
Our goal is therefore to find the coefficient of $\mathcal{P}^k$ for every $1<k<n$ for the $n$th order ladder diagram. This is a solved problem for the polynomial  $x(x+1)(x+2)\cdots (x+n-1)$, where the coefficient of $x^k$ is given by the unsigned \textit{Stirling numbers of the first kind}, and are denoted by $\left[\begin{smallmatrix} n\\k \end{smallmatrix} \right]$ \cite{knuth1989concrete}. Therefore the solution of $a_n(\mathcal{P})$ is given by
\begin{align}
    a_n(\mathcal{P}) = \sum_{k=1}^n c_k\mathcal{P}^k;\hspace{5mm} c_k=2^{n-k}\left[\begin{smallmatrix} n\\k \end{smallmatrix} \right].
\end{align}
Putting this all together, we arrive at the algorithm for calculating the $n$th order ladder diagram:
\begin{enumerate}
    \item Find the expansion of $a_n(\mathcal{P}) = \sum_{k=1}^n c_k\mathcal{P}^k$ by using $c_k=2^{n-k}\left[\begin{smallmatrix} n\\k \end{smallmatrix} \right]$\
    \item To find the coefficient of $p^k$ (which corresponds to $\frac{1}{(kq)^{n-k}}$):
    \begin{enumerate}
        \item For every $\mathcal{P}^j$ such that $j\geq k$:
        \begin{enumerate}
            \item Find the number of connections needed to be of order $p^k$, which is $j-k\equiv m$
            \item Calculate the sets of partitions for $m$, and then add 1 to every element to represent the free vertices
            \item For each set, place the elements into the lower values of multiplied choose functions, and fill the upper values with $j-\sum m_l$, where $m_l$ correspond to the lower values of any choose function multiplying to the left.
            \item Sum all these choose functions together, and then multiply by the coefficient $c_j$, call this number $h_{jk}$
        \end{enumerate}
        \item Sum together all coefficients corresponding to $p^k$, which will be $\sum_{j \geq k}^n h_{jk}$, this is the coefficient of $p^k$, call this coefficient $g_k$
    \end{enumerate}
    \item The $n$th order ladder diagram will have an expansion in terms of $p$'s as $\sum_{z=1}^n g_k p^k$.
\end{enumerate}

As a demonstration, we will calculate the first 4 coefficients of an $n$th order ladder diagram using these techniques:
\begin{center}
\begin{tabular}{|| c | c | c ||}
\hline
$\expval{11\cdots nn}$ = & $\mathcal{P}^n+2\left[\begin{smallmatrix} n\\n-1 \end{smallmatrix} \right]\mathcal{P}^{n-1}+4\left[\begin{smallmatrix} n\\n-2 \end{smallmatrix} \right]\mathcal{P}^{n-2}+8\left[\begin{smallmatrix} n\\n-3 \end{smallmatrix} \right]\mathcal{P}^{n-3} + \cdots$ & Total \\
\hline
\hline
$p^n \sim 1$ & 1 & 1 \\
\hline
$p^{n-1} \sim \frac{1}{(kq)}$ & ${n \choose 2} + \left[\begin{smallmatrix} n\\n-1 \end{smallmatrix} \right]2$ & $\frac{3}{2}n(n-1)$ \\ 
\hline
$p^{n-2} \sim \frac{1}{(kq)^2}$ & ${n \choose 3} + {n \choose 2}{n-2 \choose 2} + 2\left[\begin{smallmatrix} n\\n-1 \end{smallmatrix} \right]{n-1 \choose 2} + 4\left[\begin{smallmatrix} n\\n-2 \end{smallmatrix} \right]$ & $\frac{5}{4}n(n-1)^2(n-2)$ \\
 \hline
$p^{n-3} \sim \frac{1}{(kq)^3}$ & ${n \choose 4} + {n \choose 3}{n-3 \choose 2} + {n \choose 2}{n-2 \choose 2}{n-4 \choose 2}$  & $\frac{1}{24}n(n-1)(n-2)(n-3)$\\
 & $+ 2\left[\begin{smallmatrix} n\\n-1 \end{smallmatrix} \right]\left( {n-1 \choose 3} + {n-1 \choose 2}{n-3 \choose 2}  \right) + 4\left[\begin{smallmatrix} n\\n-2 \end{smallmatrix} \right]{n-2 \choose 2} + 8\left[\begin{smallmatrix} n\\n-3 \end{smallmatrix} \right]$ & $\times(77+n(19n-69))$ \\
 \hline
\end{tabular}
\end{center}
	
\section{Computational Specifications}\label{app:compu_specs}

All numerical simulations presented in this paper were done through the Texas Advanced Computing Center (TACC) on the Frontera supercluster \cite{frontera}. The supercluster is comprised of 8,386 nodes (of which a maximum of 200 were ever used per simulation), with each node contains 56 cores. For ease of access, compute node specifics are transcribed here:

\begin{center}
{\small
\begin{tabular}{ ||p{4.5cm}||p{7cm}|| }
    \hline
    \multicolumn{2}{||c||}{Intel Xeon Platinum 8280 ("Cascade Lake") Compute Node} \\
     \hline
     \hline
     Total cores per CLX node: &	56 cores on two sockets (28 cores/socket)\\
     \hline
     Hardware threads per core:	& 1\\
     \hline
     Clock rate:&	2.7GHz nominal\\
     \hline
     RAM:&	192GB (2933 MT/s) DDR4\\
     \hline
     Cache:	&32KB L1 data cache per core;\\
    &1MB L2 per core;\\
    &38.5 MB L3 per socket.\\
    &Each socket can cache up to 66.5 MB\\
    \hline
    Local storage:	&144GB /tmp partition on a 240GB SSD.\\
    \hline
    \end{tabular}
}
\end{center}
    
\begin{center}
    \begin{tabular}{ ||p{3cm}||p{2cm}|p{2cm}|p{2cm}|p{2cm}||  }
     \hline
     \multicolumn{5}{||c||}{Time Comparisons Average, $q=4 \hspace{3mm}\beta = 1$ (112 Samples)} \\
     \hline
     Size (N) & All-to-All & k=50 & k=10 & k=4\\
     \hline
     \hline
     16 & 00:00:02 & 00:00:02 & 00:00:00 & 00:00:00\\
     \hline
    18 & 00:00:05 & 00:00:03 & 00:00:01 & 00:00:00\\
    \hline
    20 & 00:00:16 & 00:00:05 & 00:00:01 & 00:00:01\\
    \hline
    22 & 00:00:48 & 00:00:11 & 00:00:03 & 00:00:01\\
    \hline
    24 & 00:00:48 & 00:00:14 & 00:00:04 & 00:00:01\\
    \hline
    26 & 00:02:14 & 00:00:33 & 00:00:10 & 00:00:04\\
    \hline
    28 & 00:06:22 & 00:01:19 & 00:00:23 & 00:00:10\\
    \hline
    30 & 00:17:44 & 00:03:09 & 00:00:52 & 00:00:22\\
    \hline
    32 & 00:47:42 & 00:07:15 & 00:01:55 & 00:00:48\\
    \hline
    34 & 02:06:27 & 00:16:26 & 00:04:14 & 00:01:48\\
    \hline
    36 & 05:27:22 & 00:37:26 & 00:09:28 & 00:04:02\\
     \hline
\end{tabular}
\end{center}

\begin{center}
\begin{tabular}{ ||p{3cm}||p{2cm}|p{2cm}|p{2cm}|p{2cm}|| }
     \hline
     Total (16 to 36) & 08:49:51 & 01:06:44 & 00:17:11 & 00:07:18\\
     \hline
     Time Factor & 1 & 0.126 & 0.0324 & 0.0138\\
     \hline
\end{tabular}
\end{center}

\section{Fixed Hypergraph Configuration \& Tests of Robustness}\label{app:efficiency} \label{app:robustness}

The analytical tools used in Section \ref{sec:numerics} can be applied to a fixed hypergraph configuration, results of which can be seen in Figure \ref{fig:sethyper}. For this set of figures the mean, standard deviation, and $R^2$ must be separated as their outputs lie in vastly different ranges. The lack of convergence and small value of $R^2$ demonstrate that a given fixed hypergraph exhibits complete disagreement with the symmetry based methods. These results remain true for various choices of hypergraph seeds, $k$, and $D_g$.

As a check that the results of this paper are robust to the various choices of analysis used in the main body of this work, we provide here alternative methodology to demonstrate our conclusions remain consistent. The first encountered choice which could make a significant change is which value $\mathcal{F}^*$ is chosen in the analysis. The main body of the paper utilized $\mathcal{F}^* = 0.4$, however in Figures \ref{fig:F3}, \ref{fig:F5} we demonstrate that the behavior exhibited in Figure \ref{fig:Dmaxmain} is consistent with $\mathcal{F}^* = 0.3$ and $\mathcal{F}^* = 0.5$ respectively.

The next natural choice of variation is the size of the groupings statistically analyzed for each $D_g$, as it may be argued the proper choice would be to have a consistent number of samples regardless of $D_g$. Hence, we provide plots for two different choices, the first being 10 samples (Figure \ref{fig:allerror}) and the second being 20 samples  (Figure \ref{fig:D20}). These plots are compared to Figure \ref{fig:Dmaxmain} from the main body, they exhibit a higher degree of fluctuation for smaller values of $D_g$. This result is expected, as lower values of $D_g$ (meaning a given Lyapunov exponent is extracted with low number of realizations) will inherently have higher error than for higher values of $D_g$. This inherent instability makes Lyapunov extraction more inconsistent, hence the choice of extraction used in the main body. However even given these choices, our main findings (namely $R^2 \to 1$ can be achieved at low $k$ with sufficiently high $D_g$) still hold. 

Lastly, we analyze whether including the same lower bound $N=16$ regardless of the maximum simulated $N$ is a good choice. Instead, we could keep a set number of $N$ in a given range (in this case, 11 different $N$'s) leading to a change in the lower bounds. Hence the ranges compared are $N=16$ to $N=36$, $N=18$ to $N=38$, $N=20$ to $N=40$, and $N=22$ to $N=42$, which can be found in Figure \ref{fig:Nset}. As compared to Figure \ref{fig:Nmax}, these simulations actually exhibit worse correspondence with the symmetry based model, and exhibit higher amounts of random fluctuations. This behavior is consistent with our arguments from Section \ref{subsec:largerN}; by disregarding data points from lower $N$ values, the remaining data (higher $N$'s) of the sparse models have higher discrepancy with the all-to-all model, hence requiring more averaging.

\afterpage{
\begin{figure}[H]
\centering
\begin{subfigure}{0.49\textwidth}
    \includegraphics[width=\textwidth]{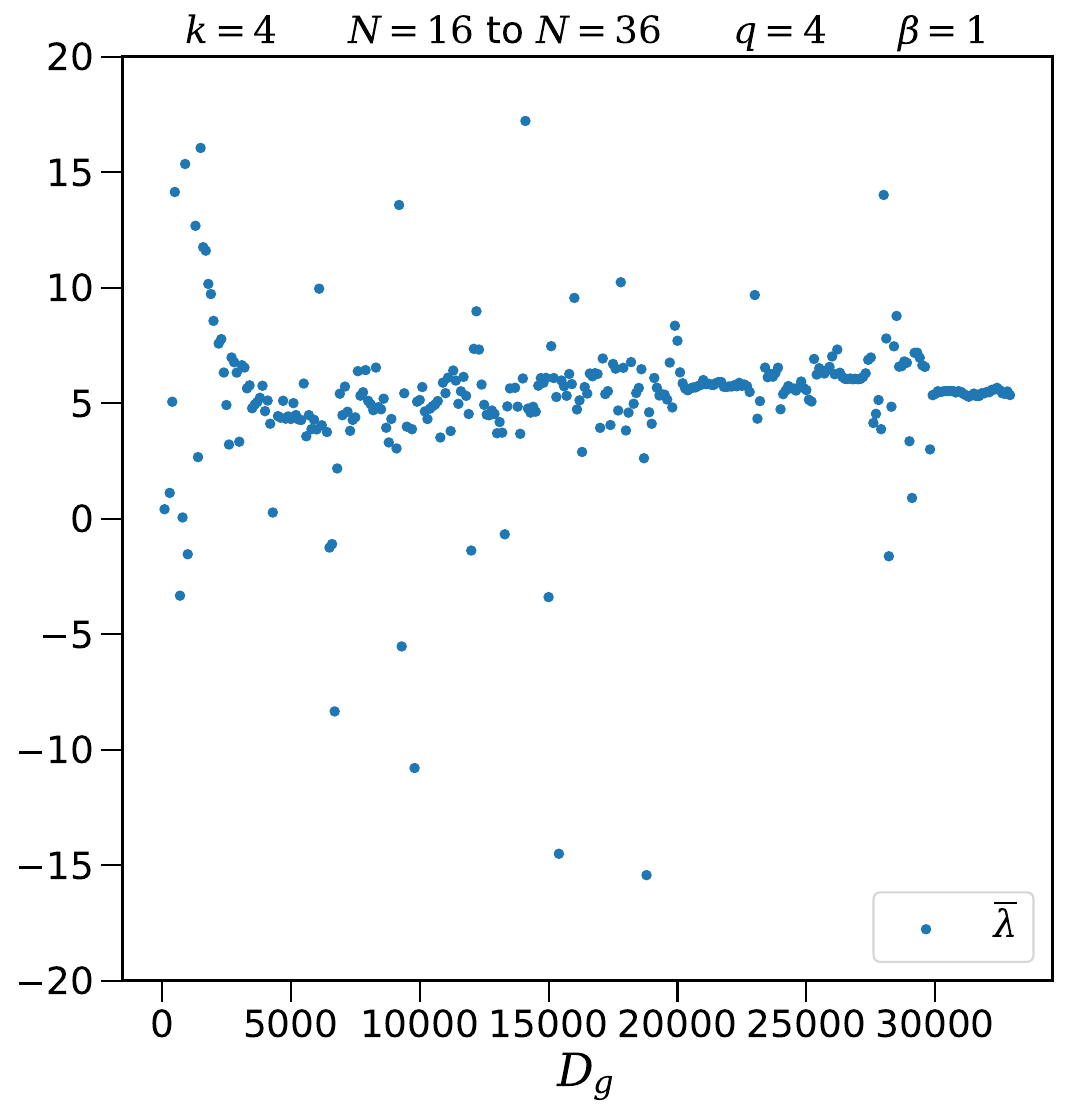}
    \caption{}
\end{subfigure}
\hfill
\begin{subfigure}{0.49\textwidth}
    \includegraphics[width=\textwidth]{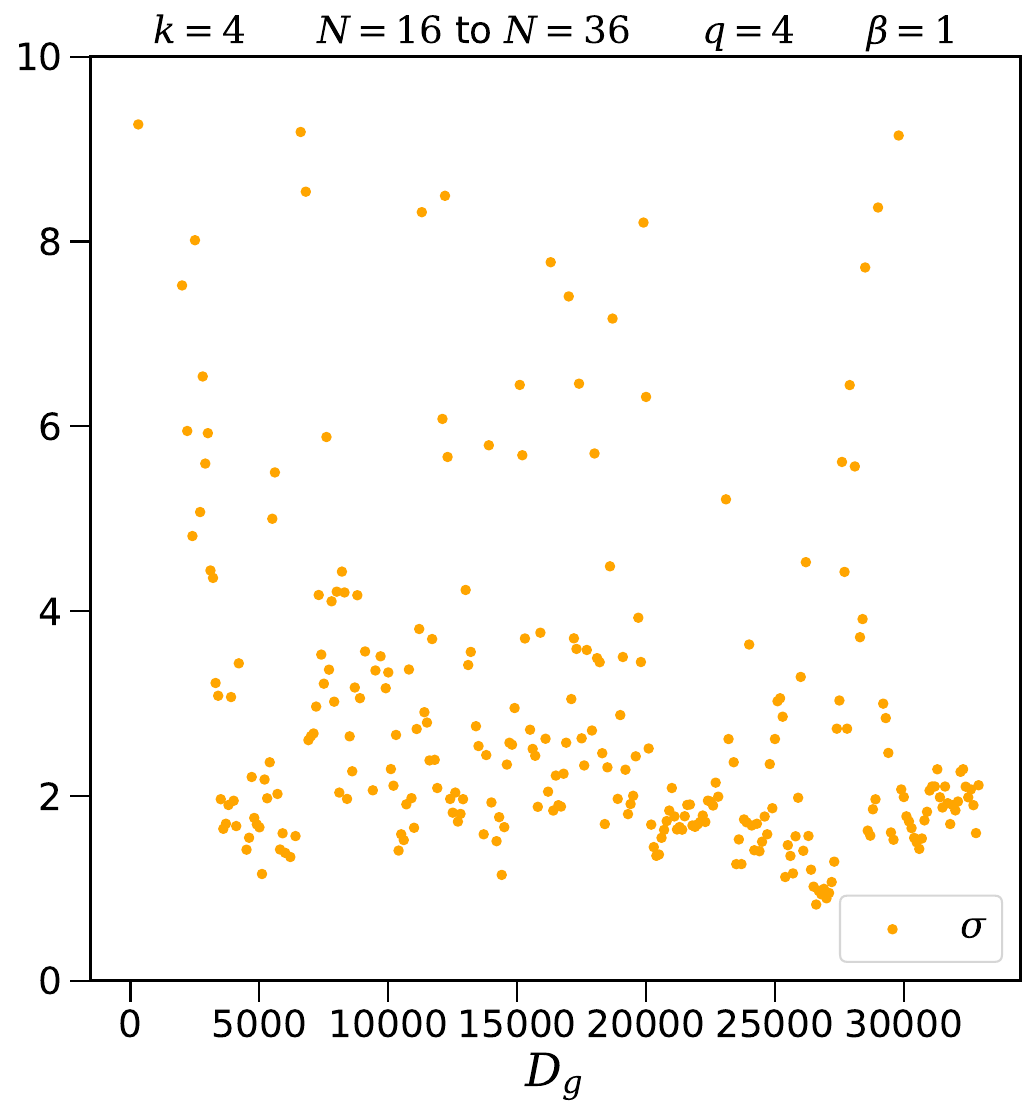}
    \caption{}
\end{subfigure}

\begin{subfigure}{0.49\textwidth}
    \includegraphics[width=\textwidth]{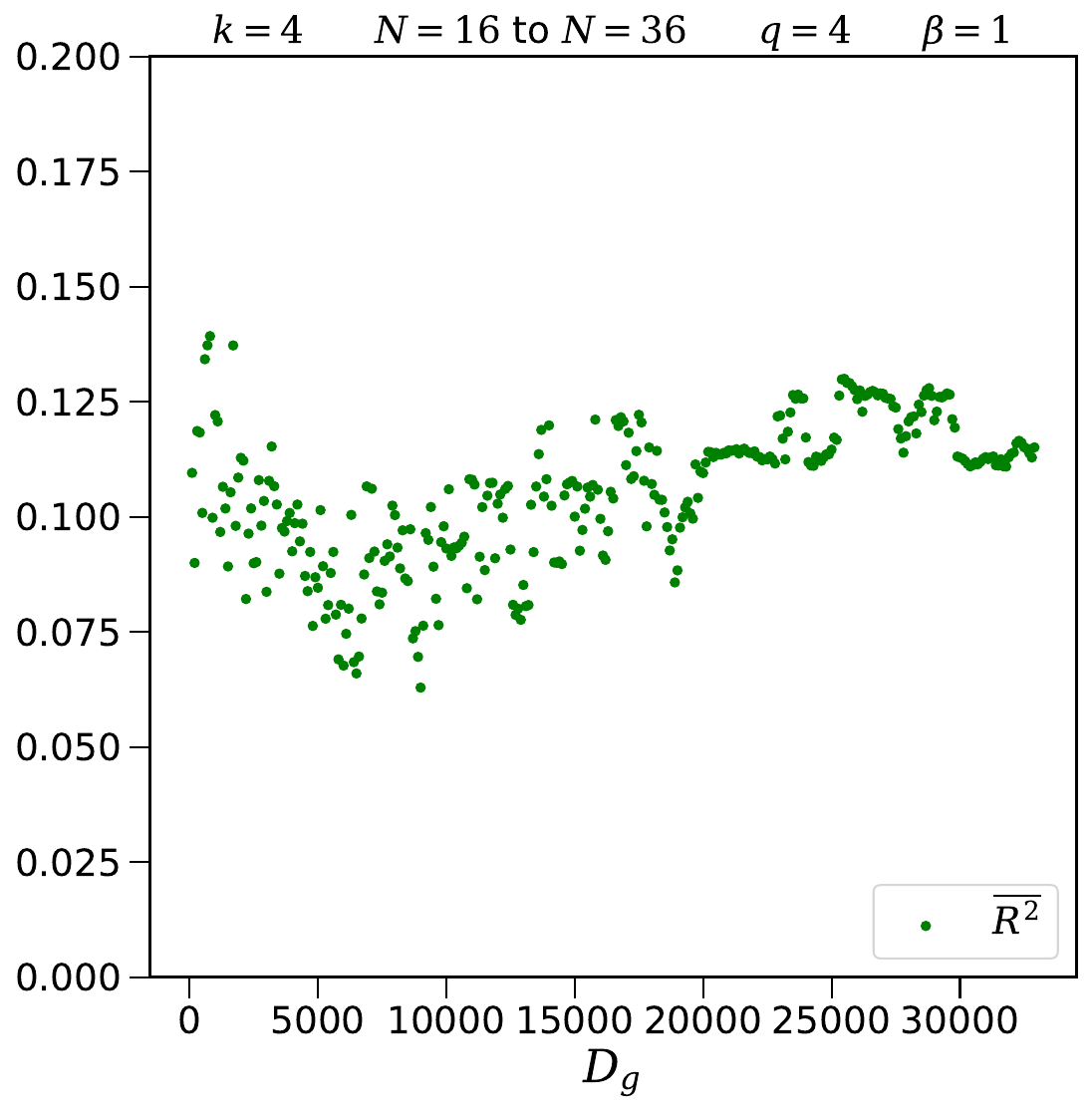}
    \caption{}
\end{subfigure}
        
\caption{\textit{Data analysis of Lyapunov extraction for a fixed hypergraph seed (meaning for each $N$, the Hamiltonian interaction configuration is fixed for each disorder realization). Random variations persist through random states used in each disorder realization. Extraction was performed on a sparse model with $k=4$ between $N=16$ to $N=36$ with $\beta = 1$. All plots demonstrate high variance and lack of convergence, suggesting the numerical derivative method of extraction is unsuitable for a fixed hypergraph seed. Each $D_g$ is averaged over 10 times. (a) The mean fitted Lyapunov exponent $\overline{\lambda_L}$ versus $D_g$ (b) The standard deviation of the fitted $\lambda_L$'s, $\sigma$, versus $D_g$ (c) The mean $R^2$ value over each set of $\lambda_L$'s. The very low values of $R^2$ suggest a very poor fit to the data. }}
\label{fig:sethyper}
\pagebreak
\end{figure}
\pagebreak
}

\afterpage{
\begin{figure}[H]
\centering
\begin{subfigure}{0.49\textwidth}
    \includegraphics[width=\textwidth]{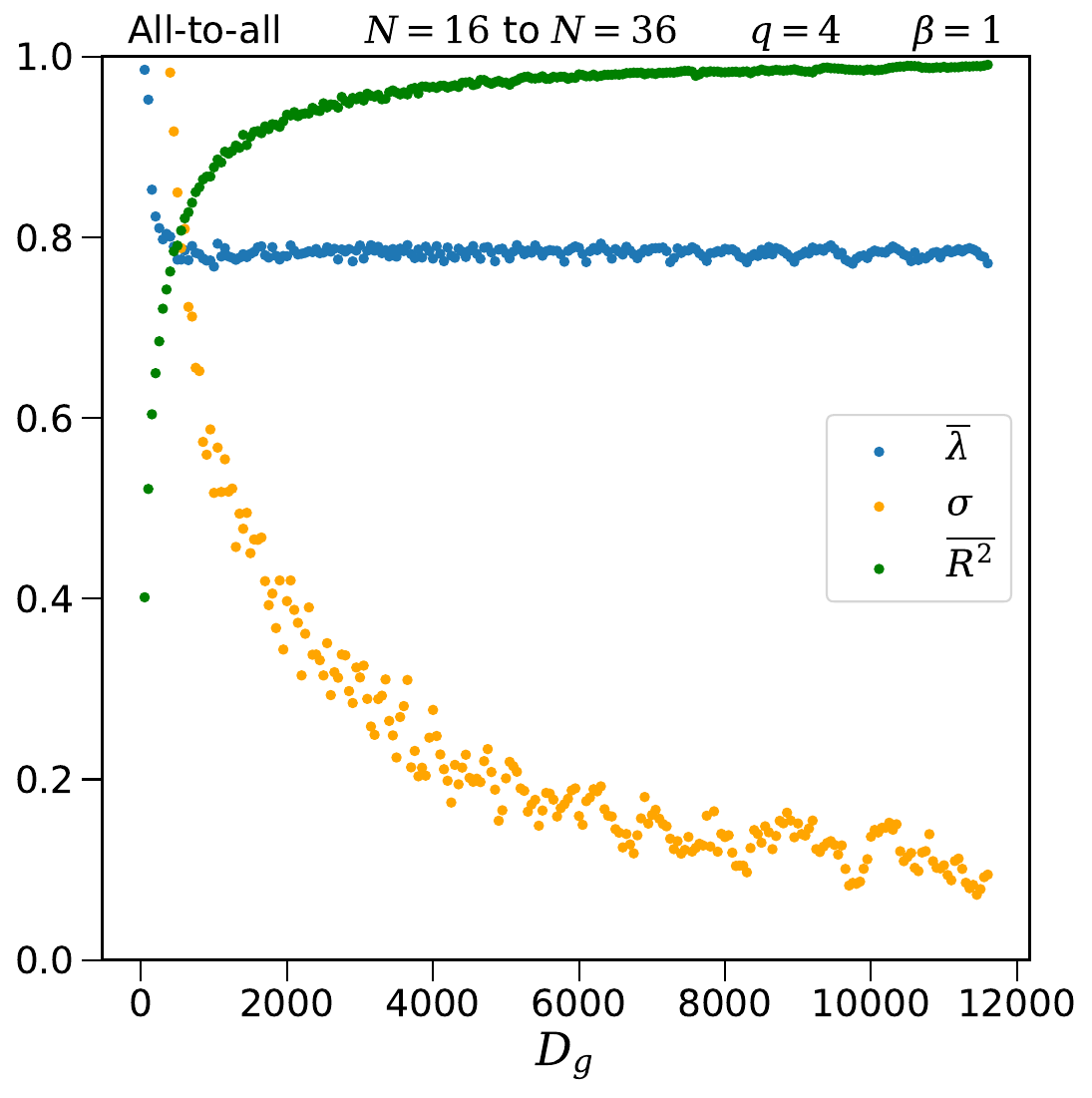}
    \caption{}
    \label{fig:1F3}
\end{subfigure}
\hfill
\begin{subfigure}{0.49\textwidth}
    \includegraphics[width=\textwidth]{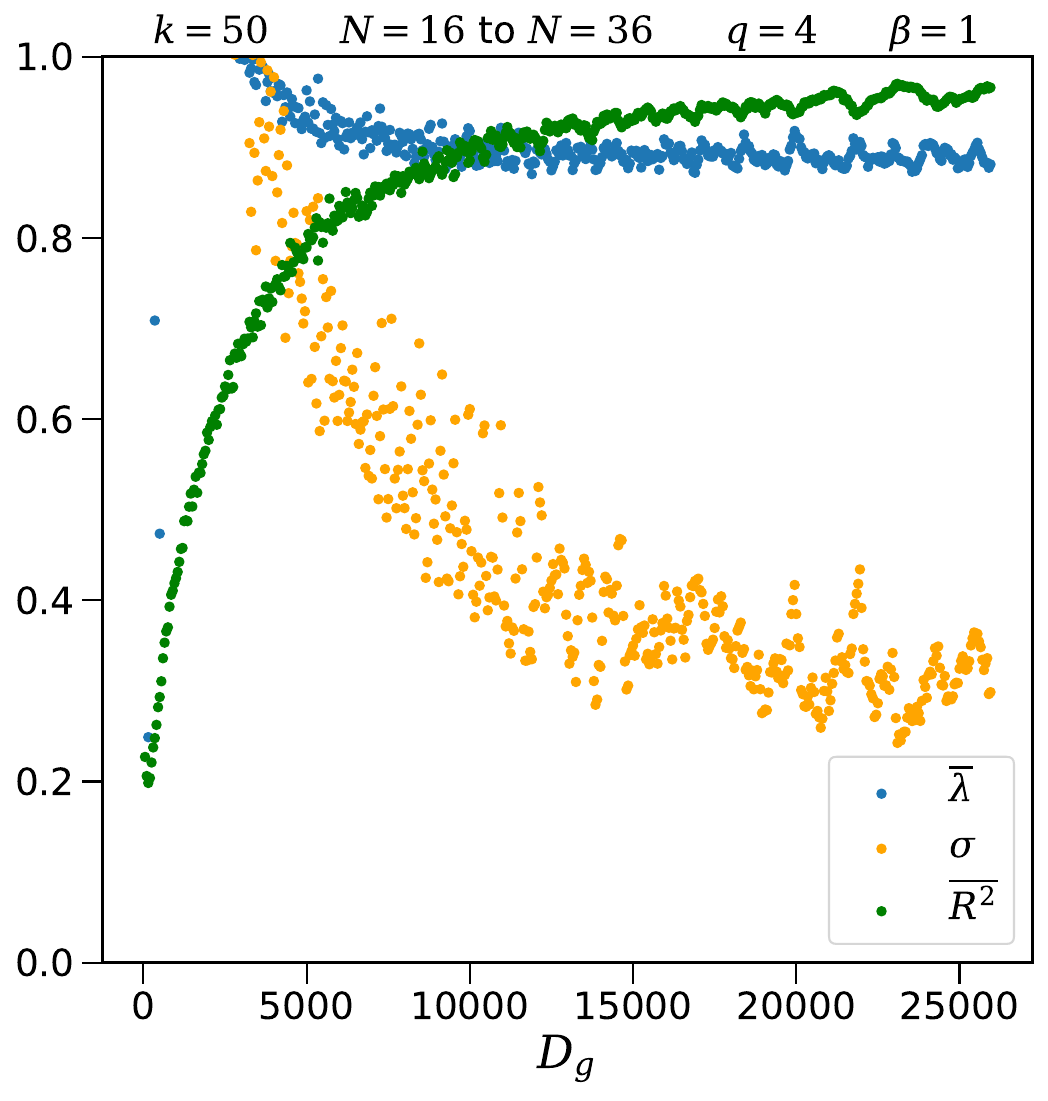}
    \caption{}
    \label{fig:2F3}
\end{subfigure}

\begin{subfigure}{0.49\textwidth}
    \includegraphics[width=\textwidth]{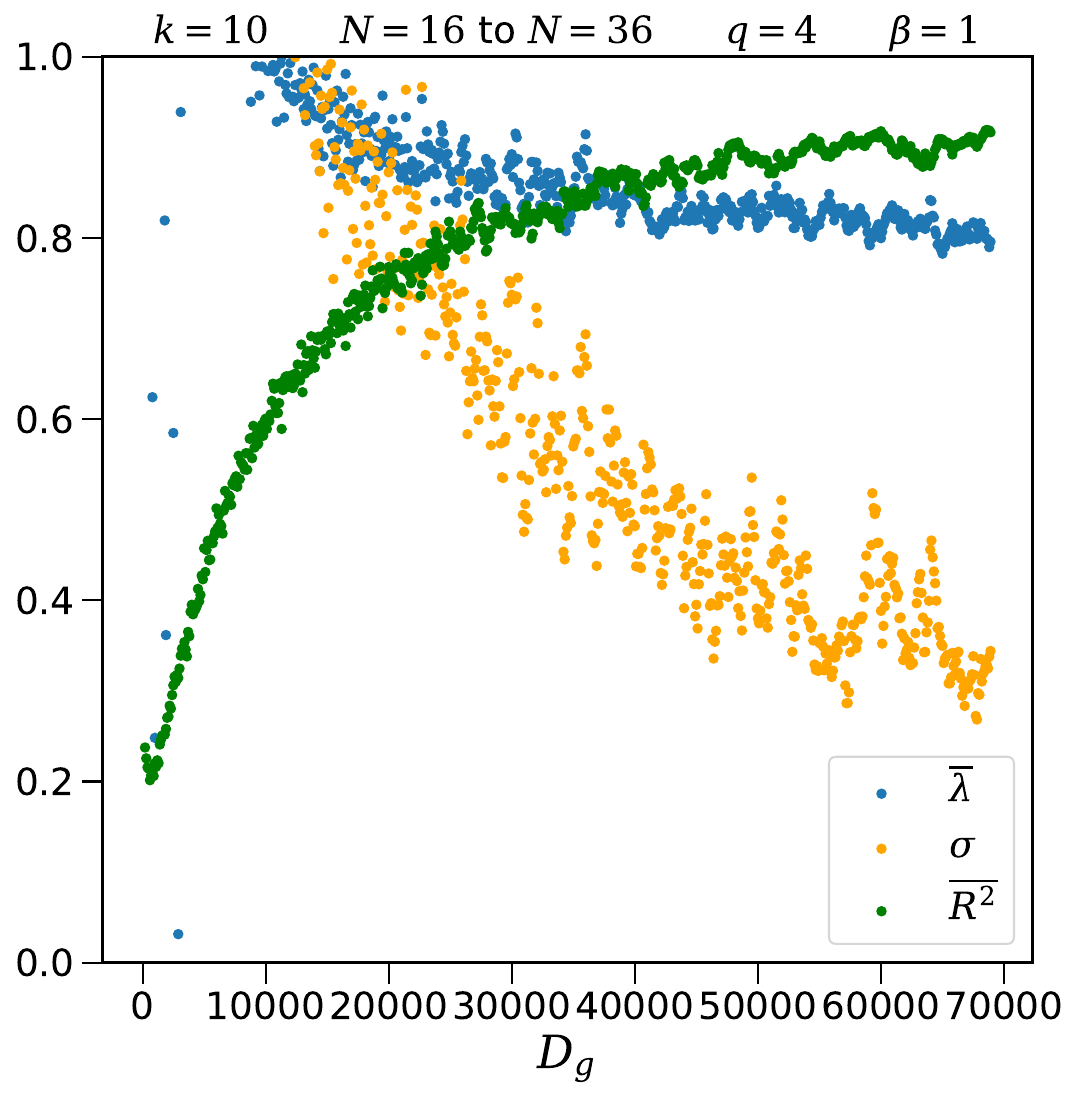}
    \caption{}
    \label{fig:3F3}
\end{subfigure}
\hfill
\begin{subfigure}{0.49\textwidth}
    \includegraphics[width=\textwidth]{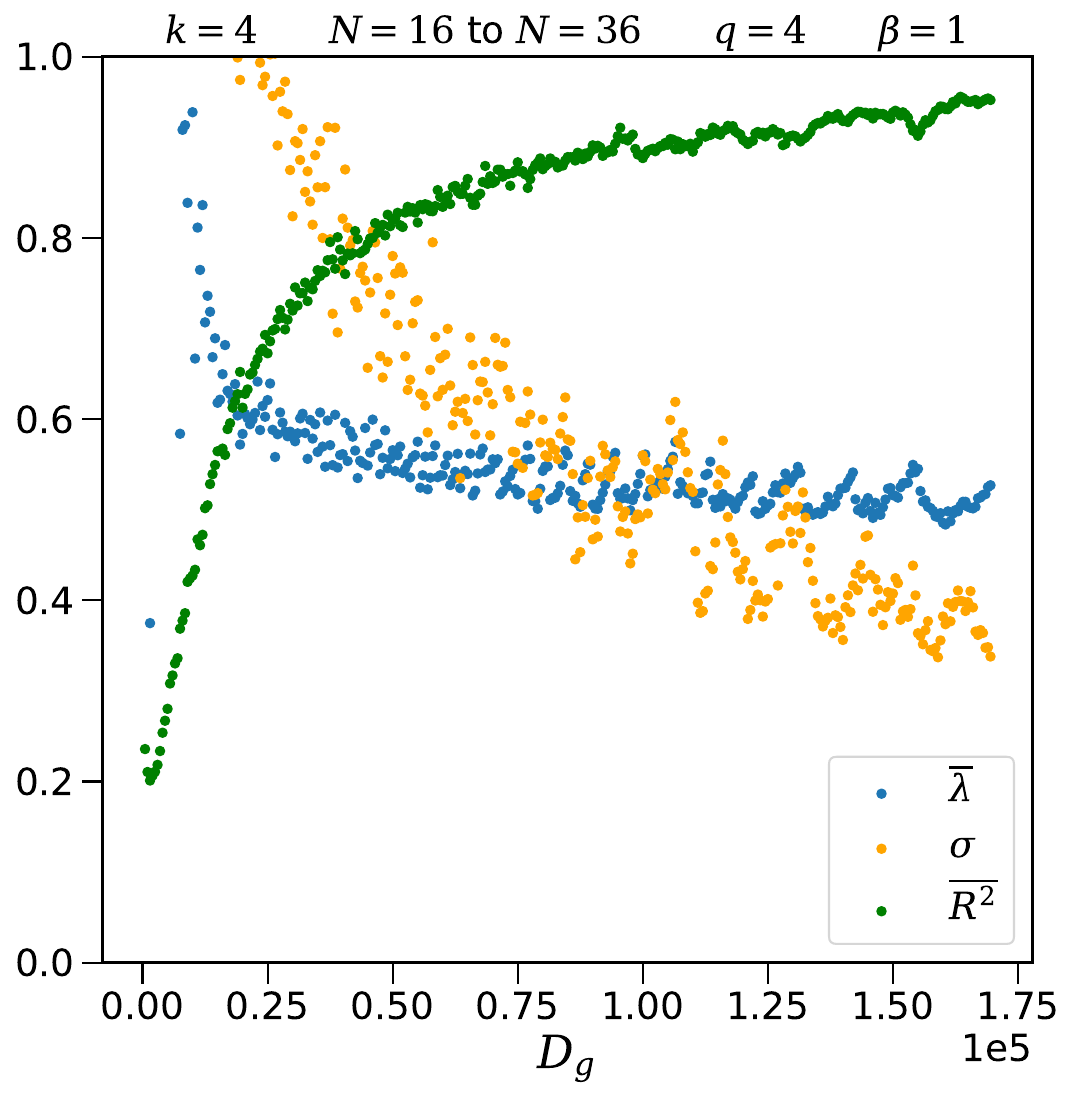}
    \caption{}
    \label{fig:4F3}
\end{subfigure}
        
\caption{\textit{A replication of Figure \ref{fig:Dmaxmain} (see corresponding caption for plot details) with a single parameter change: $\mathcal{F}^* = 0.3$. Convergence and behavior is consistent with Figure \ref{fig:Dmaxmain} across this range of $D_g$'s.}}
\label{fig:F3}
\pagebreak
\end{figure}
\pagebreak
}

\afterpage{
\begin{figure}[H]
\centering
\begin{subfigure}{0.49\textwidth}
    \includegraphics[width=\textwidth]{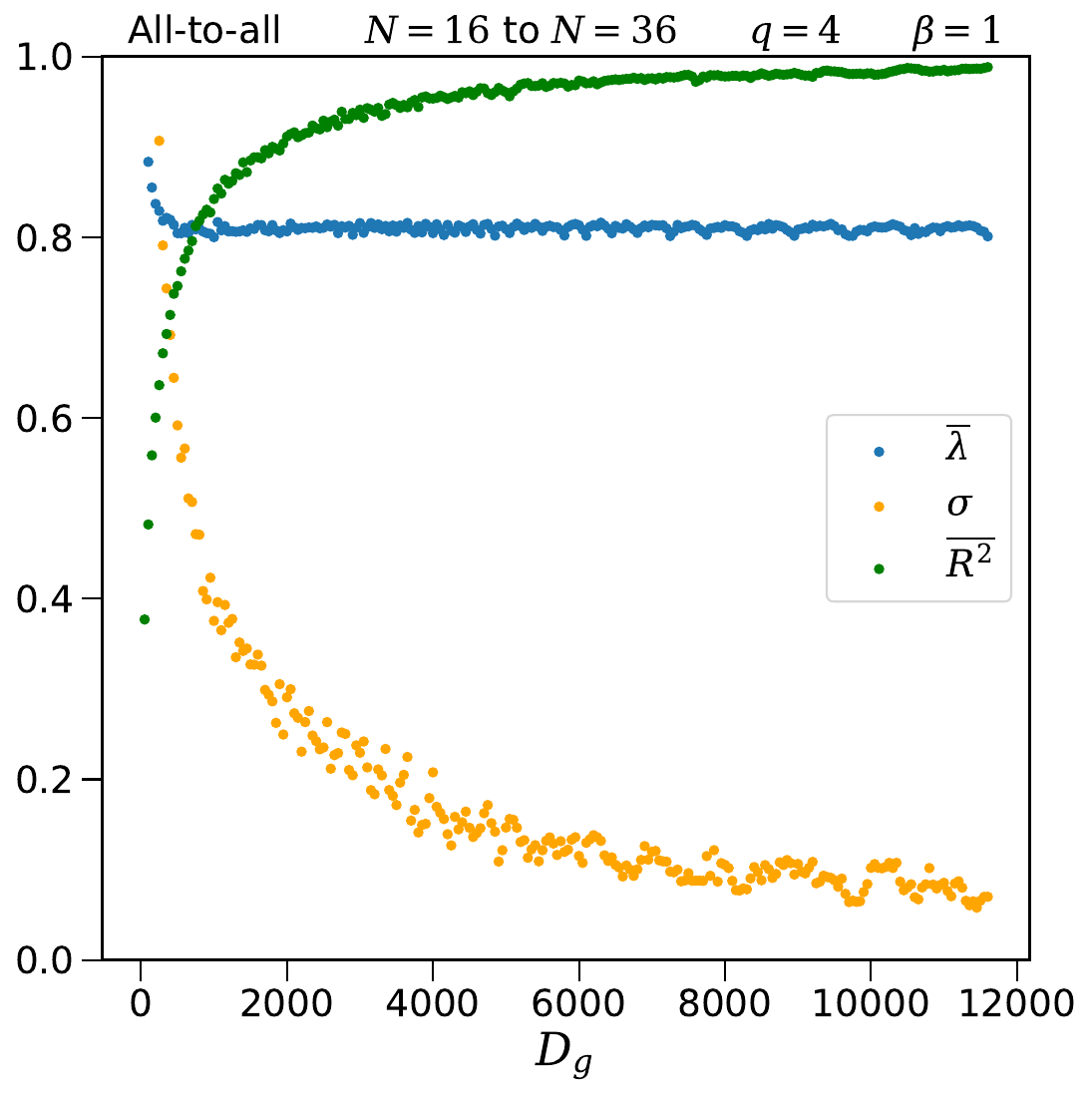}
    \caption{}
    \label{fig:1F5}
\end{subfigure}
\hfill
\begin{subfigure}{0.49\textwidth}
    \includegraphics[width=\textwidth]{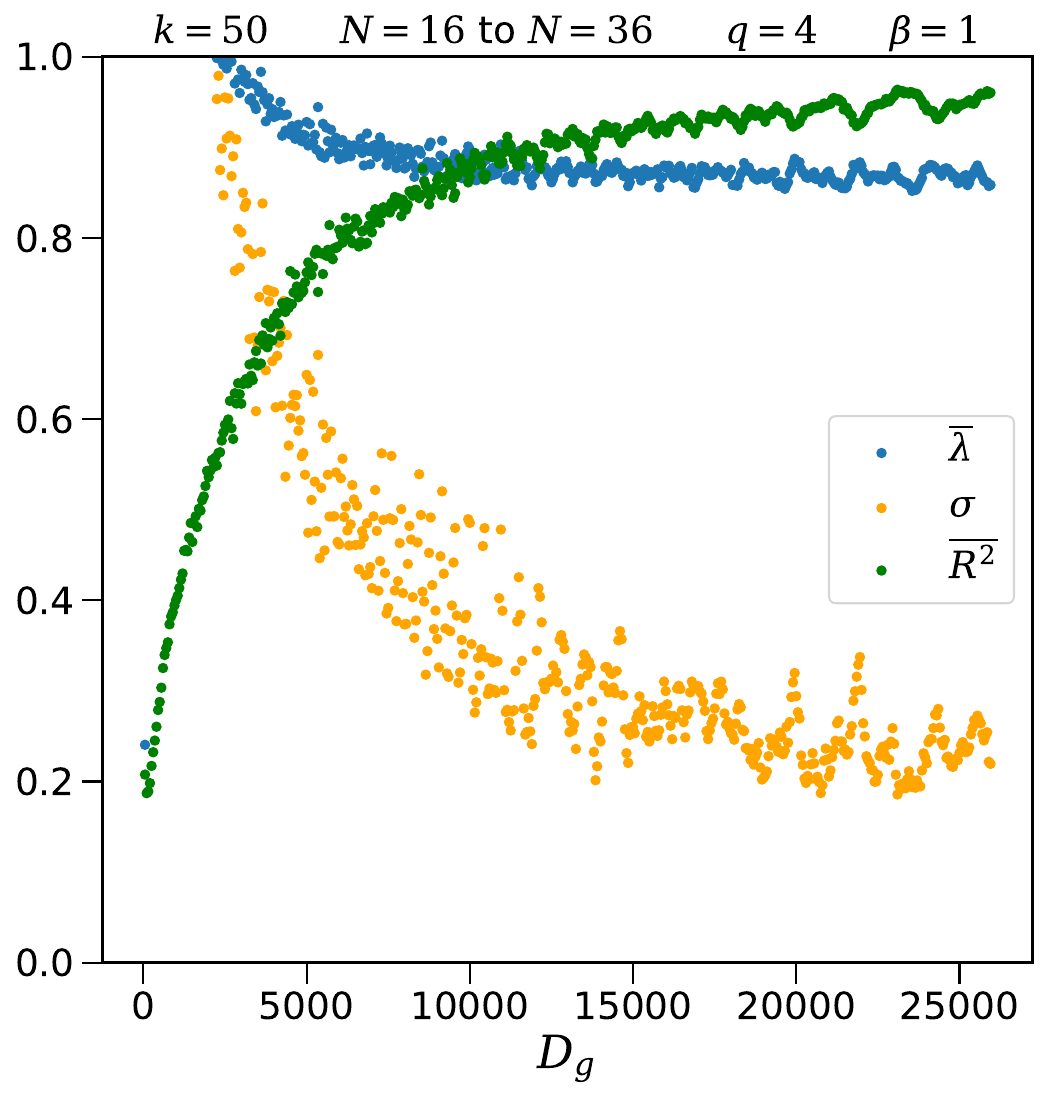}
    \caption{}
    \label{fig:2F5}
\end{subfigure}

\begin{subfigure}{0.49\textwidth}
    \includegraphics[width=\textwidth]{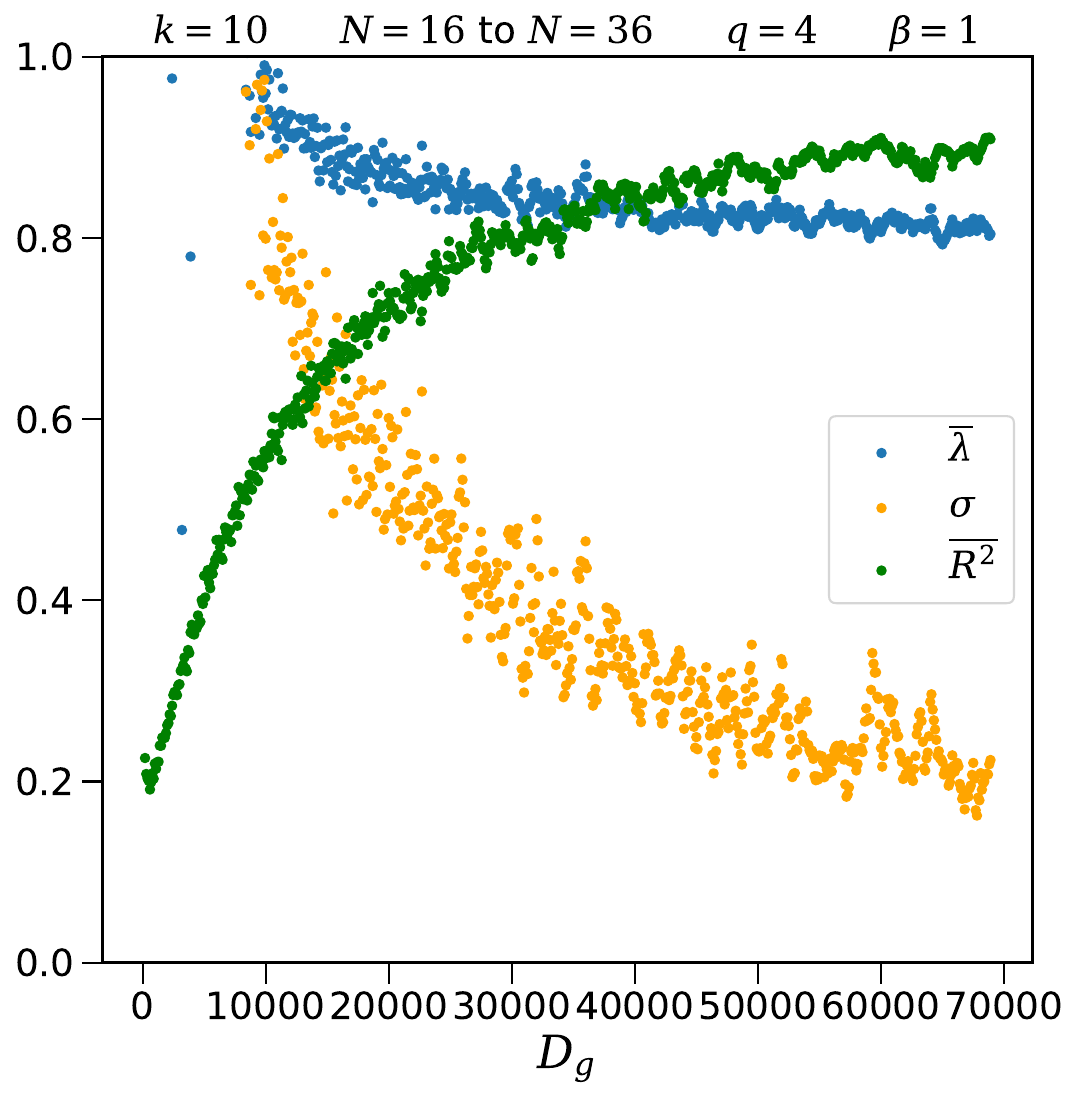}
    \caption{}
    \label{fig:3F5}
\end{subfigure}
\hfill
\begin{subfigure}{0.49\textwidth}
    \includegraphics[width=\textwidth]{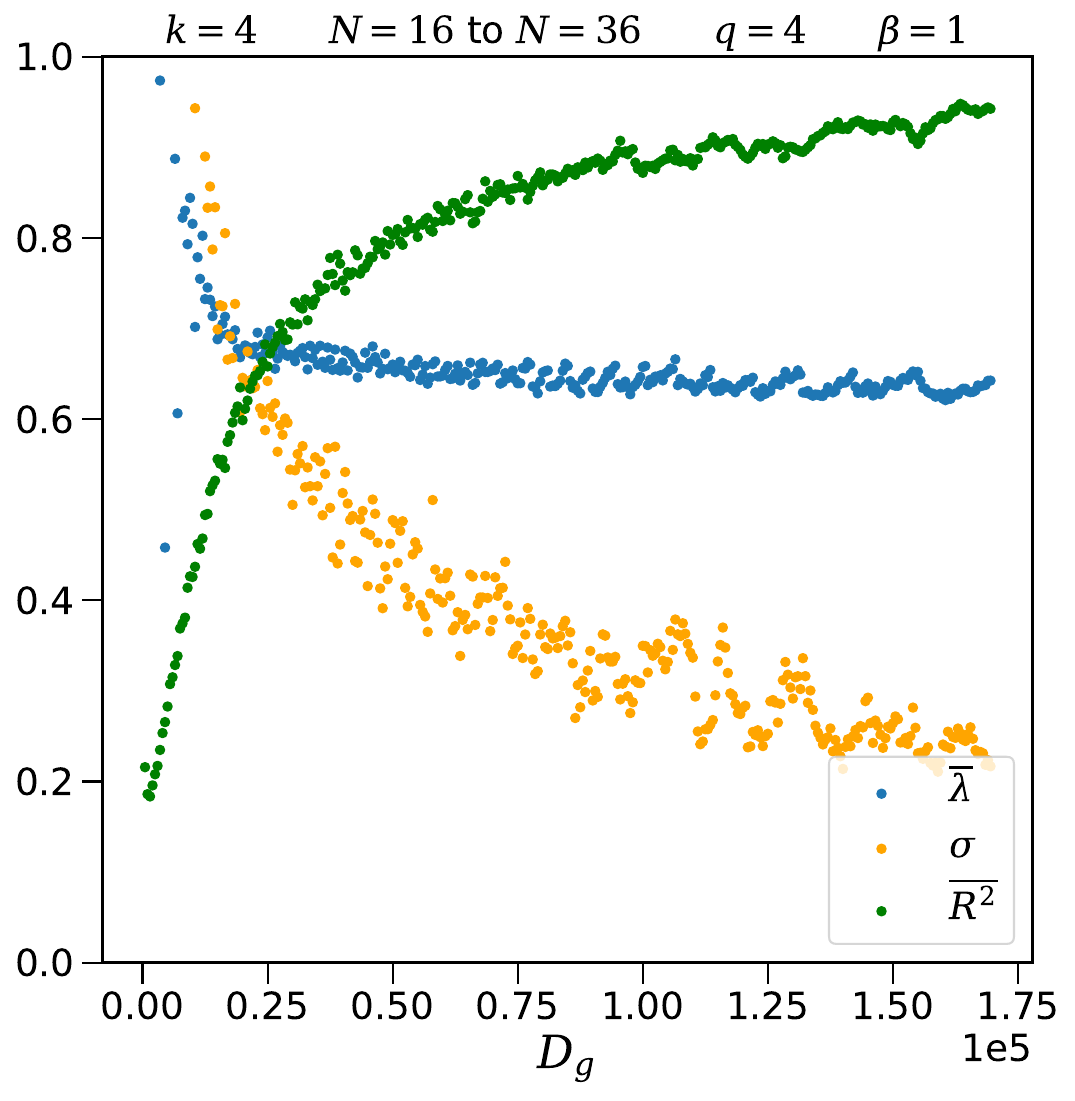}
    \caption{}
    \label{fig:4F5}
\end{subfigure}
        
\caption{\textit{A replication of Figure \ref{fig:Dmaxmain} (see corresponding caption for plot details) with a single parameter change: $\mathcal{F}^* = 0.5$. Convergence and behavior is consistent with Figure \ref{fig:Dmaxmain} across this range of $D_g$'s.}}
\label{fig:F5}
\pagebreak
\end{figure}
\pagebreak
}

\afterpage{
\newpage
\begin{figure}[H]
\centering
\begin{subfigure}{0.475\textwidth}
    \includegraphics[width=\textwidth]{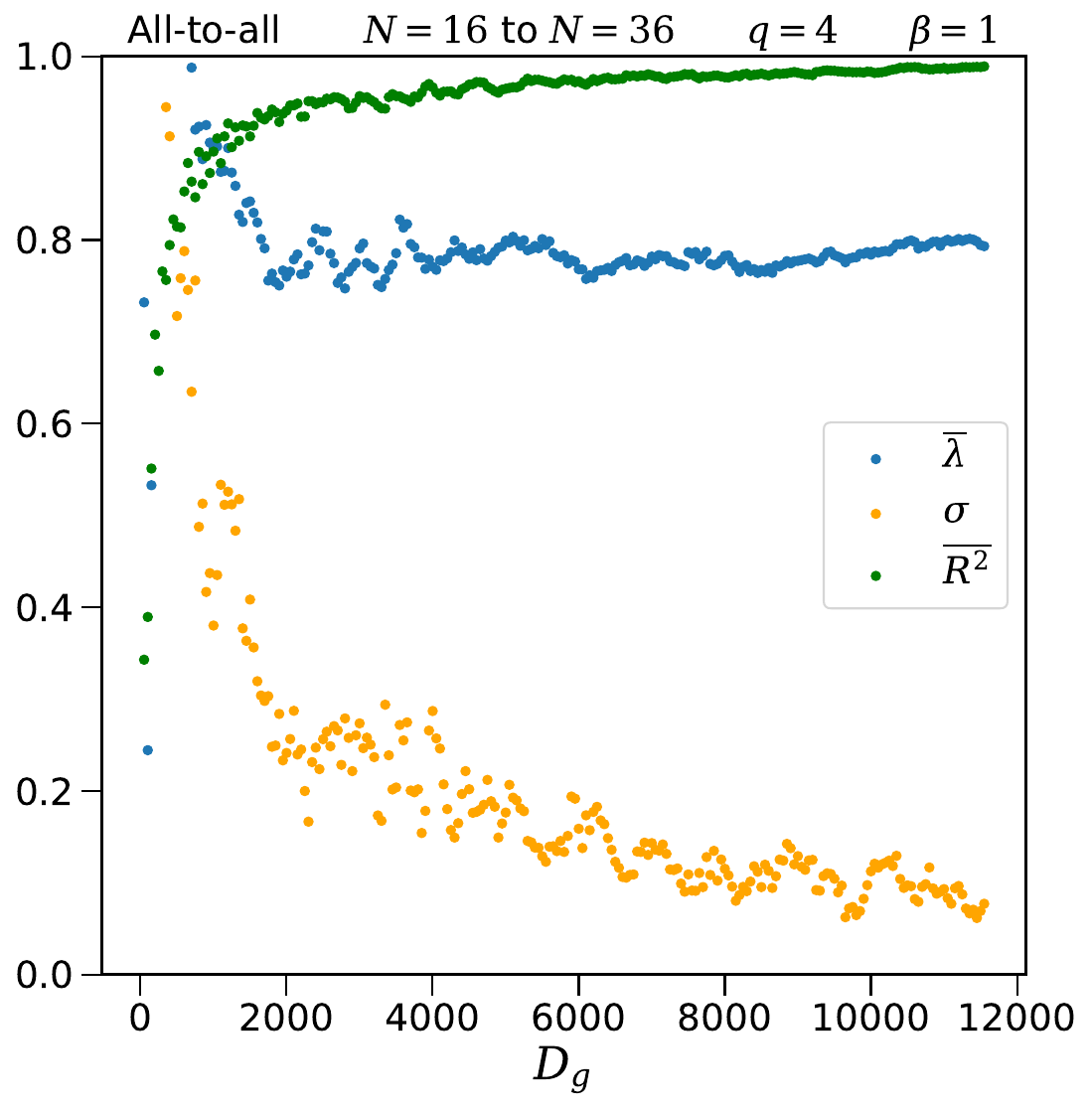}
    \caption{}
    \label{fig:firstmean}
\end{subfigure}
\hfill
\begin{subfigure}{0.475\textwidth}
    \includegraphics[width=\textwidth]{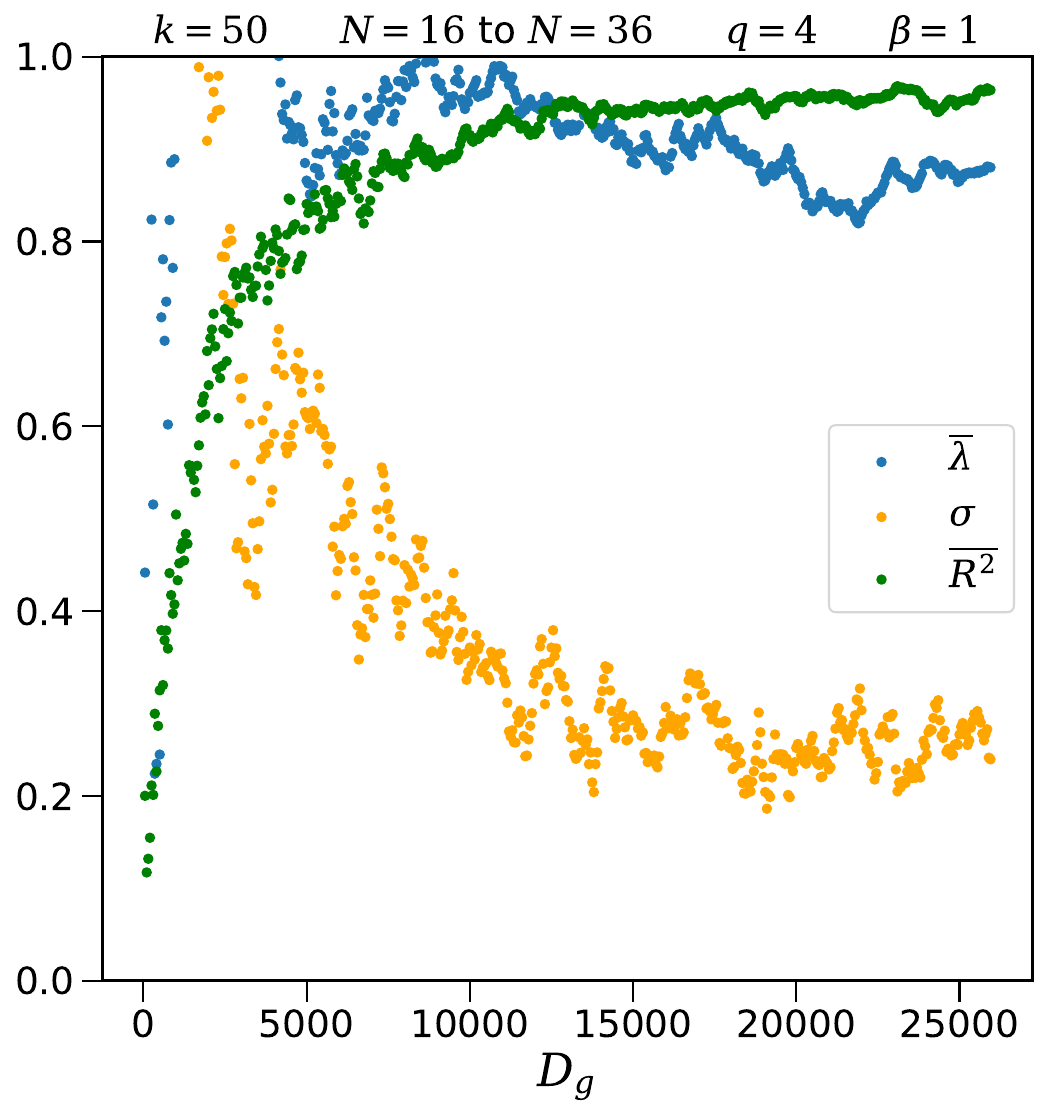}
    \caption{}
    \label{fig:secondmean}
\end{subfigure}

\begin{subfigure}{0.475\textwidth}
    \includegraphics[width=\textwidth]{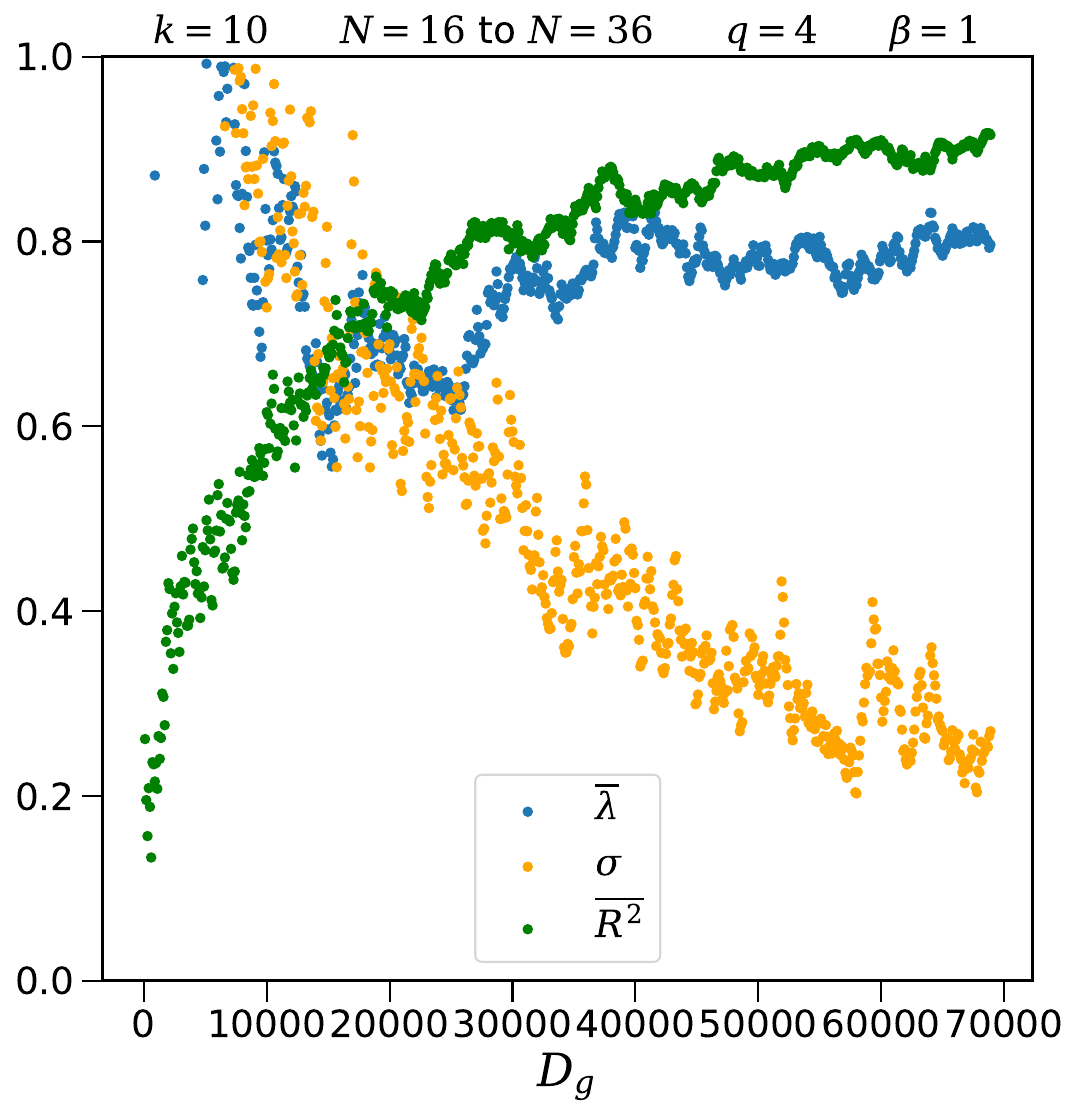}
    \caption{}
    \label{fig:thirdmean}
\end{subfigure}
\hfill
\begin{subfigure}{0.475\textwidth}
    \includegraphics[width=\textwidth]{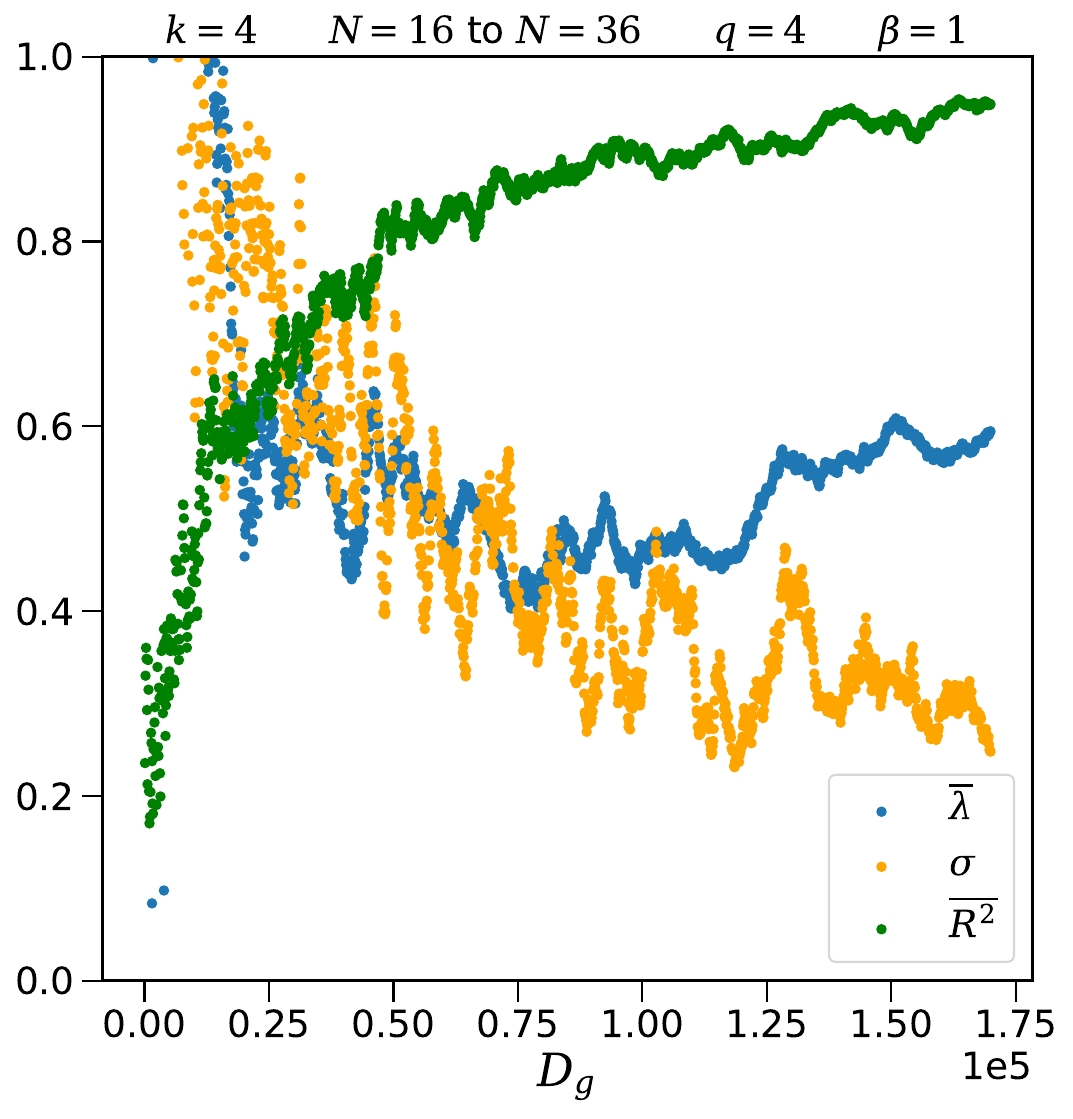}
    \caption{}
    \label{fig:fourthmean}
\end{subfigure}
        
\caption{\textit{Many realizations are subdivided into 10 samples with $D_g$ number of realizations, from which is extracted a Lyapunov exponent and $R^2$. For each choice of $D_g$, Lyapunov exponents are averaged over the 10 samples (\underline{blue}) with standard deviation (\underline{orange}). The $R^2$ factor for each $D_g$ are also averaged (\underline{green}). Resolution of plots all have $\Delta D_g = 50$. \textbf{(a)} all-to-all model. Exhibits convergence for plotted quantities; $R^2 \to 1$ shows agreement with the symmetry. \textbf{(b)} $k=50$, $R^2$ exhibits some divergence from (a), larger fluctuations. Error appears to asymptote at a higher value compared to (a). \textbf{(c,d)} k=10,4 respectively. Random fluctuations have much larger effect, very high values of $D_g$ must used to perform consistent Lyapunov extraction.}}
\label{fig:allerror}
\end{figure}
\pagebreak}

\afterpage{
\begin{figure}[H]
\centering
\begin{subfigure}{0.49\textwidth}
    \includegraphics[width=\textwidth]{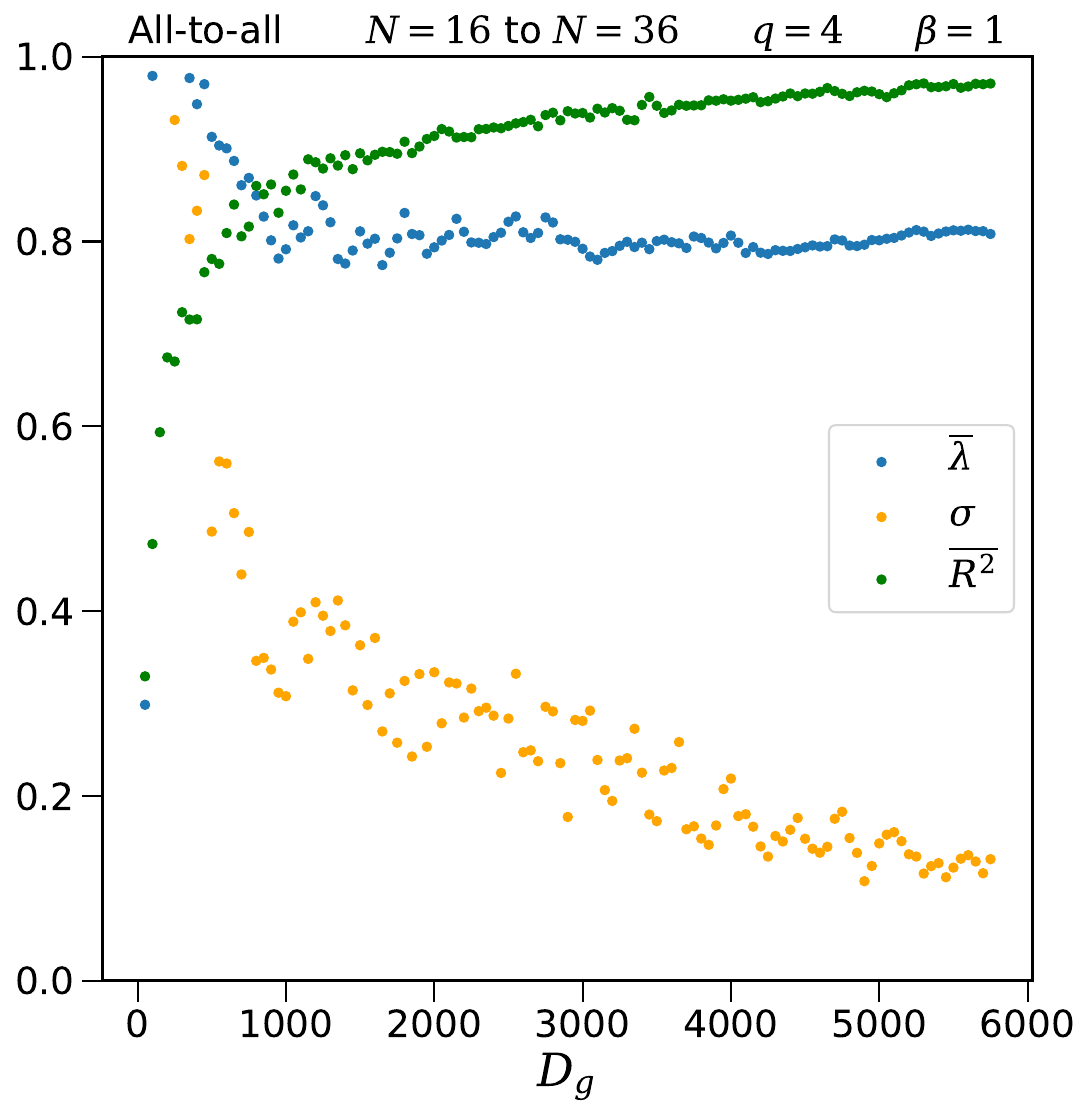}
    \caption{}
    \label{fig:1D20}
\end{subfigure}
\hfill
\begin{subfigure}{0.49\textwidth}
    \includegraphics[width=\textwidth]{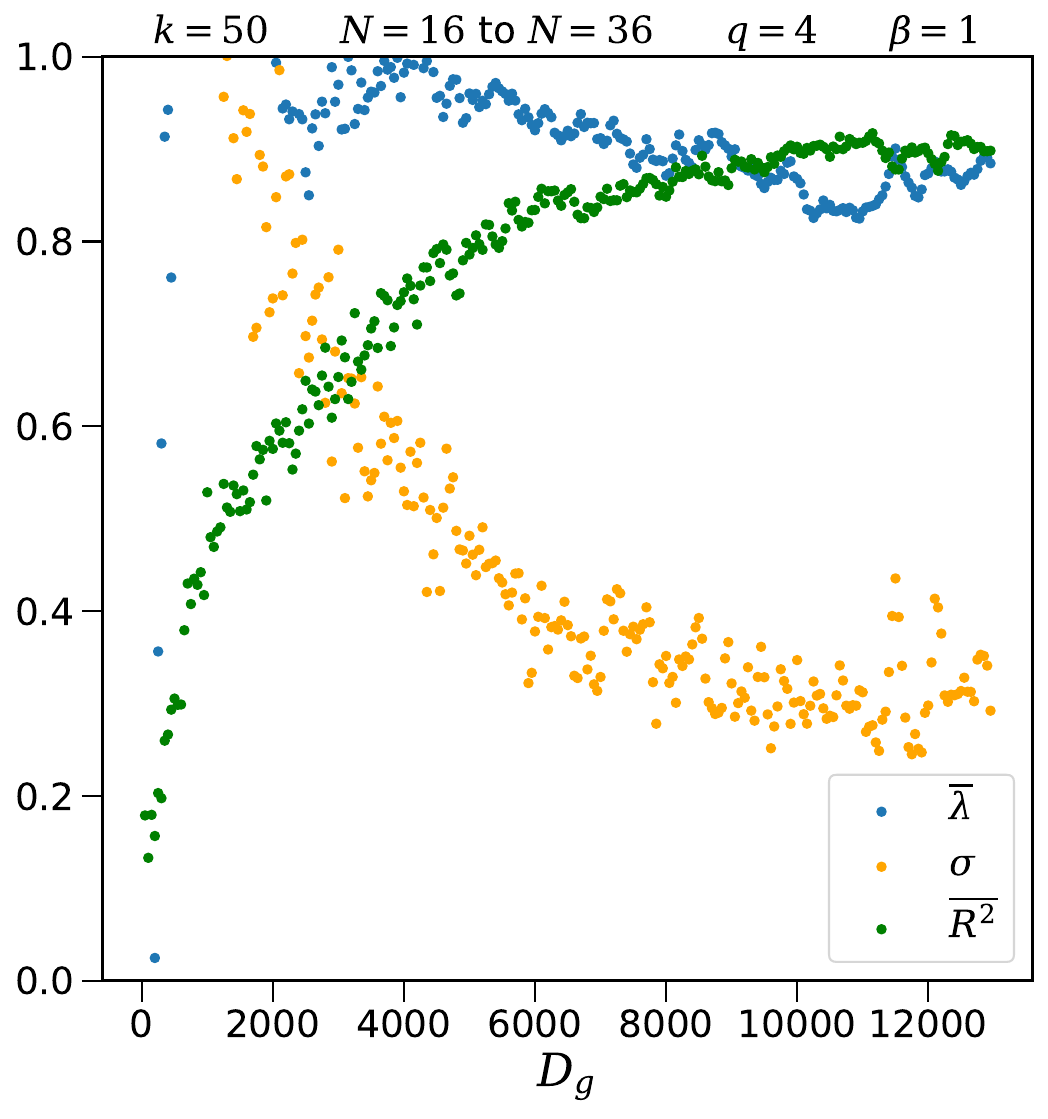}
    \caption{}
    \label{fig:2D20}
\end{subfigure}

\begin{subfigure}{0.49\textwidth}
    \includegraphics[width=\textwidth]{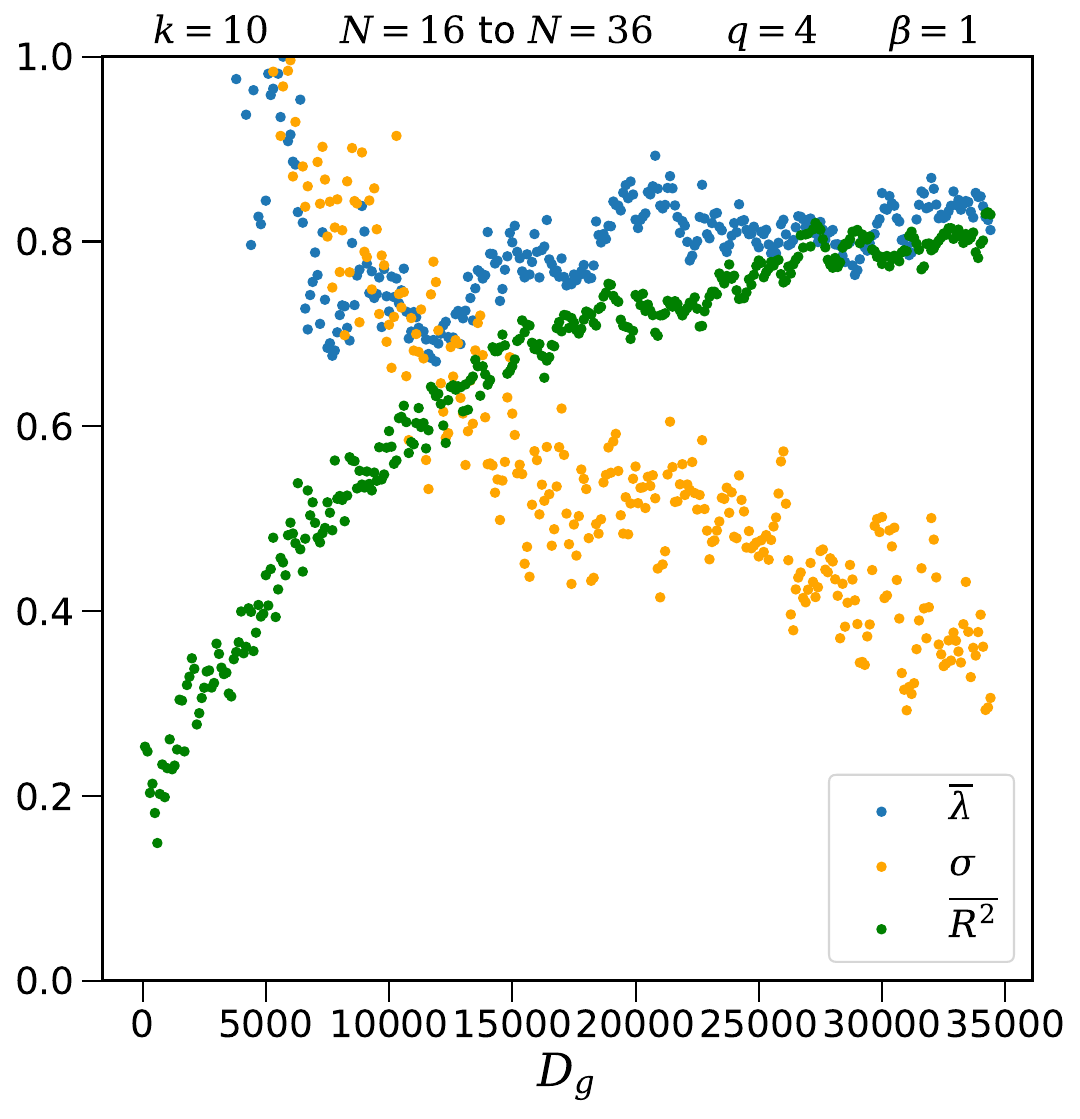}
    \caption{}
    \label{fig:3D20}
\end{subfigure}
\hfill
\begin{subfigure}{0.49\textwidth}
    \includegraphics[width=\textwidth]{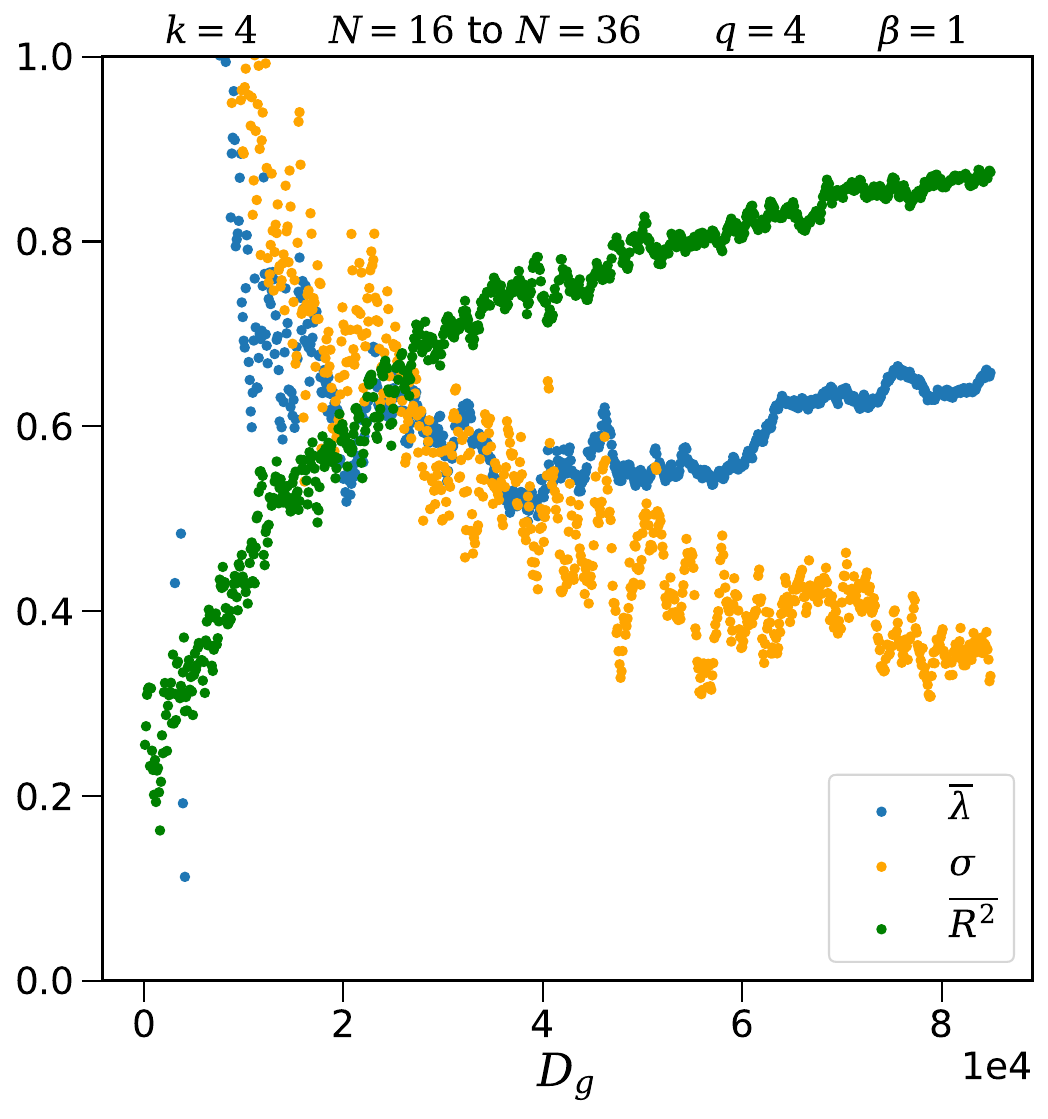}
    \caption{}
    \label{fig:4D20}
\end{subfigure}
        
\caption{\textit{A replication of Figure \ref{fig:allerror} (see corresponding caption for plot details) with a change of grouping sizes for statistics, namely 20 simulations are averaged over for each $D_g$. This reduces the maximum $D_g$ (and hence the range) by $\frac{1}{2}$. As a result, higher values of $R^2$ are unable to be attained for the sparse models (i.e. for (b), (c), (d)), requiring probing of higher maximum values of $D_g$. Within the ranges present, the behavior is consistent with Figure \ref{fig:allerror}.}}
\label{fig:D20}
\pagebreak
\end{figure}
\pagebreak
}

\afterpage{
\begin{figure}[H]
\centering
\begin{subfigure}{0.49\textwidth}
    \includegraphics[width=\textwidth]{figures/k=4Nmax36.pdf}
    \caption{}
    \label{fig:1Nset}
\end{subfigure}
\hfill
\begin{subfigure}{0.49\textwidth}
    \includegraphics[width=\textwidth]{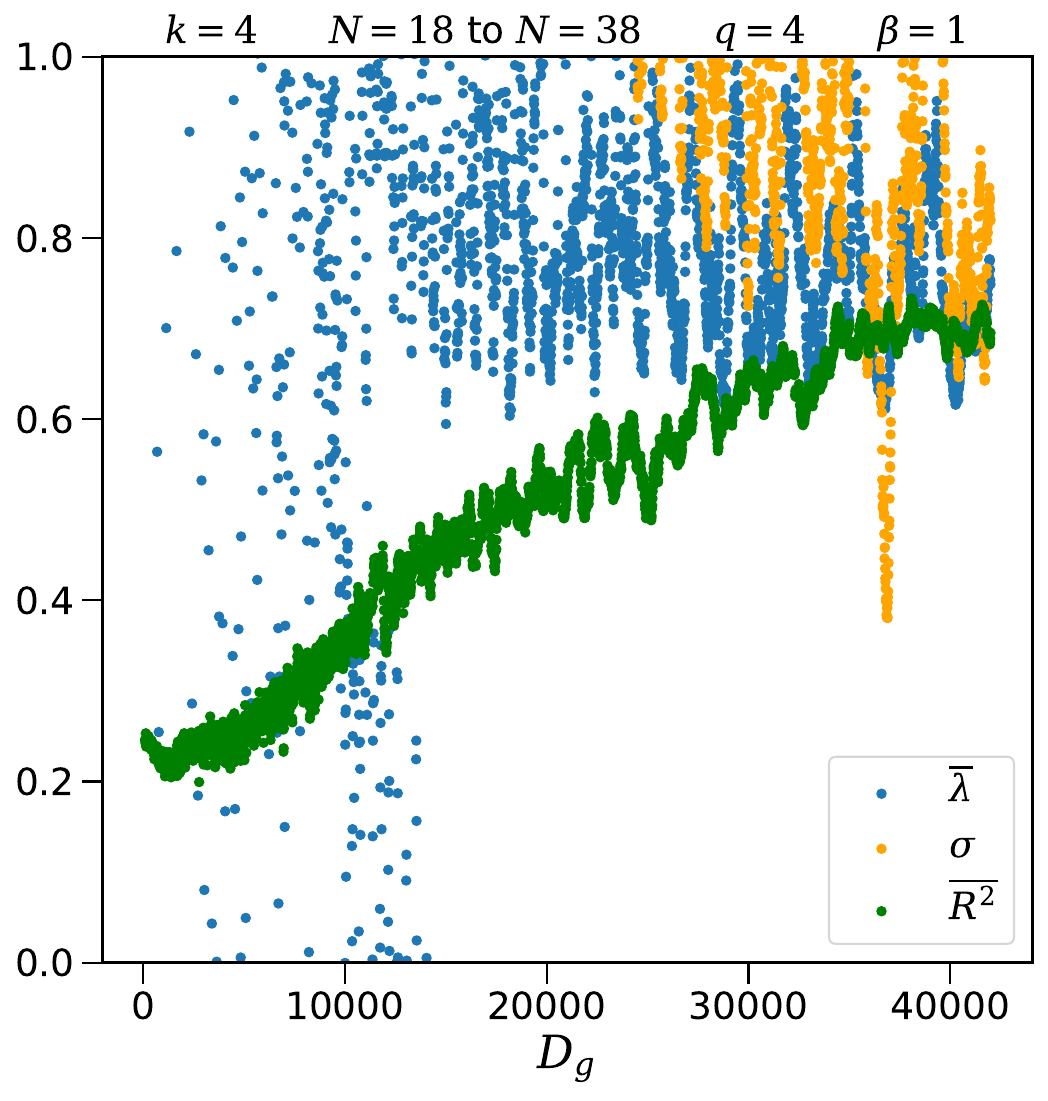}
    \caption{}
    \label{fig:2Nset}
\end{subfigure}

\begin{subfigure}{0.49\textwidth}
    \includegraphics[width=\textwidth]{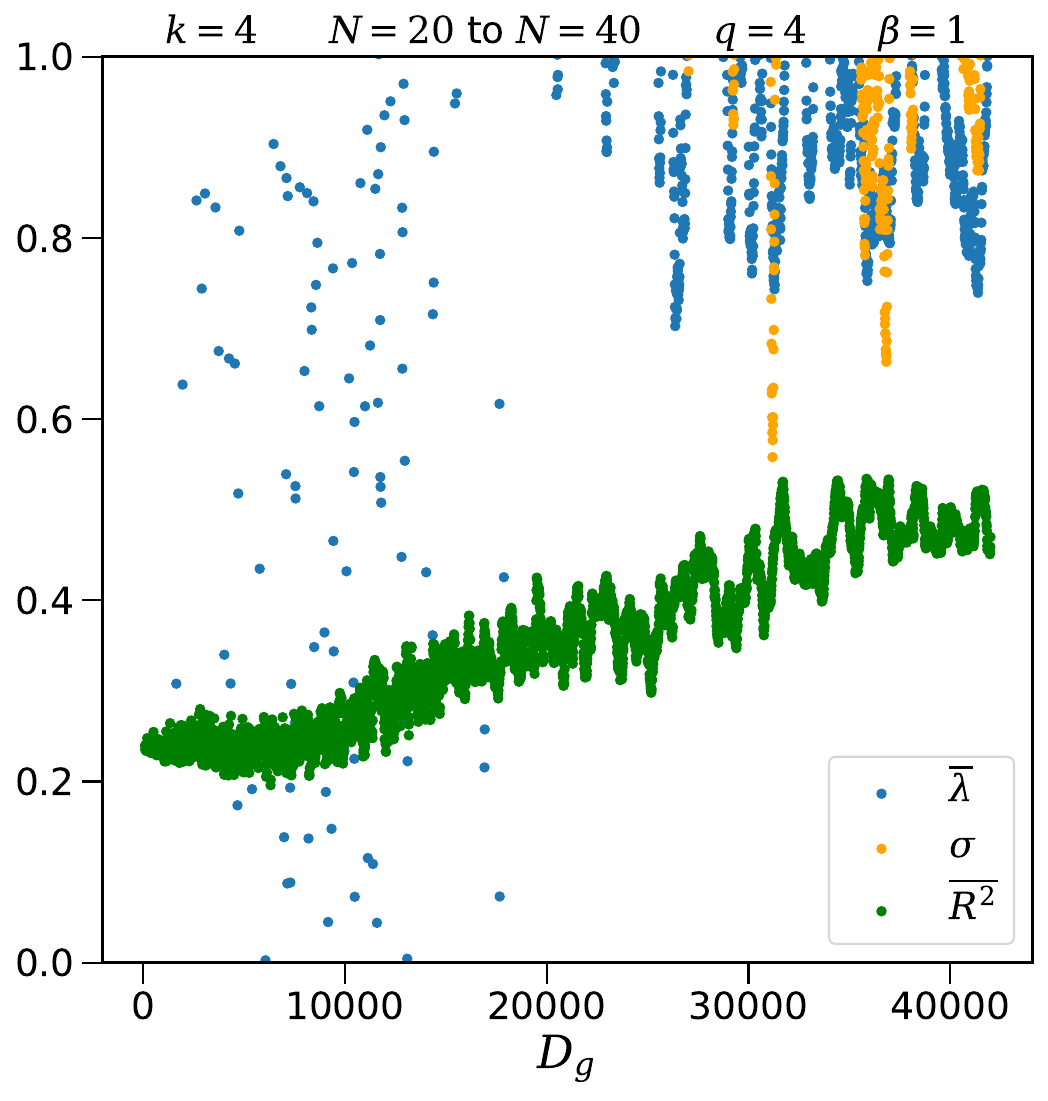}
    \caption{}
    \label{fig:3Nset}
\end{subfigure}
\hfill
\begin{subfigure}{0.49\textwidth}
    \includegraphics[width=\textwidth]{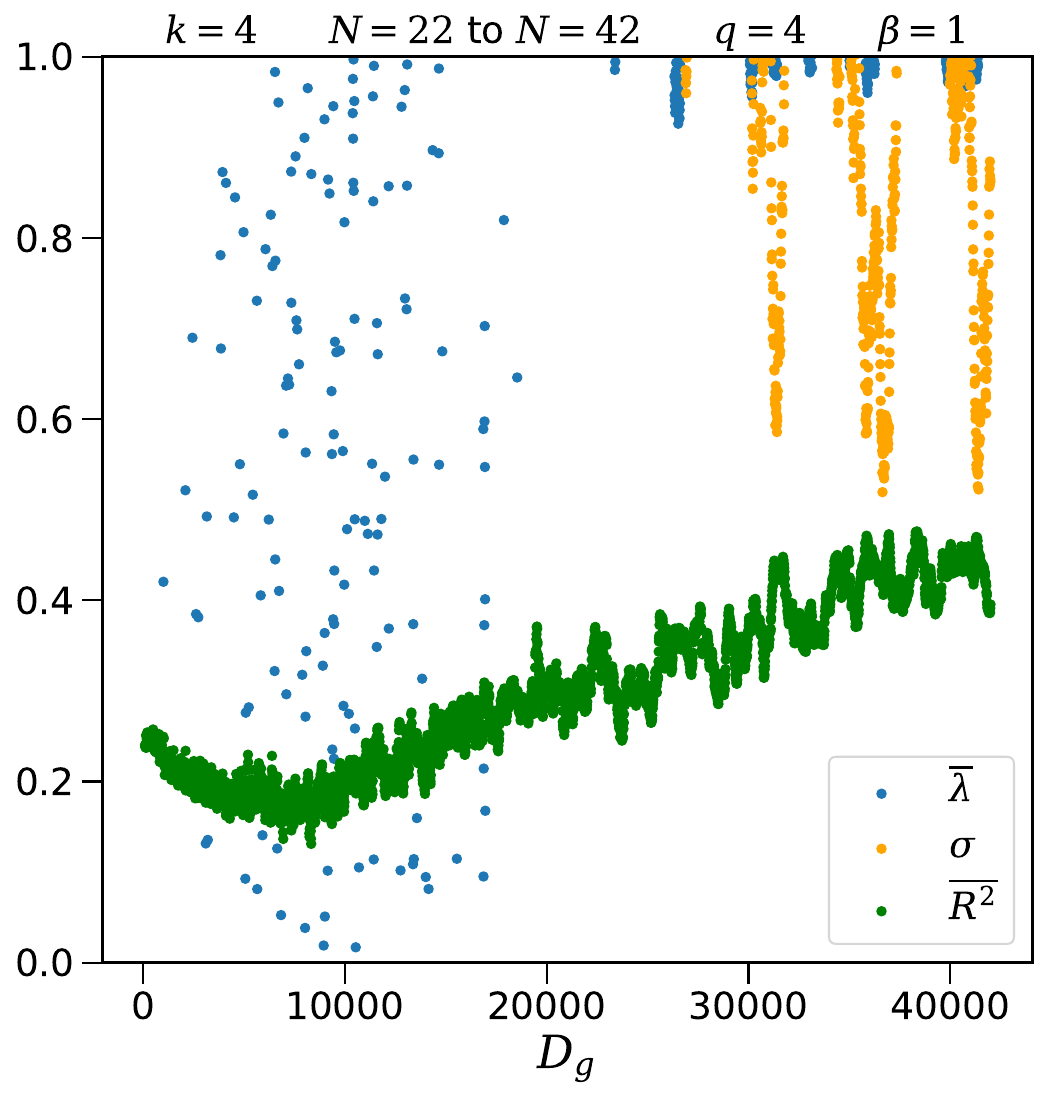}
    \caption{}
    \label{fig:4Nset}
\end{subfigure}
        
\caption{\textit{A replication of Figure \ref{fig:Nmax} (see corresponding caption for plot details) with changes in the lower bounds of $N$ used to extract the Lyapunov exponent, namely (a) $N=16$ to $N=36$ (unchanged) (b) $N=18$ to $N=38$ (c) $N=20$ to $N=40$ (d) $N=22$ to $N=42$. Despite a majority of data overflowing the plot, the choice of ranges for the axes was maintained to demonstrate even further disagreement from Figure \ref{fig:Nmax}.}}
\label{fig:Nset}
\pagebreak
\end{figure}
\pagebreak
}
\end{appendices}
\newpage
\bibliographystyle{JHEP}
\bibliography{references}

\providecommand{\href}[2]{#2}\begingroup\raggedright\begin{thebibliography}{10}

\bibitem{Shenker_2014}
S.~H. Shenker and D.~Stanford, \emph{Black holes and the butterfly effect},
  \href{https://doi.org/10.1007/jhep03(2014)067}{\emph{Journal of High Energy
  Physics} {\bfseries 2014} (2014) }.

\bibitem{Shenker:2014cwa}
S.~H. Shenker and D.~Stanford, \emph{{Stringy effects in scrambling}},
  \href{https://doi.org/10.1007/JHEP05(2015)132}{\emph{JHEP} {\bfseries 05}
  (2015) 132} [\href{https://arxiv.org/abs/1412.6087}{{\ttfamily 1412.6087}}].

\bibitem{Maldacena:2015waa}
J.~Maldacena, S.~H. Shenker and D.~Stanford, \emph{{A bound on chaos}},
  \href{https://doi.org/10.1007/JHEP08(2016)106}{\emph{JHEP} {\bfseries 08}
  (2016) 106} [\href{https://arxiv.org/abs/1503.01409}{{\ttfamily
  1503.01409}}].

\bibitem{Hosur:2015ylk}
P.~Hosur, X.-L. Qi, D.~A. Roberts and B.~Yoshida, \emph{{Chaos in quantum
  channels}}, \href{https://doi.org/10.1007/JHEP02(2016)004}{\emph{JHEP}
  {\bfseries 02} (2016) 004}
  [\href{https://arxiv.org/abs/1511.04021}{{\ttfamily 1511.04021}}].

\bibitem{Stanford:2015owe}
D.~Stanford, \emph{{Many-body chaos at weak coupling}},
  \href{https://doi.org/10.1007/JHEP10(2016)009}{\emph{JHEP} {\bfseries 10}
  (2016) 009} [\href{https://arxiv.org/abs/1512.07687}{{\ttfamily
  1512.07687}}].

\bibitem{Jensen:2016pah}
K.~Jensen, \emph{{Chaos in AdS$_2$ Holography}},
  \href{https://doi.org/10.1103/PhysRevLett.117.111601}{\emph{Phys. Rev. Lett.}
  {\bfseries 117} (2016) 111601}
  [\href{https://arxiv.org/abs/1605.06098}{{\ttfamily 1605.06098}}].

\bibitem{Maldacena:2016hyu}
J.~Maldacena and D.~Stanford, \emph{{Remarks on the Sachdev-Ye-Kitaev model}},
  \href{https://doi.org/10.1103/PhysRevD.94.106002}{\emph{Phys. Rev. D}
  {\bfseries 94} (2016) 106002}
  [\href{https://arxiv.org/abs/1604.07818}{{\ttfamily 1604.07818}}].

\bibitem{Roberts:2016hpo}
D.~A. Roberts and B.~Yoshida, \emph{{Chaos and complexity by design}},
  \href{https://doi.org/10.1007/JHEP04(2017)121}{\emph{JHEP} {\bfseries 04}
  (2017) 121} [\href{https://arxiv.org/abs/1610.04903}{{\ttfamily
  1610.04903}}].

\bibitem{Swingle:2018ekw}
B.~Swingle, \emph{{Unscrambling the physics of out-of-time-order correlators}},
  \href{https://doi.org/10.1038/s41567-018-0295-5}{\emph{Nature Phys.}
  {\bfseries 14} (2018) 988}.

\bibitem{Garcia-Garcia:2017bkg}
A.~M. Garc\'\i{}a-Garc\'\i{}a, B.~Loureiro, A.~Romero-Berm\'udez and M.~Tezuka,
  \emph{{Chaotic-Integrable Transition in the Sachdev-Ye-Kitaev Model}},
  \href{https://doi.org/10.1103/PhysRevLett.120.241603}{\emph{Phys. Rev. Lett.}
  {\bfseries 120} (2018) 241603}
  [\href{https://arxiv.org/abs/1707.02197}{{\ttfamily 1707.02197}}].

\bibitem{Gu_2019}
Y.~Gu and A.~Kitaev, \emph{On the relation between the magnitude and exponent
  of {OTOCs}}, \href{https://doi.org/10.1007/jhep02(2019)075}{\emph{Journal of
  High Energy Physics} {\bfseries 2019} (2019) }.

\bibitem{Fischler:2021rxy}
W.~Fischler, T.~Guglielmo and P.~Nguyen, \emph{{Quantum chaos in a
  weakly-coupled field theory with nonlocality}},
  \href{https://doi.org/10.1007/JHEP09(2022)097}{\emph{JHEP} {\bfseries 09}
  (2022) 097} [\href{https://arxiv.org/abs/2111.10895}{{\ttfamily
  2111.10895}}].

\bibitem{Xu:2022vko}
S.~Xu and B.~Swingle, \emph{{Scrambling Dynamics and Out-of-Time Ordered
  Correlators in Quantum Many-Body Systems: a Tutorial}},
  \href{https://arxiv.org/abs/2202.07060}{{\ttfamily 2202.07060}}.

\bibitem{Garcia-Mata:2022voo}
I.~Garc\'\i{}a-Mata, R.~A. Jalabert and D.~A. Wisniacki,
  \emph{{Out-of-time-order correlators and quantum chaos}},
  \href{https://arxiv.org/abs/2209.07965}{{\ttfamily 2209.07965}}.

\bibitem{Larkin1969}
A.~I. {Larkin} and Y.~N. {Ovchinnikov}, \emph{{Quasiclassical Method in the
  Theory of Superconductivity}}, {\emph{Soviet Journal of Experimental and
  Theoretical Physics} {\bfseries 28} (1969) 1200}.

\bibitem{Sekino:2008he}
Y.~Sekino and L.~Susskind, \emph{{Fast Scramblers}},
  \href{https://doi.org/10.1088/1126-6708/2008/10/065}{\emph{JHEP} {\bfseries
  10} (2008) 065} [\href{https://arxiv.org/abs/0808.2096}{{\ttfamily
  0808.2096}}].

\bibitem{Kitaev:2015}
A.~Kitaev, \emph{{A simple model of quantum holography}}.
  \url{http://online.kitp.ucsb.edu/online/entangled15/kitaev/},
  \url{http://online.kitp.ucsb.edu/online/entangled15/kitaev2/}, 2015.

\bibitem{Almheiri:2014cka}
A.~Almheiri and J.~Polchinski, \emph{{Models of AdS$_{2}$ backreaction and
  holography}}, \href{https://doi.org/10.1007/JHEP11(2015)014}{\emph{JHEP}
  {\bfseries 11} (2015) 014} [\href{https://arxiv.org/abs/1402.6334}{{\ttfamily
  1402.6334}}].

\bibitem{Xu:2020shn}
S.~Xu, L.~Susskind, Y.~Su and B.~Swingle, \emph{{A Sparse Model of Quantum
  Holography}},  \href{https://arxiv.org/abs/2008.02303}{{\ttfamily
  2008.02303}}.

\bibitem{Garcia-Garcia:2020cdo}
A.~M. Garc\'\i{}a-Garc\'\i{}a, Y.~Jia, D.~Rosa and J.~J.~M. Verbaarschot,
  \emph{{Sparse Sachdev-Ye-Kitaev model, quantum chaos and gravity duals}},
  \href{https://doi.org/10.1103/PhysRevD.103.106002}{\emph{Phys. Rev. D}
  {\bfseries 103} (2021) 106002}
  [\href{https://arxiv.org/abs/2007.13837}{{\ttfamily 2007.13837}}].

\bibitem{Caceres:2022kyr}
E.~C\'aceres, A.~Misobuchi and A.~Raz, \emph{{Spectral form factor in sparse
  SYK models}}, \href{https://doi.org/10.1007/JHEP08(2022)236}{\emph{JHEP}
  {\bfseries 08} (2022) 236}
  [\href{https://arxiv.org/abs/2204.07194}{{\ttfamily 2204.07194}}].

\bibitem{Caceres:2021nsa}
E.~Caceres, A.~Misobuchi and R.~Pimentel, \emph{{Sparse SYK and traversable
  wormholes}}, \href{https://doi.org/10.1007/JHEP11(2021)015}{\emph{JHEP}
  {\bfseries 11} (2021) 015}
  [\href{https://arxiv.org/abs/2108.08808}{{\ttfamily 2108.08808}}].

\bibitem{Jafferis:2022crx}
D.~Jafferis, A.~Zlokapa, J.~D. Lykken, D.~K. Kolchmeyer, S.~I. Davis, N.~Lauk
  et~al., \emph{{Traversable wormhole dynamics on a quantum processor}},
  \href{https://doi.org/10.1038/s41586-022-05424-3}{\emph{Nature} {\bfseries
  612} (2022) 51}.

\bibitem{Kobrin:2020xms}
B.~Kobrin, Z.~Yang, G.~D. Kahanamoku-Meyer, C.~T. Olund, J.~E. Moore,
  D.~Stanford et~al., \emph{{Many-Body Chaos in the Sachdev-Ye-Kitaev Model}},
  \href{https://doi.org/10.1103/PhysRevLett.126.030602}{\emph{Phys. Rev. Lett.}
  {\bfseries 126} (2021) 030602}
  [\href{https://arxiv.org/abs/2002.05725}{{\ttfamily 2002.05725}}].

\bibitem{Tezuka:2022mrr}
M.~Tezuka, O.~Oktay, E.~Rinaldi, M.~Hanada and F.~Nori, \emph{{Binary-coupling
  sparse Sachdev-Ye-Kitaev model: An improved model of quantum chaos and
  holography}}, \href{https://doi.org/10.1103/PhysRevB.107.L081103}{\emph{Phys.
  Rev. B} {\bfseries 107} (2023) L081103}
  [\href{https://arxiv.org/abs/2208.12098}{{\ttfamily 2208.12098}}].

\bibitem{Dumitriu2021}
I.~Dumitriu and Y.~Zhu, \emph{Spectra of random regular hypergraphs},
  \href{https://doi.org/10.37236/8741}{\emph{The Electronic Journal of
  Combinatorics} {\bfseries 28} (2021) }.

\bibitem{Yao:2016ayk}
N.~Y. Yao, F.~Grusdt, B.~Swingle, M.~D. Lukin, D.~M. Stamper-Kurn, J.~E. Moore
  et~al., \emph{{Interferometric Approach to Probing Fast Scrambling}},
  \href{https://arxiv.org/abs/1607.01801}{{\ttfamily 1607.01801}}.

\bibitem{Lantagne-Hurtubise:2019svg}
E.~Lantagne-Hurtubise, S.~Plugge, O.~Can and M.~Franz, \emph{{Diagnosing
  quantum chaos in many-body systems using entanglement as a resource}},
  \href{https://doi.org/10.1103/PhysRevResearch.2.013254}{\emph{Phys. Rev.
  Res.} {\bfseries 2} (2020) 013254}
  [\href{https://arxiv.org/abs/1907.01628}{{\ttfamily 1907.01628}}].

\bibitem{Shen:2017kez}
H.~Shen, P.~Zhang, R.~Fan and H.~Zhai, \emph{{Out-of-Time-Order Correlation at
  a Quantum Phase Transition}},
  \href{https://doi.org/10.1103/PhysRevB.96.054503}{\emph{Phys. Rev. B}
  {\bfseries 96} (2017) 054503}
  [\href{https://arxiv.org/abs/1608.02438}{{\ttfamily 1608.02438}}].

\bibitem{Keles:2018akp}
A.~Kele\c{s}, E.~Zhao and W.~V. Liu, \emph{{Scrambling dynamics and many-body
  chaos in a random dipolar spin model}},
  \href{https://doi.org/10.1103/PhysRevA.99.053620}{\emph{Phys. Rev. A}
  {\bfseries 99} (2019) 053620}
  [\href{https://arxiv.org/abs/1810.03815}{{\ttfamily 1810.03815}}].

\bibitem{dynamite}
G.~D. Kahanamoku-Meyer and J.~Wei, \emph{Gregdmeyer/dynamite: v0.3.1},  Mar.,
  2023.
\newblock 10.5281/zenodo.7706785.

\bibitem{park1986unitary}
T.~J. Park and J.~Light, \emph{Unitary quantum time evolution by iterative
  lanczos reduction}, {\emph{The Journal of chemical physics} {\bfseries 85}
  (1986) 5870}.

\bibitem{Goldstein:2005aib}
S.~Goldstein, J.~L. Lebowitz, R.~Tumulka and N.~Zanghi, \emph{{Canonical
  Typicality}},
  \href{https://doi.org/10.1103/PhysRevLett.96.050403}{\emph{Phys. Rev. Lett.}
  {\bfseries 96} (2006) 050403}
  [\href{https://arxiv.org/abs/cond-mat/0511091}{{\ttfamily
  cond-mat/0511091}}].

\bibitem{Luitz:2016kqa}
D.~J. Luitz and Y.~B. Lev, \emph{{The ergodic side of the many-body
  localization transition}},
  \href{https://doi.org/10.1002/andp.201600350}{\emph{Annalen Phys.} {\bfseries
  529} (2017) 1600350} [\href{https://arxiv.org/abs/1610.08993}{{\ttfamily
  1610.08993}}].

\bibitem{Garcia-Garcia:2016mno}
A.~M. Garc\'\i{}a-Garc\'\i{}a and J.~J.~M. Verbaarschot, \emph{{Spectral and
  thermodynamic properties of the Sachdev-Ye-Kitaev model}},
  \href{https://doi.org/10.1103/PhysRevD.94.126010}{\emph{Phys. Rev. D}
  {\bfseries 94} (2016) 126010}
  [\href{https://arxiv.org/abs/1610.03816}{{\ttfamily 1610.03816}}].

\bibitem{Shankar:2023pax}
A.~S. Shankar, M.~Fremling, S.~Plugge and L.~Fritz, \emph{{Tunable Lyapunov
  exponent in a Sachdev-Ye-Kitaev-type model}},
  \href{https://arxiv.org/abs/2302.08876}{{\ttfamily 2302.08876}}.

\bibitem{knuth1989concrete}
R.~L. Graham, D.~E. Knuth and O.~Patashnik, \emph{Concrete Mathematics: A
  Foundation for Computer Science}. Addison-Wesley, Reading, 1989.

\bibitem{frontera}
D.~Stanzione, J.~West, R.~T. Evans, T.~Minyard, O.~Ghattas and D.~K. Panda,
  \emph{Frontera: The evolution of leadership computing at the national science
  foundation},  in \emph{Practice and Experience in Advanced Research
  Computing}, PEARC '20, (New York, NY, USA), p.~106–111, Association for
  Computing Machinery, 2020,
  \href{https://doi.org/10.1145/3311790.3396656}{DOI}.

\end{thebibliography}\endgroup
    
\end{document}